\documentclass[prb,onecolumn,twoside,showpacs,superscriptaddress]{revtex4}
\usepackage{epsfig}
\usepackage{graphicx} 
\usepackage{amsmath}
\usepackage{longtable}
\usepackage{bm}         
\usepackage{color}      
\usepackage[normalem]{ulem}

\begin{document}
  \date{\today}


\preprint{}

\title{Radiation-induced segregation in dilute Re-W solid solutions.}


\author{Jan S. Wr\'obel}
\affiliation{CCFE, Culham Science Centre, Abingdon, Oxon OX14 3DB, UK}
\affiliation{Faculty of Materials Science and Engineering, Warsaw University of Technology, Wo\l{}oska 141, 02-811 Warsaw, Poland}
\author{Duc Nguyen-Manh}
\email[]{duc.nguyen@ukaea.uk}
\affiliation{CCFE, Culham Science Centre, Abingdon, Oxon OX14 3DB, UK}
\author{Krzysztof J. Kurzyd\l{}owski}
\affiliation{Faculty of Materials Science and Engineering, Warsaw University of Technology, Wo\l{}oska 141, 02-811 Warsaw, Poland} 
\author{Sergei L. Dudarev}
\affiliation{CCFE, Culham Science Centre, Abingdon, Oxon OX14 3DB, UK}

\begin{abstract}
The occurrence of segregation in highly dilute alloys under irradiation is an unusual phenomenon that has so far eluded theoretical explanation. The fact that solute atoms segregate in alloys that, according to thermodynamics, exhibit full solubility, has significant practical implications, as the formation of precipitates strongly affects physical and mechanical properties. Using {\it ab initio} calculations, we are able to explain the origin of radiation-induced rhenium segregation in dilute tungsten-rhenium alloys. The model treats rhenium atoms and vacancies in tungsten as components of a ternary alloy. The phase stability of ternary W-Re-Vac alloys is evaluated using a combination of Density Functional Theory (DFT) calculations, performed for more than 200 alloy structures, and cluster expansion (CE). The accuracy of CE parametrization is assessed against the DFT data, where the cross-validation error is found to be less than 4.2 meV/atom. The formation free energy of W-Re-Vac ternary alloys is evaluated as a function of temperature by means of quasi-canonical Monte Carlo simulations, using effective two, three and four-body cluster interaction parameters. In the low solute concentration range ($<$ 5 at$\%$Re), solute segregation is found in the form of Re atoms decorating vacancy clusters. These clusters remain stable over a broad temperature range from 800K to 1600K. At lower temperatures, simulations predict the formation of 30 to 50 at.$\%$ Re-rich precipitates. The origin of the anomalous vacancy-mediated segregation of Re atoms in W can be rationalized using {\it ab initio} data on binding energies as functions of Re to vacancy ratio as well as from the perspective of ``chemical'' effective pair-interaction between the rhenium atoms and vacancies. DFT analysis shows that binding energies can be as high as 1.5 eV if the rhenium to vacancy ratio is in the range from 2.4 to 6.6. Predictions derived from Monte Carlo simulations of Re precipitates are in surprisingly good agreement with experimental observations performed using  Atom Probe Tomography of self-ion irradiated W-2at.$\%$ Re alloys, as well as with Transmission Electron Microscopy investigations of neutron-irradiated W samples containing up to 1.4 at.$\%$ Re.
\end{abstract}


\maketitle


\section{Introduction\protect\\}
In the design of fusion power plants and fusion experimental devices, such as ITER, tungsten (W) is a candidate material
for various plasma-facing components because of its high melting temperature, high thermal conductivity, high resistance to sputtering and erosion \cite{Rieth2013}. W is also expected to be used as a material for divertor armour and other components of the first wall \cite{Maisonnier2006}. In relation to fusion applications, it is important to consider how tungsten behaves under neutron irradiation. In addition to the generation of radiation defects, neutrons initiate nuclear transmutation reactions, which change the atomic number and/or mass number of nuclei to their near-neighbours in the Periodic Table \cite{Gilbert2011}. Transmutation reactions can be detrimental or beneficial to mechanical and engineering properties of the original material. Rhenium (Re) is the main solid transmutation element formed in W under neutron irradiation over the entire expected period of divertor operation \cite{Gilbert2011}. It has long been recognized that room-temperature brittleness of tungsten can be alleviated by alloying it with rhenium \cite{Raffo1969}. Microstructural evolution, stimulated by neutron irradiation to 1.5 dpa and high temperature, in W, W-Re and W-Re-Os alloys were investigated in Refs. \cite{Hasegawa2014,He2006,Tanno2011,Hasegawa2011,Fukuda2013}. It was found that nucleation and growth of voids dominates evolution of microstructure in pure W, but the addition of Re clearly suppresses void formation. Hardening is caused by radiation-induced precipitation of $\sigma$ phase (WRe) and $\chi$ phase (WRe3) which form in W-5Re and W-10Re alloys after irradiation to 0.5-0.7 dpa at 600-1500$^{\circ}$C, see Ref. \cite{Tanno2011}. This happens despite the fact that the solubility limit of Re in W is high, close to 30 at.$\%$ \cite{Ekman2000}. Precipitation of $\alpha$-Mn phase in W-25Re alloy, resulting from exposure to neutron dose of several dpa, was reported in \cite{Sikka1974}. Platelet-like precipitates in neuron-irradiated W-10Re and W-25Re alloys were found using a combination of Field-Ion Microscopy (FIM) and Atom Probe Tomography (ATP) in Refs. \cite{Herschitz1984a,Herschitz1984b}. A recent TEM investigation \cite{Klimenkov2015} of neutron-irradiated polycrystalline and single crystal tungsten exposed to the neutron dose of 1.6 dpa at 900$^{\circ}$C, where transmutation reactions resulted in the accumulation of 1.4at.$\%$Re, shows the formation of rhenium clusters as well as voids decorated by rhenium.

Ion irradiation also provides means for investigating radiation damage effects, allowing for the accumulation of high damage dose, and at the same time avoiding sample activation \cite{Was2007,Yi2013}. Experimental examination of W-2 at.$\%$Re and W-1 at.$\%$Re-1 at.$\%$Os alloys, irradiated with 2MeV W$^{+}$ self-ions \cite{Armstrong2013,Armstrong2011} by Atom Probe Tomography (APT) and nano-hardness measurements, provide a direct evidence for the occurrence of radiation-induced segregation in under-saturated solid solutions \cite{Xu2015}. Experimental observations show the formation of Re-rich clusters, approximately 3nm in diameters, in the atom maps of Re solutes in a highly dilute 2 at$\%$ binary alloy irradiated at 773K to the dose of 33 dpa. In W-1 at.$\%$Re-1 at.$\%$Os alloy, just one percent rhenium concentration also appears sufficient to form Re-rich clusters under irradiation. The presence of Os suppresses the formation of Re precipitates, with Os-rich clusters forming instead. In addition to W alloys, radiation-induced segregation in under-saturated solid solutions was found also in Ni-Si, Ni-Ge and Al-Zn \cite{Martin1980,Silvestre1975,Barbu1977,Cauvin1979}. Segregation of Cr was found, using APT, in Fe-5at.$\%$Cr alloy, following self-ion irradiation at 400$^{\circ}$C\cite{Hardie2013}. This latter finding is particularly striking and significant since binary Fe-Cr alloys are a well known model alloy family \cite{Nguyen-Manh2008}, the thermodynamic properties of which are linked to the properties of ferritic/martensitic steels, which are the low-activation structural materials for fusion as well as for the new generation fission power plants \cite{Boutard2008}. Hence, we have extensive experimental evidence \cite{Klimenkov2015,Xu2015,Hardie2013} suggesting that irradiation stimulates the formation of secondary-phase precipitates in under-saturated iron and tungsten alloys that otherwise exhibit no evidence of precipitation under normal thermodynamic equilibrium conditions.

Interpretation of experimental observations of solute segregation effects in the context of theoretical models, linking solute segregation to binding between the defects and solute atoms, and/or different rates of diffusion of solute and solvent atoms, have so far focused primarily on kinetic approaches to the treatment of phase stability \cite{Martin1980,Okamoto1979,Cauvin1981,Krasnochtchekov2007}. Such models, treating solute fluctuations in solid solution under irradiation in combination with density functional theory (DFT) calculations
\cite{Becquart2012,Nguyen-Manh2012,Crocombette2012,Dudarev2013} provide insight into the highly complex phenomena of solute-defect trapping, solute segregation, point-defect recombination, dislocation interactions, nucleation and growth of voids. Similar models have recently been applied to W-based alloys \cite{Becquart2006,Becquart2007,Fitzgerald2008,Romaner2010,Muzyk2011,Kong2014,Zhou2014,Hofmann2015,Nguyen-Manh2015,Suzudo2015},
and iron alloys and steels \cite{Kuriplach2006,Olsson2007,Marinica2012,Ventelon2015,Schuler2015,Senninger2016}.

In this work, we adopt a different approach to modelling radiation-stimulated precipitation of solutes in irradiated alloys, where we treat point defects as extra ``chemical'' components within the framework of equilibrium thermodynamics of solid solutions. This makes it possible to understand what drives radiation-induced precipitation, using a thermodynamic viewpoint of free energy minimization, which is applied to an alloy containing defects produced by irradiation. The treatment overcomes a major difficulty encountered in the context of a kinetic approach, associated with ascertaining that a kinetic model, which requires information not only about energies of various configurations but also about the transition rates and defect mobilities, goes beyond an {\it ad hoc} explanation of experimental observations. In this paper, as a proof of concept, we consider vacancies as an additional component of a binary W-Re alloy, mapping the alloy onto a ternary alloy system containing tungsten and rhenium atoms, as well as vacancies.

Formation energies of interstitial defects in tungsten are about three times the vacancy formation energy \cite{Muzyk2011,Nguyen-Manh2015,Nguyen-Manh2006}. Hence it is natural to first consider vacancies when evaluating the free energy of an alloy subjected to irradiation. Earlier models for the free energy of W-Re alloys, taking into account a contribution from point defects were either based on the relatively simple effective pair interaction model of Bragg and Williams \cite{Bragg1934,Kaufman1977}
or on higher-order approximations taking into account multisite correlations, including for example the Cluster Variation Method (CVM)\cite{Kikuchi1966,Bocquet1979}.

In this paper, we show that radiation-induced segregation of Re solute atoms in W can be analyzed using a combination of first-principles calculations and statistical mechanics simulations based on a generalized Ising alloy model, known as Cluster Expansion (CE) \cite{Sanchez1984,vandeWalle2009}. CE makes it possible to explore the phase stability of magnetic fcc and bcc Fe-Cr-Ni ternary alloys \cite{Wrobel2015} as well as to evaluate the free energy of five-component W-Ta-Mo-Nb-V high-entropy alloys
\cite{Toda-Caraballo2015}. A ternary CE model based on DFT calculations has been applied to investigate the thermodynamic properties of vacancies in fcc Cu-Ni at equilibrium \cite{Zhang2015}. Under irradiation, microstructural evolution associated with the formation and growth of voids in supersaturated solutions of vacancies was investigated not only in the context of classical phase-field approach \cite{Semenov2012} but also using first-principles approaches to heterogeneous nucleation \cite{Kato2011}, where in the latter case the vacancy content was close to 0.1 at$\%$.

The paper is organized as follows. In Section II, the CE formalism, which includes the treatment of vacancies as an alloy component, is applied to W-Re-Vac ternary alloy. In Section III, DFT data for binding energies of Re-Vac clusters in tungsten are analyzed
as functions of the Re/Vac ratio. Section IV focuses on finite-temperature effects, where rhenium clustering is investigated as a function of Re and vacancy solute concentrations using quasi-canonical Monte Carlo simulations. In Section V, we discuss the origin of radiation-induced segregation in an undersaturated Re-W-Vac alloy and compare theoretical predictions with experimental APT and TEM observations. 

\section{Computational methodology}

\subsection{Cluster Expansion formalism for ternary W-Re-Vac alloys}

We describe the state of a solute solution under irradiation by three concentrations: $x_{A}$, $x_{B}$, $x_{C}$ which are, respectively, the solvent, solute and vacancy concentrations in the number per lattice site units. The three concentrations are subject to condition
\begin{equation}
x_{A} + x_{B} + x_{C}  = 1 .
\label{eq:conc_conservation}
\end{equation}

In Cluster Expansion (CE), the configuration enthalpy of mixing of a multi-component alloy is defined as \cite{vandeWalle2009}
\begin{equation}
\Delta H_{CE}(\vec{\sigma}) = \sum_{\omega}m_{\omega}J_{\omega}\left\langle \Gamma_{\omega'}(\vec{\sigma})\right\rangle_\omega ,
\label{eq:CE_1}
\end{equation}
where summation is performed over all the clusters $\omega$ that are distinct under group symmetry operations applied to the underlying lattice, $m^{lat}_\omega$ are the multiplicities indicating the number of clusters equivalent to $\omega$ by symmetry, divided by the number of lattice sites, and $\left\langle \Gamma_{\omega'}(\vec{\sigma})\right\rangle$ are the cluster functions defined as products of \textit{functions} of occupation variables on a specific cluster $\omega$ averaged over all the clusters $\omega'$ that are equivalent by
symmetry to cluster $\omega$. $J_{\omega}$ are the concentration-independent Effective Cluster Interaction (ECI) parameters, derived from a set of \textit{ab-initio} calculations using the structure inversion method (SIM) \cite{Connolly1983}. A cluster $\omega$ is defined by its size (i.e. the number of lattice points) $|\omega|$ and relative positions of points. For clarity, each cluster $\omega$ is described by two parameters $(|\omega|,n)$, where $|\omega|$ is the cluster size and $n$ is a label, defined in Table \ref{tab:ECI_ternary} for bcc lattice.

In CE developed for a $K$-component system, a cluster function is not a simple product of occupation variables, $\{\sigma_i\}$. Instead, it is defined as a product of orthogonal point functions $\gamma_{j_i,K}(\sigma_i)$,
\begin{equation}
\Gamma_{\omega,n}^{(s)}(\vec{\sigma}) = \gamma_{j_1,K}(\sigma_1)\gamma_{j_2,K}(\sigma_2)\ldots\gamma_{j_{|\omega|},K}(\sigma_{|\omega|}),
\label{eq:cluster_function}
\end{equation}
where sequence $(s) =(j_{1} j_{2} \ldots\ j_{|\omega|})$ is the decoration \cite{Sandberg2007} of a cluster by point functions. We use the following definition of point functions for a $K$-component system
\begin{equation}
\gamma_{j,K}\left(\sigma_i\right)=\begin{cases}
 1 & \textrm{ if }j=0\textrm{ }, \\
 -\cos\left(2\pi\lceil\frac{j}{2}\rceil\frac{\sigma_i}{K}\right) & \textrm{ if }j>0\textrm{ and odd},  \\
 -\sin\left(2\pi\lceil\frac{j}{2}\rceil\frac{\sigma_i}{K}\right) & \textrm{ if }j>0\textrm{ and even},
\end{cases}
\label{eq:point_function}
\end{equation}
where $\sigma_i = 0,1,2,\ldots,\left(K-1\right)$, $j$ is the index of point functions ($j=0,1,2,\ldots,(K-1)$). In a ternary alloy, index $K$ equals 3 and occupation variables are defined as $\sigma=0,1,2$, referring to the constituent components of the alloy \textit{A}, \textit{B} and \textit{C}, which here correspond to W, Re and Vac.

Following the mathematical derivation given in\cite{Wrobel2015}, the configuration enthalpy of mixing of a ternary alloy can be expressed
analytically up to the three-body interactions as a function of concentrations, $x_i$, and the average pair and 3-body probabilities, $y_n^{ij}$ and  $y_n^{ijk}$ as
\begin{widetext}
\begin{eqnarray}
\Delta H_{CE}(\vec{\sigma}) &=& J_1^{(0)}+J_1^{(1)}\left(1-3x_A\right) + J_1^{(2)}\frac{\sqrt{3}}{2}\left(x_C-x_B\right) \nonumber \\
&+&\sum_{n}^{pairs} \left[\frac{1}{4}m_{2,n}^{(11)}J_{2,n}^{(11)}\left(1+3y_n^{AA}-6y_n^{AB}-6y_n^{AC}\right)  \right. \nonumber \\
&+& \frac{\sqrt{3}}{4}m_{2,n}^{(12)}J_{2,n}^{(12)}\left(-y_n^{BB}+y_n^{CC}+2y_n^{AB}-2y_n^{AC}\right) + \left.\frac{3}{4}m_{2,n}^{(22)}J_{2,n}^{(22)}\left(y_n^{BB}+y_n^{CC}-2y_n^{BC}\right) \right] \nonumber \\
&+& \sum_{n}^{triples} \left[ \frac{1}{8}m_{3,n}^{(111)}J_{3,n}^{(111)}\left(-8y_n^{AAA}+12y_n^{AAB}+12y_n^{AAC}
\right.\right. \nonumber \\
&-& \left. 6y_n^{ABB}-6y_n^{ABC}-6y_n^{ACC}+y_n^{BBB}+3y_n^{BBC}+3y_n^{BCC}+y_n^{CCC}\right) \nonumber \\
&+& \frac{\sqrt{3}}{8}\left(m_{3,n}^{(112)}J_{3,n}^{(112)}+m_{3,n}^{(121)}J_{3,n}^{(121)}+m_{3,n}^{(211)}J_{3,n}^{(211)}\right)  \nonumber \\
&\cdot& \left(-4y_n^{AAB}+4y_n^{AAC} + 4y_n^{ABB}-4y_n^{ACC}-y_n^{BBB}-y_n^{BBC}+y_n^{BCC}+y_n^{CCC}\right) \nonumber \\
&+& \frac{3}{8}\left(m_{3,n}^{(122)}J_{3,n}^{(122)}+m_{3,n}^{(212)}J_{3,n}^{(212)}+m_{3,n}^{(221)}J_{3,n}^{(221)}\right) \nonumber \\
&\cdot & \left(-2y_n^{ABB}+2y_n^{ABC}-2y_n^{ACC}+y_n^{BBB}-y_n^{BBC}-y_n^{BCC}+y_n^{CCC}\right) \nonumber \\
&+& \left. \frac{3\sqrt{3}}{8}m_{3,n}^{(222)}J_{3,n}^{(222)}\left(-y_n^{BBB}+3y_n^{BBC}-3y_n^{BCC}+y_n^{CCC} \right)
 \right] \nonumber \\
&+& \sum_{n}^{multibody} \ldots
\label{eq:CE_expanded}
\end{eqnarray}
\end{widetext}

We now apply the above treatment to the problem of radiation-induced precipitation in bcc W-rich alloys where anomalous segregation of Re atoms was investigated experimentally \cite{Klimenkov2015,Xu2015}. The enthalpy of mixing of any W-Re-Vac structure is found by subtracting the total enthalpy from the pure end terms as:
\begin{eqnarray}
\Delta H_{mix}\left(x_W,x_{Re},x_{Vac}\right) &=& E_{tot}\left(x_W,x_{Re},x_{Vac}\right) - x_{W}E_{tot}\left(W\right) - x_{Re}E_{tot}\left(Re\right) - x_{vac}E_{tot}\left(Vac\right) \nonumber \\
&=& E_{tot}\left(x_W,x_{Re},x_{Vac}\right) - x_{W}E_{tot}\left(W\right) - x_{Re}E_{tot}\left(Re\right) ,
\label{eq:mixing_enthalpy}
\end{eqnarray}
where the reference enthalpy of a vacancy is assumed to be zero, $E_{tot}\left(Vac\right)=0$.

\begin{figure}
  \includegraphics[width=\linewidth]{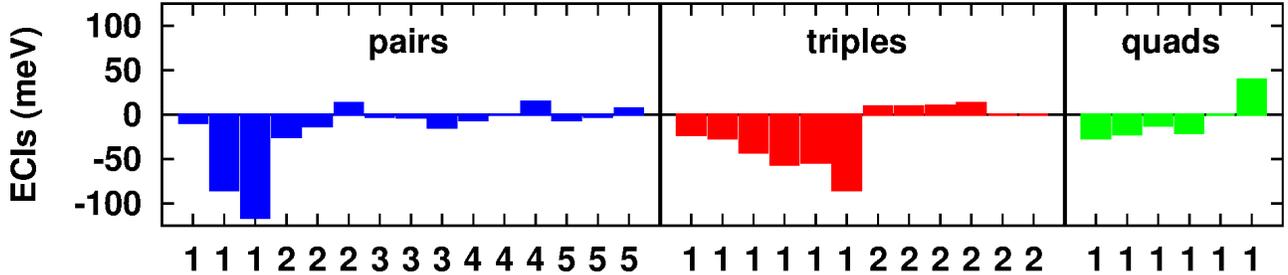}
                        \caption{
                Effective cluster interactions in ternary W-Re-Vac alloy. }
                \label{fig:ECI_WReVac}
                \end{figure}

The values of ECIs ($J_{|\omega|,n}^{(s)}$) for ternary W-Re-Vac bcc alloys are derived by mapping DFT energies onto CE for 224 structures, which are described in the next sections of this paper. According to the definition of cluster functions (Eq. \ref{eq:cluster_function}) and point functions (Eq. \ref{eq:point_function}) for ternary alloys, each cluster can be decorated in various ways for each nearest-neighbour shell of interactions. Here, we have used a set of 15 two-body, 12 three-body, and 6 four-body clusters on bcc lattice. Values of all the optimized ECIs for ternary W-Re-Vac alloys are shown in Fig. \ref{fig:ECI_WReVac} and given in the last column of Table \ref{tab:ECI_ternary}. Section III shows that the inclusion of five nearest neighbour shells in two-body interactions is essential for understanding the origin of Re-vacancy binding energy trends characterizing under-saturated solid solutions. Cross-validation error between DFT and CE has been assessed in comparison with DFT data for the interactions between vacancies and between Re and vacancy clusters, and the cross-validation error is 4.12 meV/atom for the present set of ECIs.

\begin{table*}
\caption{Size $|\omega|$, label $n$, decoration $(s)$, multiplicity $m_{|\omega|,n}^{(s)}$ and coordinates of points in clusters on bcc lattice. $J_{|\omega|,n}^{(s)}$ (in meV) are effective cluster interactions (ECIs) calculated in the framework of CE for bcc ternary W-Re-vacancy system. Index $(s)$ is the same as the sequence of points in the corresponding cluster.
        \label{tab:ECI_ternary}}
\begin{ruledtabular}
    \begin{tabular}{cccccc}
    $|\omega|$ & $n$ & ($s$) & Coordinates & $m_{|\omega|,n}^{(s)}$ & $J_{|\omega|,n}^{(s)}$  \\
    \hline
    0     &       & (0)   &       & 1     & 761.945 \\
    1     &       & (1)   & $(0,0,0)$ & 1     & 605.955 \\
    1     &       & (2)   &       & 1     & 1623.449 \\
    2     & 1     & (1,1) & $(0,0,0; \frac{1}{2},\frac{1}{2},\frac{1}{2})$ & 4     & -1.259 \\
          &       & (1,2) &       & 8     & -89.783 \\
          &       & (2,2) &       & 4     & -174.961 \\
    2     & 2     & (1,1) & $(0,0,0; 1,0,0)$ & 3     & -11.955 \\
          &       & (1,2) &       & 6     & -3.381 \\
          &       & (2,2) &       & 3     & 17.514 \\
    2     & 3     & (1,1) & $(0,0,0; 1,0,1)$ & 6     & 2.212 \\
          &       & (1,2) &       & 12    & 5.341 \\
          &       & (2,2) &       & 6     & -1.272 \\
    2     & 4     & (1,1) & $(0,0,0; 1\frac{1}{2},\frac{1}{2},\frac{1}{2})$ & 12    & -4.697 \\
          &       & (1,2) &       & 24    & 1.652 \\
          &       & (2,2) &       & 12    & 16.947 \\
    2     & 5     & (1,1) & $(0,0,0; 1,1,1)$ & 4     & -6.121 \\
          &       & (1,2) &       & 8     & -2.155 \\
          &       & (2,2) &       & 4     & 6.446 \\
    3     & 1     & (1,1,1) & $(1,0,0; \frac{1}{2},\frac{1}{2},\frac{1}{2}$; & 12    & -12.552 \\
          &       & (2,1,1) & $0,0,0)$ & 24    & -16.271 \\
          &       & (1,2,1) &       & 12    & -26.455 \\
          &       & (2,2,1) &       & 24    & -46.238 \\
          &       & (2,1,2) &       & 12    & -44.102 \\
          &       & (2,2,2) &       & 12    & -92.177 \\
    3     & 2     & (1,1,1) & $(\frac{1}{2},-\frac{1}{2},-\frac{1}{2}$; 0,0,0; & 12    & 7.038 \\
          &       & (2,1,1) &  $-\frac{1}{2},-\frac{1}{2},\frac{1}{2})$ & 24    & 5.672 \\
          &       & (1,2,1) &       & 12    & 0.541 \\
          &       & (2,2,1) &       & 24    & -4.069 \\
          &       & (2,1,2) &       & 12    & -7.373 \\
          &       & (2,2,2) &       & 12    & -31.539 \\
    4     & 1     & (1,1,1,1) & $(1,0,0; \frac{1}{2},-\frac{1}{2},\frac{1}{2}$; & 6     & -18.802 \\
          &       & (2,1,1,1) & $\frac{1}{2},\frac{1}{2},\frac{1}{2}; 0,0,0)$ & 24    & -12.881 \\
          &       & (2,2,1,1) &       & 24    & -4.480 \\
          &       & (1,2,2,1) &       & 12    & -8.310 \\
          &       & (2,2,2,1) &       & 24    & 0.000 \\
          &       & (2,2,2,2) &       & 6     & 5.120 \\
    \end{tabular}%
\end{ruledtabular}
\end{table*}

\subsection{Computational details}

DFT calculations were performed using Vienna Ab-initio Simulation Package (VASP) with the interaction between ions and electrons described using the projector augmented waves (PAW) method \cite{Kresse1996a, Kresse1996b}. Exchange and correlation were treated in the generalized gradient approximation GGA-PBE \cite{Perdew1996}, with PAW potentials containing semi-core $p$ electron contributions. Supercell calculations were performed considering vacancy clusters interacting with Re atoms in bcc W lattice under constant pressure conditions, with structures optimized by relaxing both atomic positions as well as the shape and volume of the supercell. To treat clusters containing from 2 to 47 sites with Re atoms and vacancies, orthogonal supercells containing 128 and 250 atoms were used in calculations. Total energies were calculated using the Monkhorst-Pack mesh\cite{Monkhorst1976} of $k$-points in the Brillouin zone, with the $k$-mesh spacing of 0.15 $\AA^{-1}$.
This corresponds to 4$\times$4$\times$4 or 3$\times$3$\times$3 $k$-point meshes for a bcc supercell of 4$\times$4$\times$4 or 5$\times$5$\times$5 bcc structural units, respectively. The plane wave cut-off energy was 400 eV. The total energy convergence criterion was set to 10$^{-6}$ eV/cell, and force components were relaxed to 10$^{-3}$ eV/$\AA$. Vacancy and self-interstitial-atom (crowdion)
formation energies calculated using the above conditions for pure bcc W were 3.307 eV and 10.917 eV, respectively.

\begin{figure}
  \includegraphics[width=\linewidth]{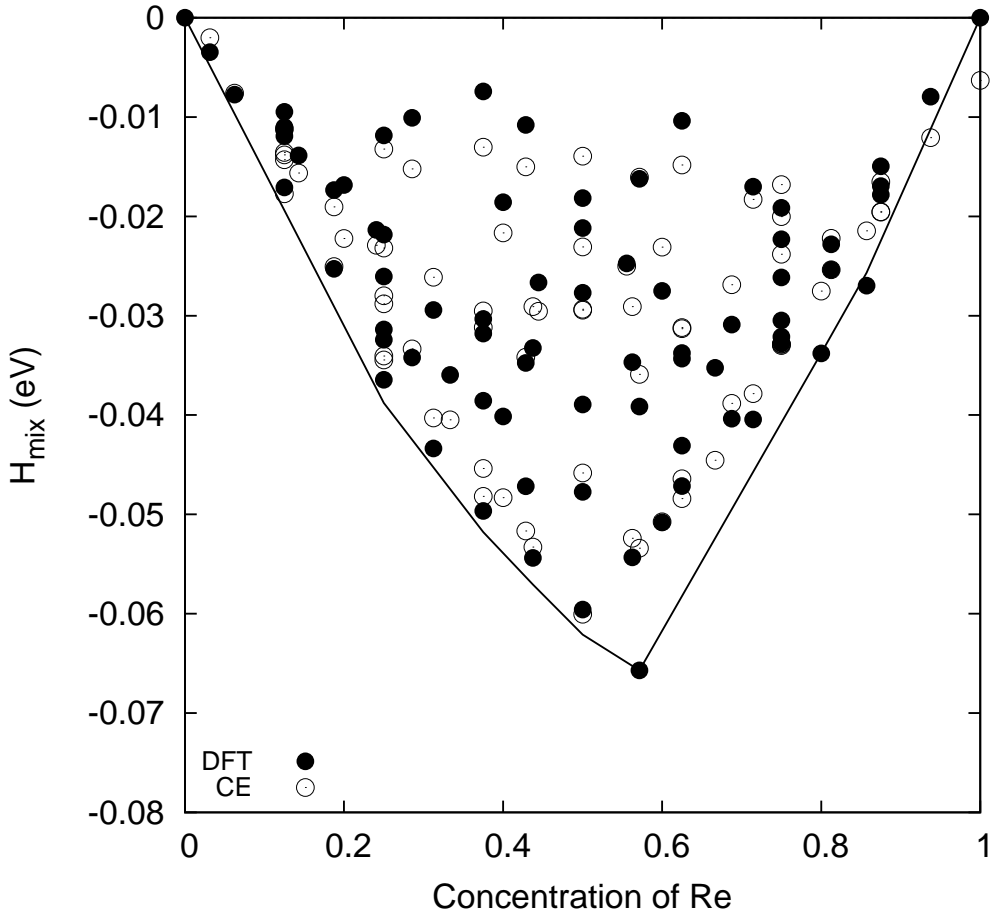}	
			\caption{
		Enthalpy of mixing of W-Re structures calculated using DFT and CE. }
		\label{fig:Hmix_WRe}
		\end{figure}

Mapping of DFT energies to CE was performed using the ATAT package\cite{Walle2002}. For binary bcc alloys we used 58 structures from Ref. \cite{Nguyen-Manh2007}, and the corresponding results for the enthalpy of mixing in the binary W-Re are shown in Figure \ref{fig:Hmix_WRe} and Table \ref{tab:Hmix_results}. We find that the  negative values of enthalpy of mixing of bcc W-Re alloys were smaller than those obtained using similar DFT/CE calculations for binary W-Ta and W-V alloys \cite{Muzyk2011}. The most stable configurations predicted for W-Re binary alloys remain the same in the extended treatment of ternary W-Re-Vac alloy system. This is discussed in the last section of the paper (see Table \ref{tab:Vij_ternary}), while for the first nearest neighbour (1NN) shell the effective chemical interaction between
W and Re is attractive  but is more than 5 times smaller than those between Re and vacancy. For the second nearest neighbour (2NN) shell the W-Re effective interaction becomes repulsive as opposed to very strong Re-Vac attractive interaction. Therefore, we conclude that clustering of Re in bcc W is driven primarily by binding between Re atoms and vacancies. 

The DFT database of structures used in the present study uses data not only on W-Re binary system but also the data on vacancy clusters in W (W-Vac) \cite{Muzyk2011}, as well as data on Re-vacancy cluster interactions in W. 

Quasi-canonical MC simulations were carried out using the ATAT package \cite{Walle2002}. All the simulations described in Section IV used a 30$\times$30$\times$30 bcc supercell containing 54000 lattice sites. To investigate the formation free energy for each composition, simulations were performed starting from a disordered high-temperature state, corresponding to $T$ = 3000 K. Configurations were then cooled down with the temperature step of $\Delta T$ = 10K, with 3000 MC steps per atom at both thermalization and accumulation stages. It was found that it was not necessary to use a higher number of MC steps to achieve convergence.

\begin{table*}
\caption{Enthalpies of mixing of binary W-Re structures $E_{mix}^{DFT}$, in eV, calculated using DFT, and $E_{mix}^{CE}$ derived using CE. $x_{Re}$ is Re solution concentration.
        \label{tab:Hmix_results}}
\begin{ruledtabular}
    \begin{tabular}{ccc|ccc}
    $x_{Re}$ & $E_{mix}^{DFT}$ & $E_{mix}^{CE}$ & $c_{Re}$ & $E_{mix}^{DFT}$ & $E_{mix}^{CE}$ \\
    \hline
    0.000 & 0.000 & 0.001 & 0.500 & -0.048 & -0.046 \\
    1.000 & 0.000 & -0.006 & 0.500 & -0.028 & -0.029 \\
    0.500 & -0.060 & -0.060 & 0.625 & -0.043 & -0.048 \\
    0.333 & -0.036 & -0.040 & 0.625 & -0.010 & -0.015 \\
    0.667 & -0.035 & -0.045 & 0.625 & -0.034 & -0.031 \\
    0.500 & -0.021 & -0.023 & 0.750 & -0.019 & -0.017 \\
    0.500 & -0.039 & -0.029 & 0.750 & -0.030 & -0.033 \\
    0.750 & -0.022 & -0.020 & 0.875 & -0.017 & -0.020 \\
    0.750 & -0.032 & -0.033 & 0.444 & -0.027 & -0.030 \\
    0.750 & -0.033 & -0.033 & 0.556 & -0.025 & -0.025 \\
    0.250 & -0.031 & -0.034 & 0.313 & -0.043 & -0.040 \\
    0.250 & -0.036 & -0.034 & 0.375 & -0.050 & -0.048 \\
    0.250 & -0.032 & -0.028 & 0.063 & -0.008 & -0.008 \\
    0.400 & -0.040 & -0.048 & 0.438 & -0.054 & -0.053 \\
    0.400 & -0.019 & -0.022 & 0.563 & -0.054 & -0.052 \\
    0.600 & -0.051 & -0.051 & 0.125 & -0.017 & -0.018 \\
    0.600 & -0.028 & -0.023 & 0.625 & -0.047 & -0.046 \\
    0.800 & -0.034 & -0.028 & 0.688 & -0.040 & -0.039 \\
    0.200 & -0.017 & -0.022 & 0.813 & -0.025 & -0.025 \\
    0.286 & -0.010 & -0.015 & 0.875 & -0.018 & -0.020 \\
    0.286 & -0.034 & -0.033 & 0.938 & -0.008 & -0.012 \\
    0.429 & -0.035 & -0.034 & 0.188 & -0.025 & -0.025 \\
    0.429 & -0.047 & -0.052 & 0.125 & -0.011 & -0.014 \\
    0.429 & -0.011 & -0.015 & 0.188 & -0.017 & -0.019 \\
    0.571 & -0.039 & -0.036 & 0.250 & -0.022 & -0.023 \\
    0.571 & -0.066 & -0.053 & 0.313 & -0.029 & -0.026 \\
    0.571 & -0.016 & -0.016 & 0.375 & -0.032 & -0.031 \\
    0.714 & -0.017 & -0.018 & 0.438 & -0.033 & -0.029 \\
    0.714 & -0.040 & -0.038 & 0.563 & -0.035 & -0.029 \\
    0.143 & -0.014 & -0.016 & 0.625 & -0.034 & -0.031 \\
    0.857 & -0.027 & -0.021 & 0.688 & -0.031 & -0.027 \\
    0.250 & -0.012 & -0.013 & 0.750 & -0.026 & -0.024 \\
    0.250 & -0.026 & -0.029 & 0.813 & -0.023 & -0.022 \\
    0.375 & -0.039 & -0.045 & 0.875 & -0.015 & -0.017 \\
    0.375 & -0.007 & -0.013 & 0.241 & -0.021 & -0.023 \\
    0.375 & -0.030 & -0.030 & 0.031 & -0.003 & -0.002 \\
    0.125 & -0.011 & -0.014 & 0.125 & -0.009 & -0.011 \\
    0.500 & -0.018 & -0.014 & 0.125 & -0.012 & -0.014 \\
    \end{tabular}%
\end{ruledtabular}
\end{table*}

\section{Re-Vacancy binding energies at 0 K}

Binding energy of a cluster containing $n$ vacancies and $m$ Re atoms in W matrix is the energy required to dissolve the cluster into $n$ separate vacancies and $m$ separate Re atoms in W matrix. It can be defined as \cite{Domain2005}:

\begin{equation}
E_{bind}\left(n\cdot Vac,m\cdot Re\right) = nE_{tot}\left(1\cdot Vac\right) + mE_{tot}\left(1\cdot Re\right) - E_{tot}\left(n\cdot Vac,m\cdot Re\right) - \left(n+m-1\right)E_{tot}\left(W\right) ,
\label{eq:binding_energy_2}
\end{equation}
where $E_{tot}\left(n\cdot Vac,m\cdot Re\right)$ is the total energy of a supercell of W atoms containing a cluster with $n$ vacancies and $m$ Re atoms, $E_{tot}\left(1\cdot Vac\right)$ is the total energy of a supercell with one vacancy, $E_{tot}\left(1\cdot Re\right)$ is the energy of a supercell with one Re atom, and $E_{tot}\left(W\right)$ is the total energy of a supercell of containing only W atoms. All the supercells contain the same number of sites.

\begin{figure*}
                        \centering
                        \begin{minipage}{.22\textwidth}
                                \centering
                                a) \includegraphics[width=.9\linewidth]{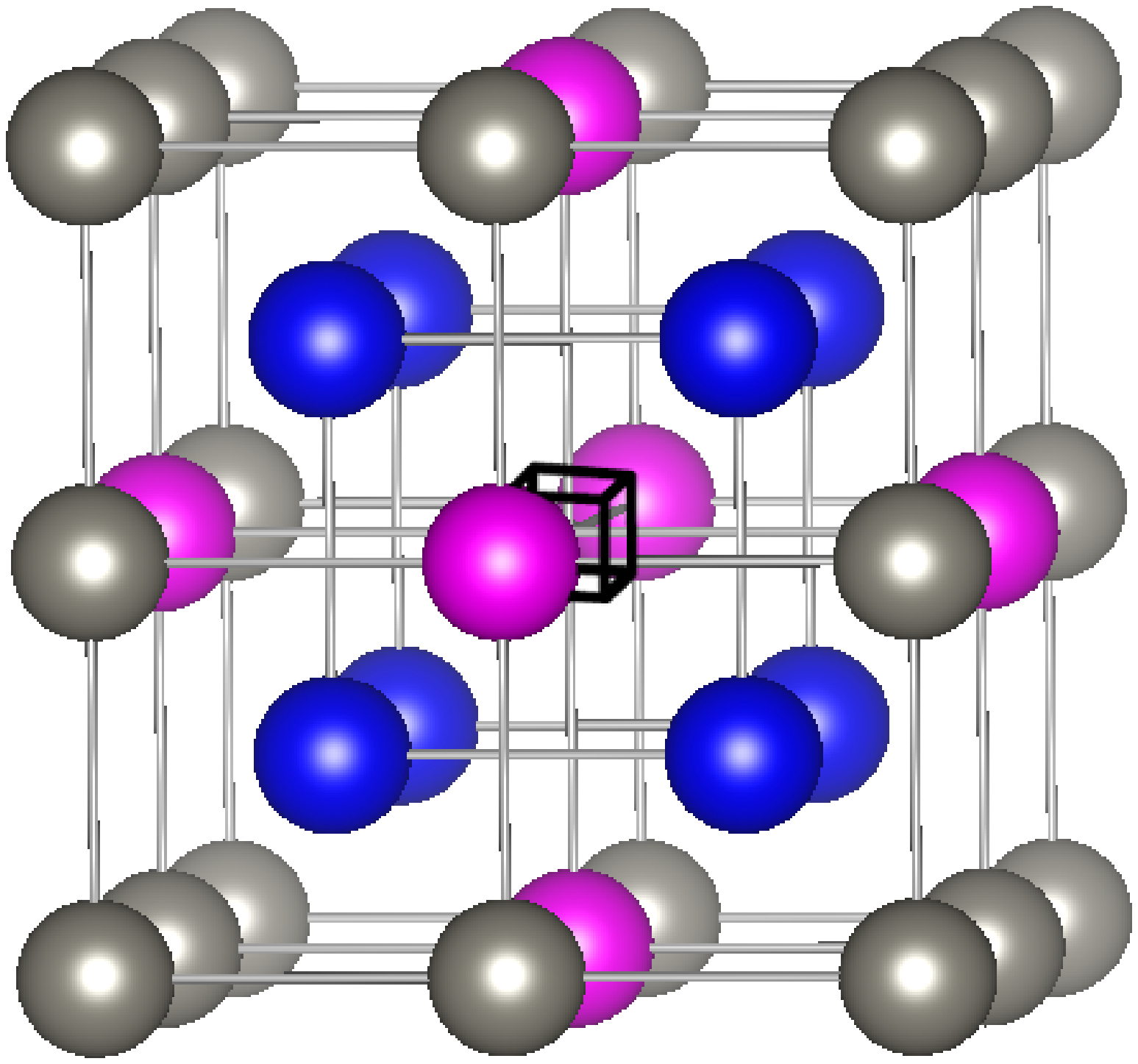}
                        \end{minipage}%
                        \begin{minipage}{.23\textwidth}
                                \centering
                                b) \includegraphics[width=.9\linewidth]{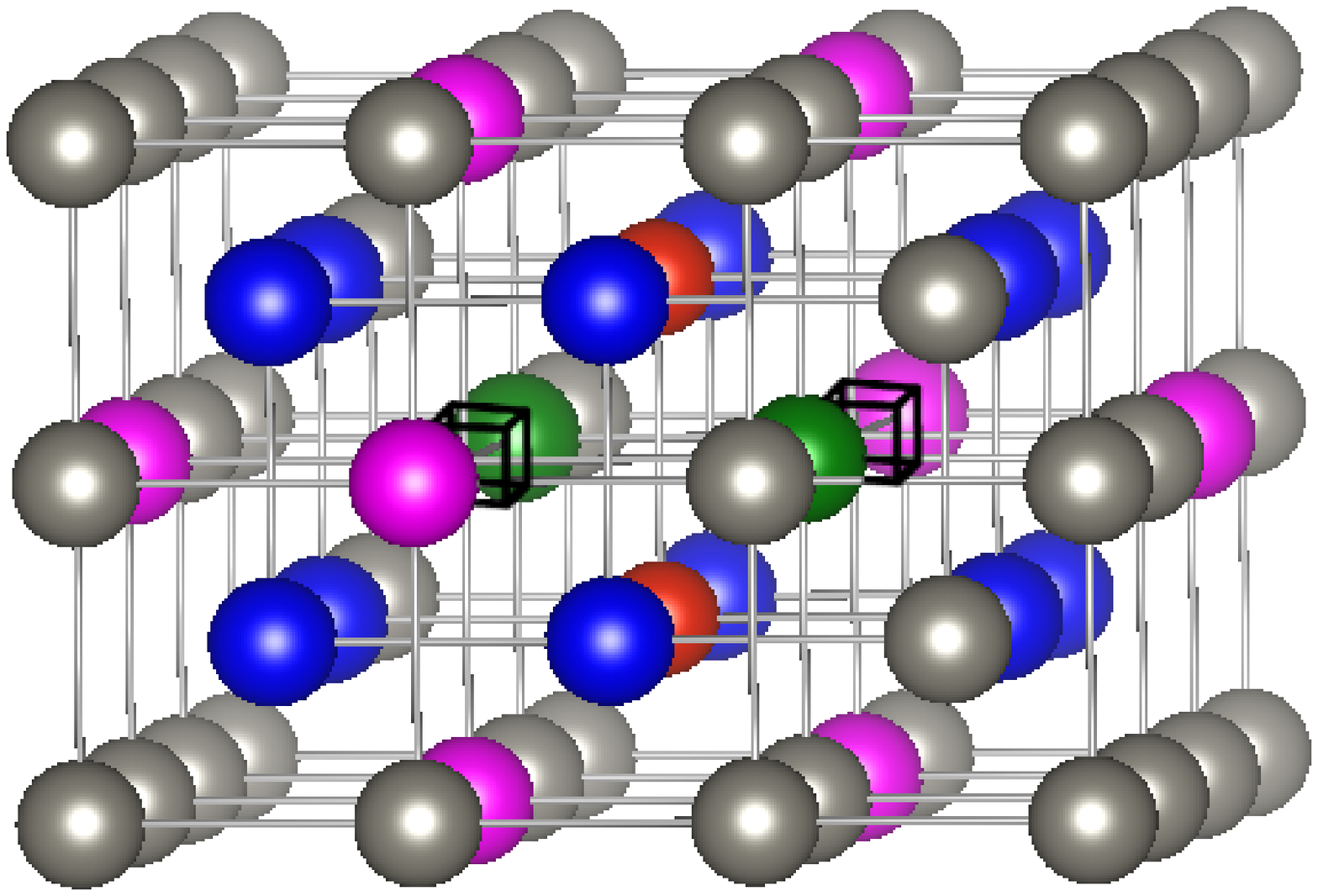}
                        \end{minipage}
                        \begin{minipage}{.22\textwidth}
                                \centering
                                c) \includegraphics[width=.9\linewidth]{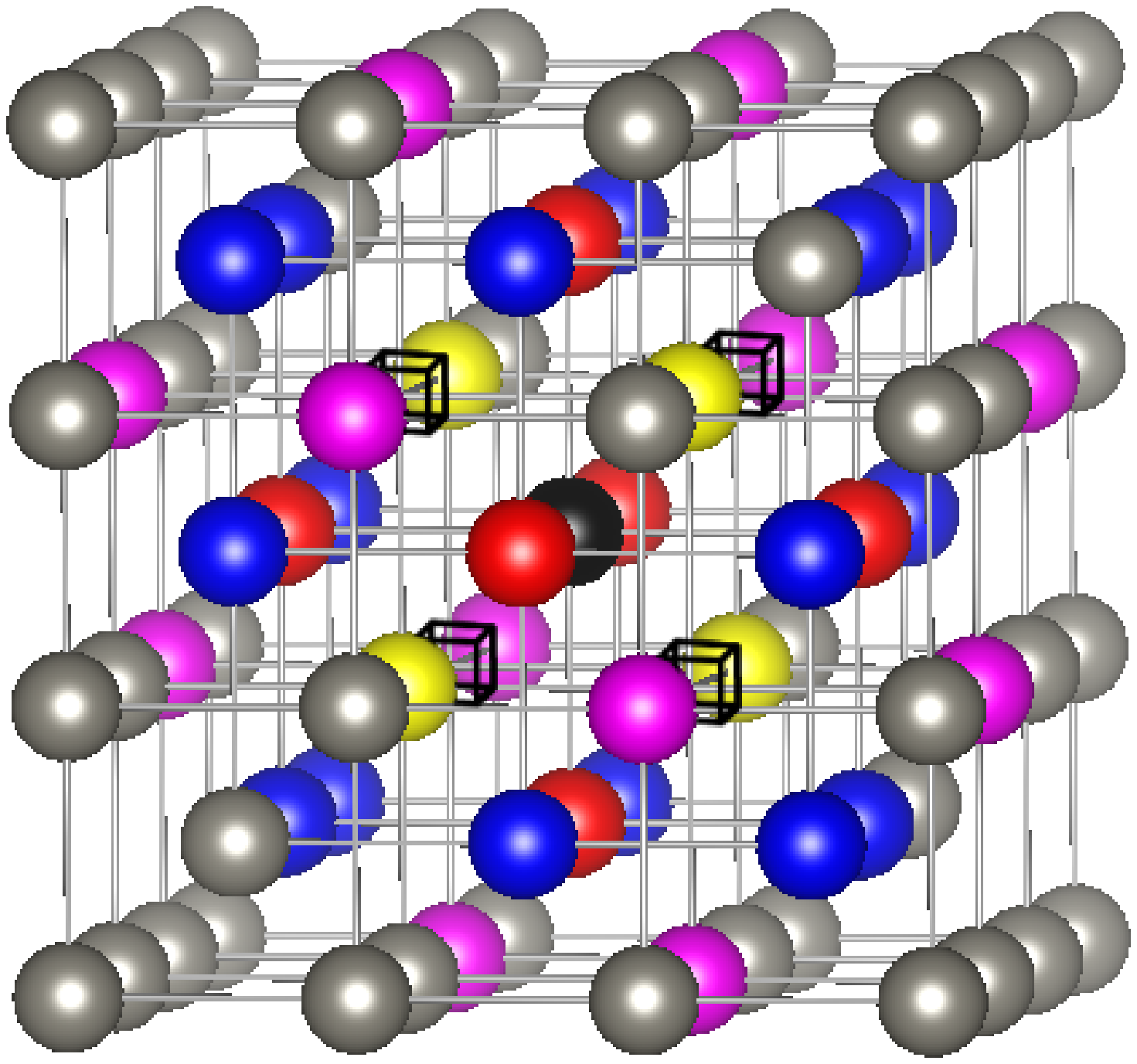}
                        \end{minipage}%
                        \begin{minipage}{.23\textwidth}
                                \centering
                                d) \includegraphics[width=.9\linewidth]{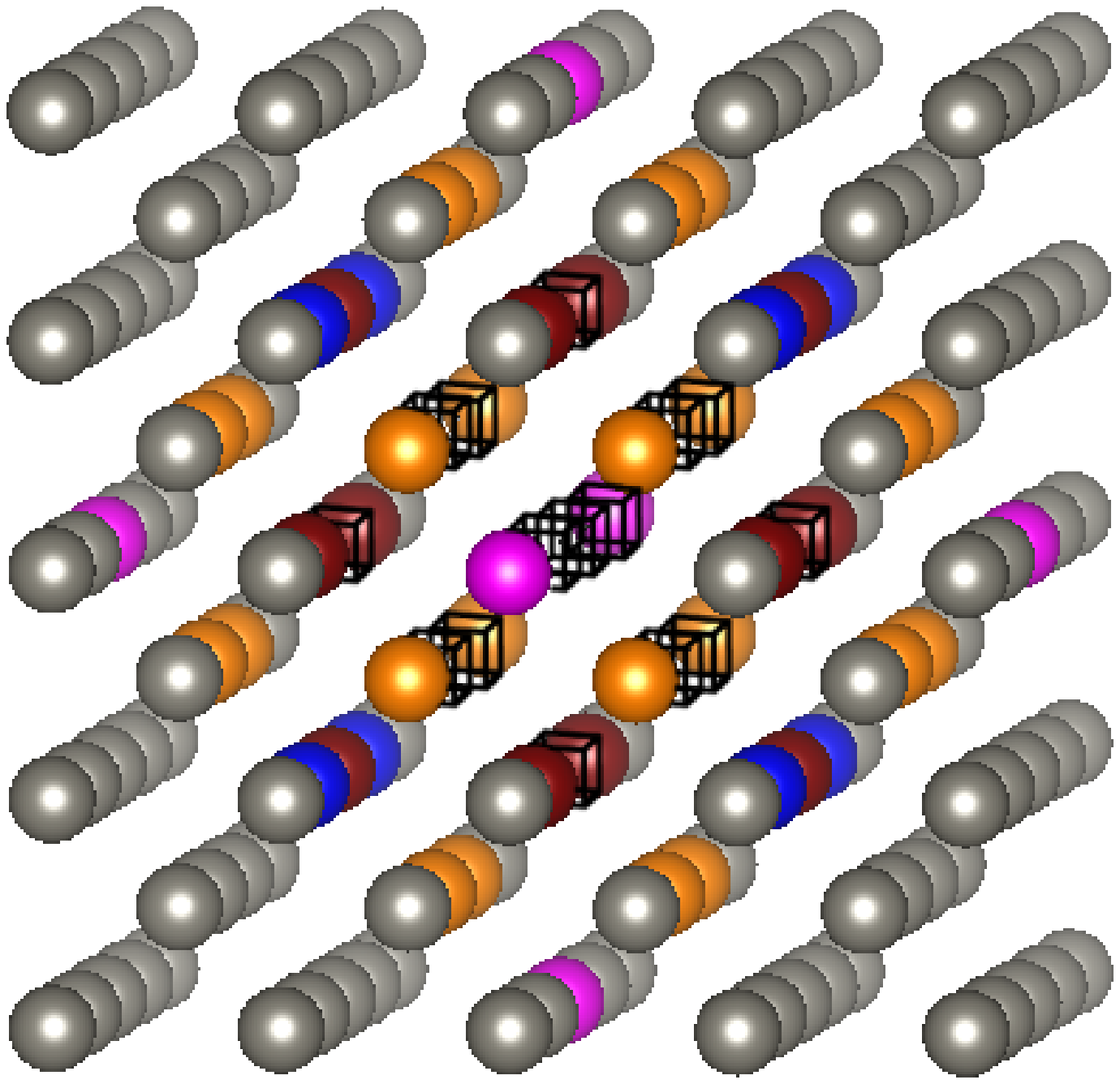}
                        \end{minipage}
                        \caption{
                (Color online) Schematic representation of atoms in the neighbourhood of (a) a vacancy, (b) a di-vacancy in the third nearest neighbourhood, a four-vacancy cluster (Fig. \ref{fig:Schematic_vacancy_clusters}h) and (d) a void (Fig. \ref{fig:Schematic_vacancy_clusters}h) . Colours indicate the number of vacancies in the first (1NN) and second (2NN) neighbourhood of a given atom: black - four 1NN and zero 2NN; brown - two 1NN and two 2NN; red - two 1NN and zero 2NN; orange - one 1NN and one 2NN; blue - one 1NN and zero 2NN; yellow - zero 1NN and three 2NN; green - zero 1NN and two 2NN; pink - zero 1NN and one 2NN.}
                \label{fig:Schematic shells}
                \end{figure*}

\subsection{Interaction of a mono-vacancy with Re}

\begin{figure*}
  \includegraphics[width=\linewidth]{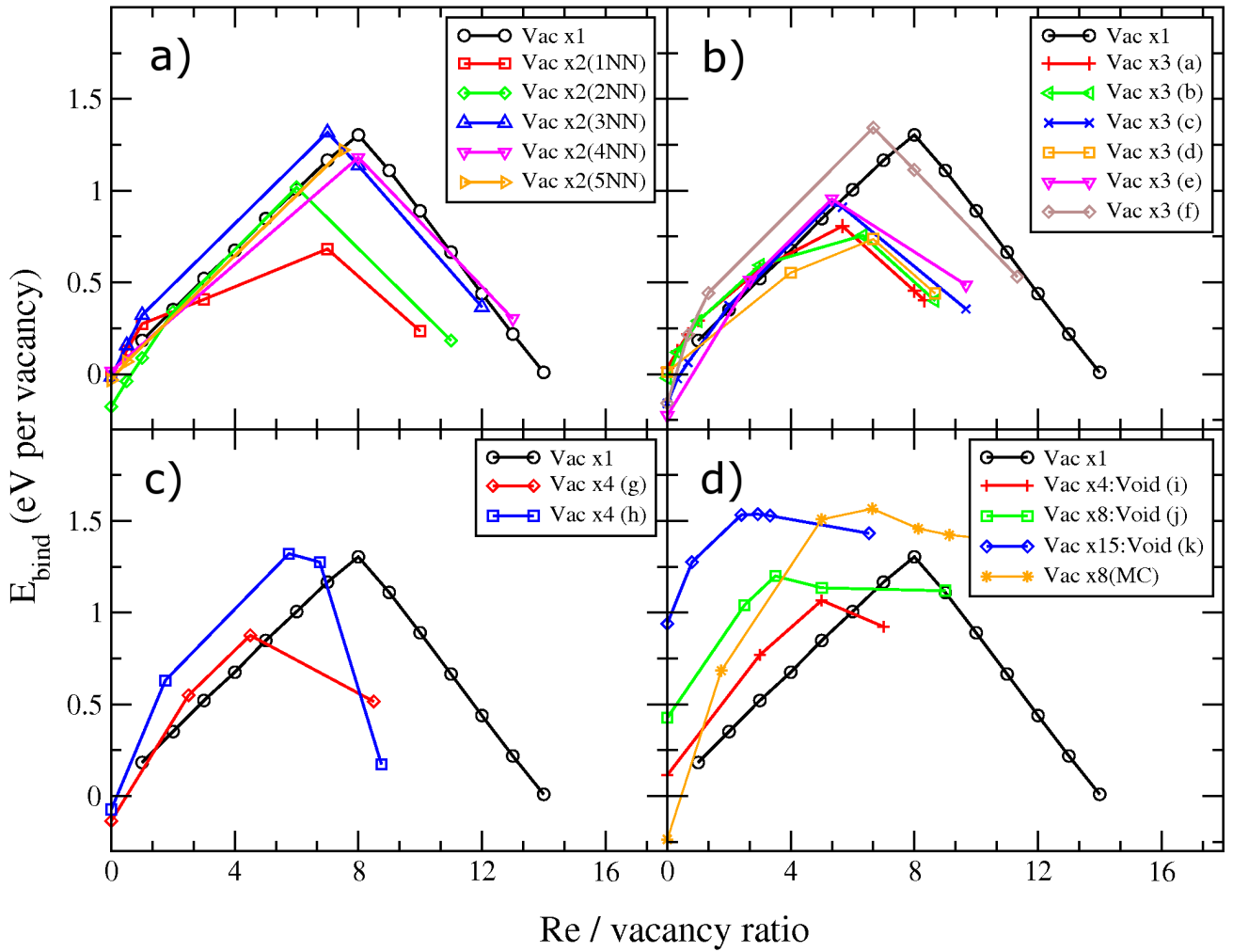}	
			\caption{
		(Color online) Binding energy between Re atoms and vacancy clusters shown as a function of Re to vacancy ratio for di-vacancy (a), tri-vacancy (b), quarto-vacancy (c) clusters as well as for voids, the most compact vacancy clusters (d). Explanation of abbreviations and the schematic representations of vacancy clusters are given in Fig. \ref{fig:Schematic_vacancy_clusters}. Vac x8(MC) are results referring to the configuration of 8 vacancies obtained from MC simulations with 2\% at. Re and 0.05\% at. vacancies (see Table \ref{tab:MC_results_v1})}
		\label{fig:Binding_energy}
		\end{figure*}

Calculations of binding energies of Re atoms with a mono-vacancy at 0K as a function of the ratio of the number of Re atoms to vacancies, 
were performed by replacing W atoms by Re atoms in the first and then in the second coordination shell of a vacancy, see blue and pink spheres in Fig. \ref{fig:Schematic shells}a. Results are given in Fig. \ref{fig:Binding_energy} and Table \ref{tab:Binding_energy}. The binding energy between a vacancy and a Re atom is 0.183 eV, which is slightly smaller than the value of approximately 0.2 eV found earlier \cite{Kong2014}. Since the value of enthalpy of mixing between W and Re atoms is small, see Fig. \ref{fig:Hmix_WRe}, the strength of binding in vacancy-Re clusters depends mainly on the number of Re atoms in the neighbourhood of a vacancy, and does not depend significantly on the specific configuration of Re atoms in the first or second shell around a vacancy. As a result, the binding energy increases almost linearly as a function of Re atoms up to the value of 1.304 eV when Re atoms occupy eight positions in the first nearest neighbour coordination shell of a vacancy. Binding energy decreases when Re atoms start occupying positions in the second nearest neighbour coordination shell, see
Fig. \ref{fig:Binding_energy}.

\begin{figure*}
                        \centering
                        \begin{minipage}{.22\textwidth}
                                \centering
                                a) \includegraphics[width=.9\linewidth]{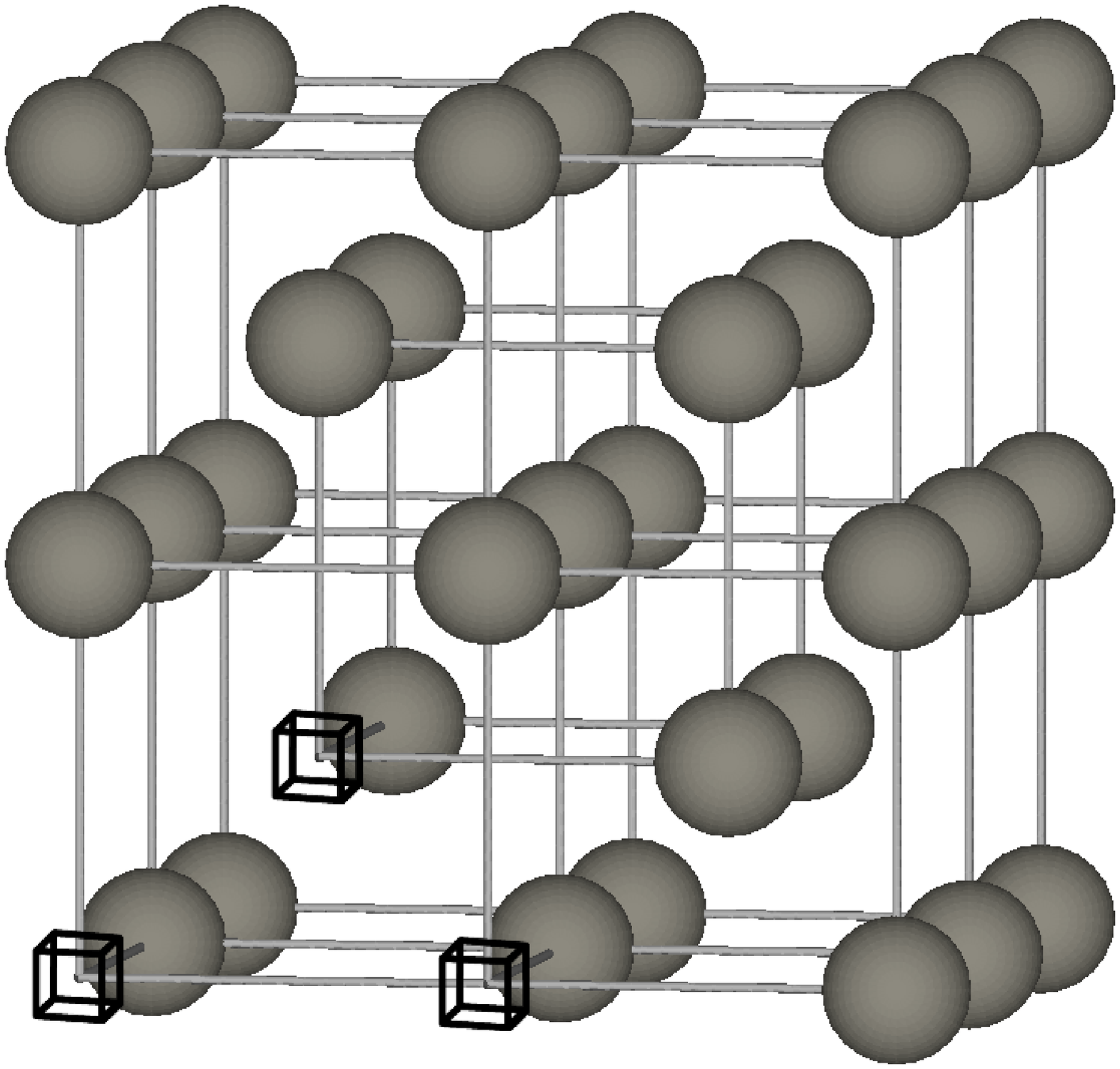}
                        \end{minipage}%
                        \begin{minipage}{.23\textwidth}
                                \centering
                                b) \includegraphics[width=.9\linewidth]{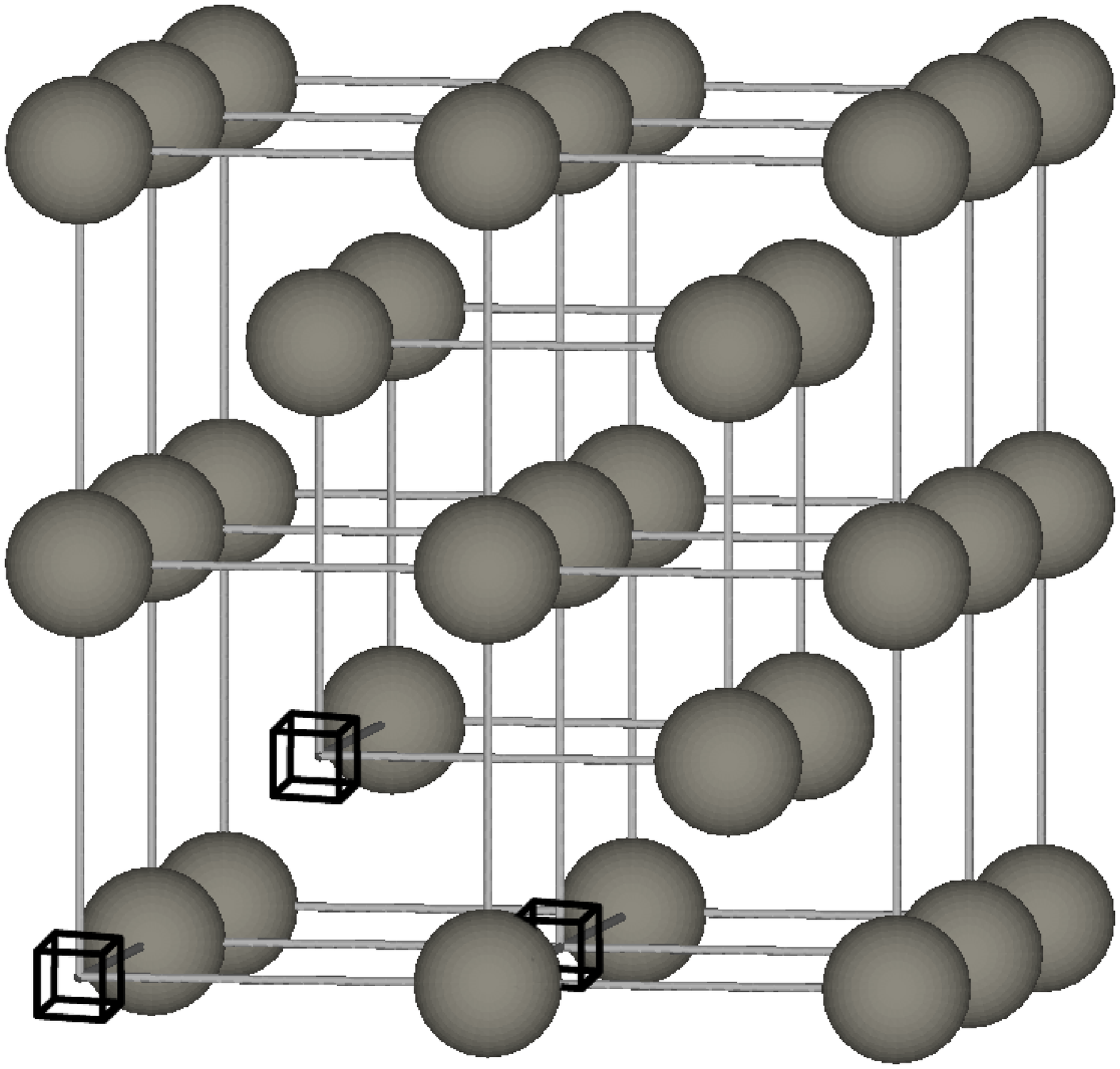}
                        \end{minipage}
                        \begin{minipage}{.22\textwidth}
                                \centering
                                c) \includegraphics[width=.9\linewidth]{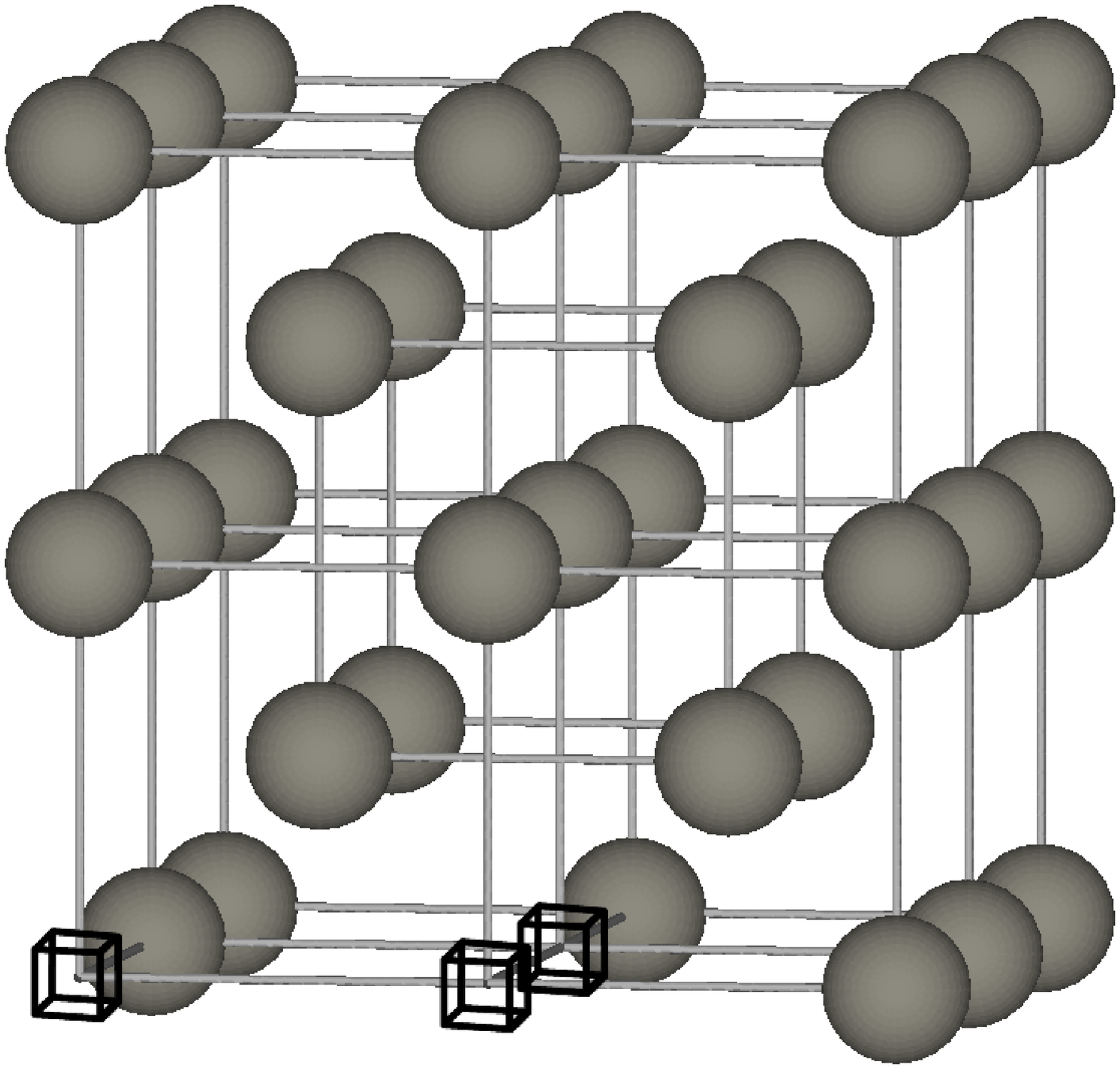}
                        \end{minipage}%
                        \begin{minipage}{.23\textwidth}
                                \centering
                                d) \includegraphics[width=.9\linewidth]{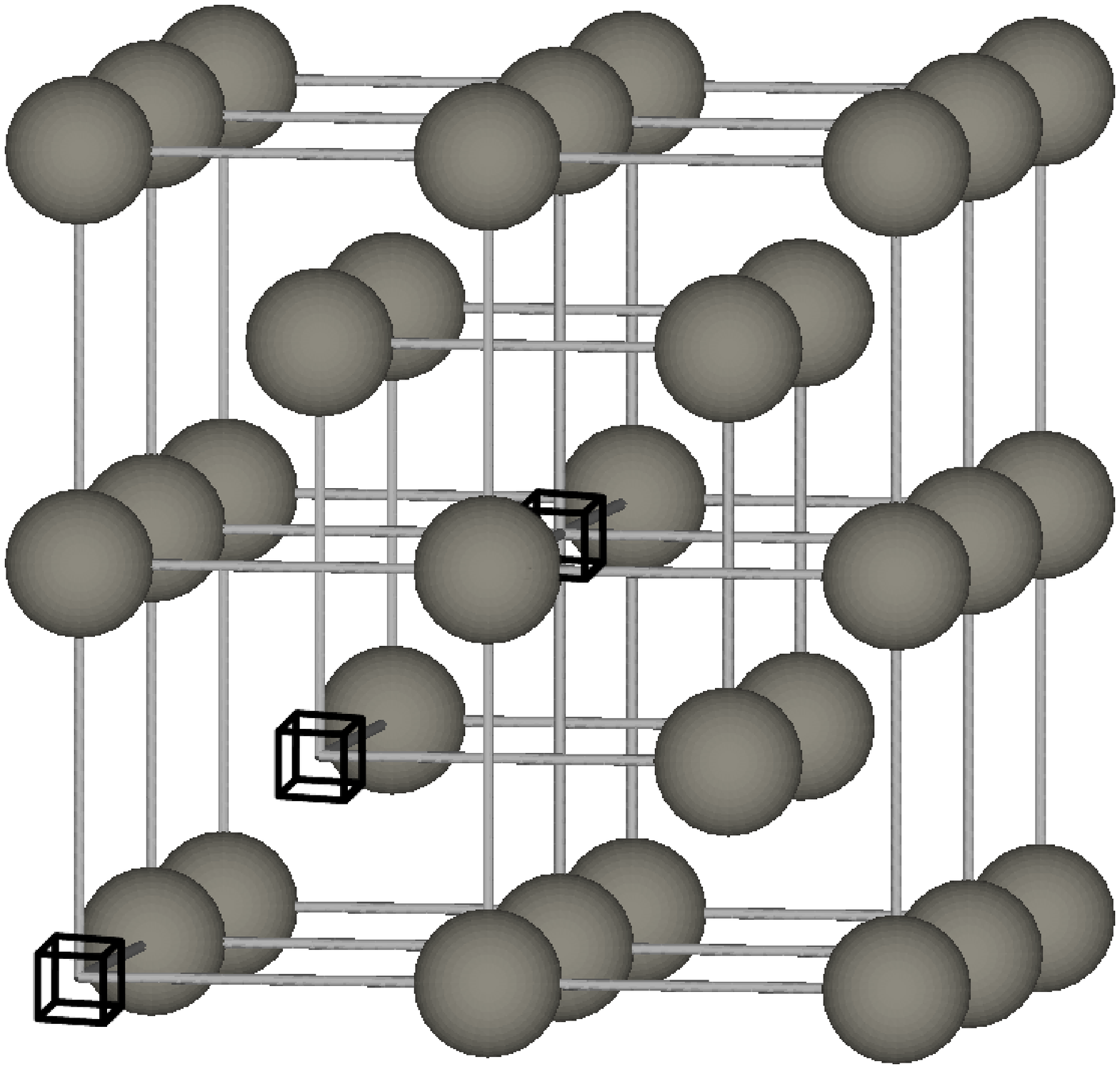}
                        \end{minipage}
                        \newline
                        \begin{minipage}{.22\textwidth}
                                \centering
                                e) \includegraphics[width=.9\linewidth]{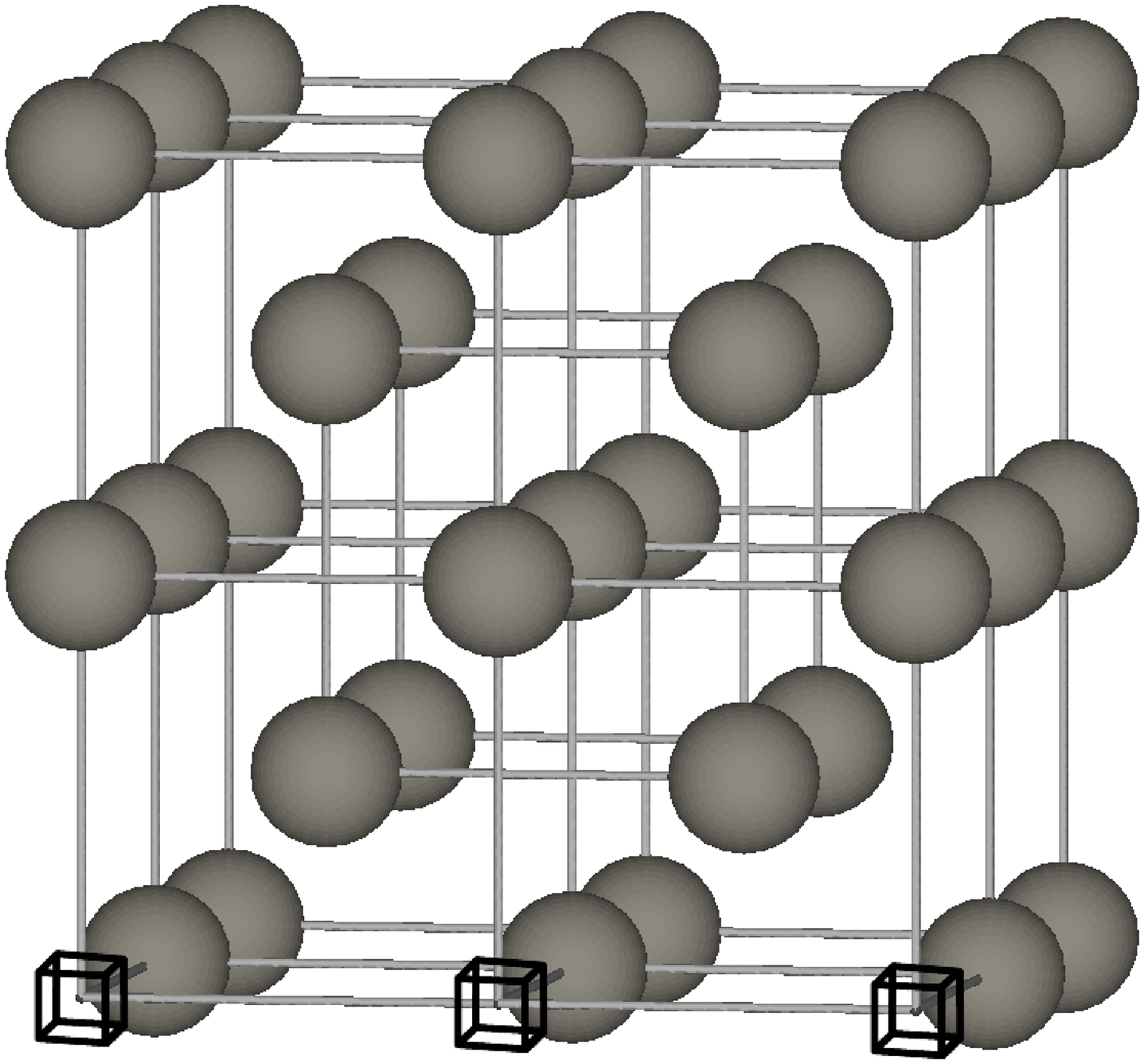}
                        \end{minipage}%
                        \begin{minipage}{.22\textwidth}
                                \centering
                                f) \includegraphics[width=.9\linewidth]{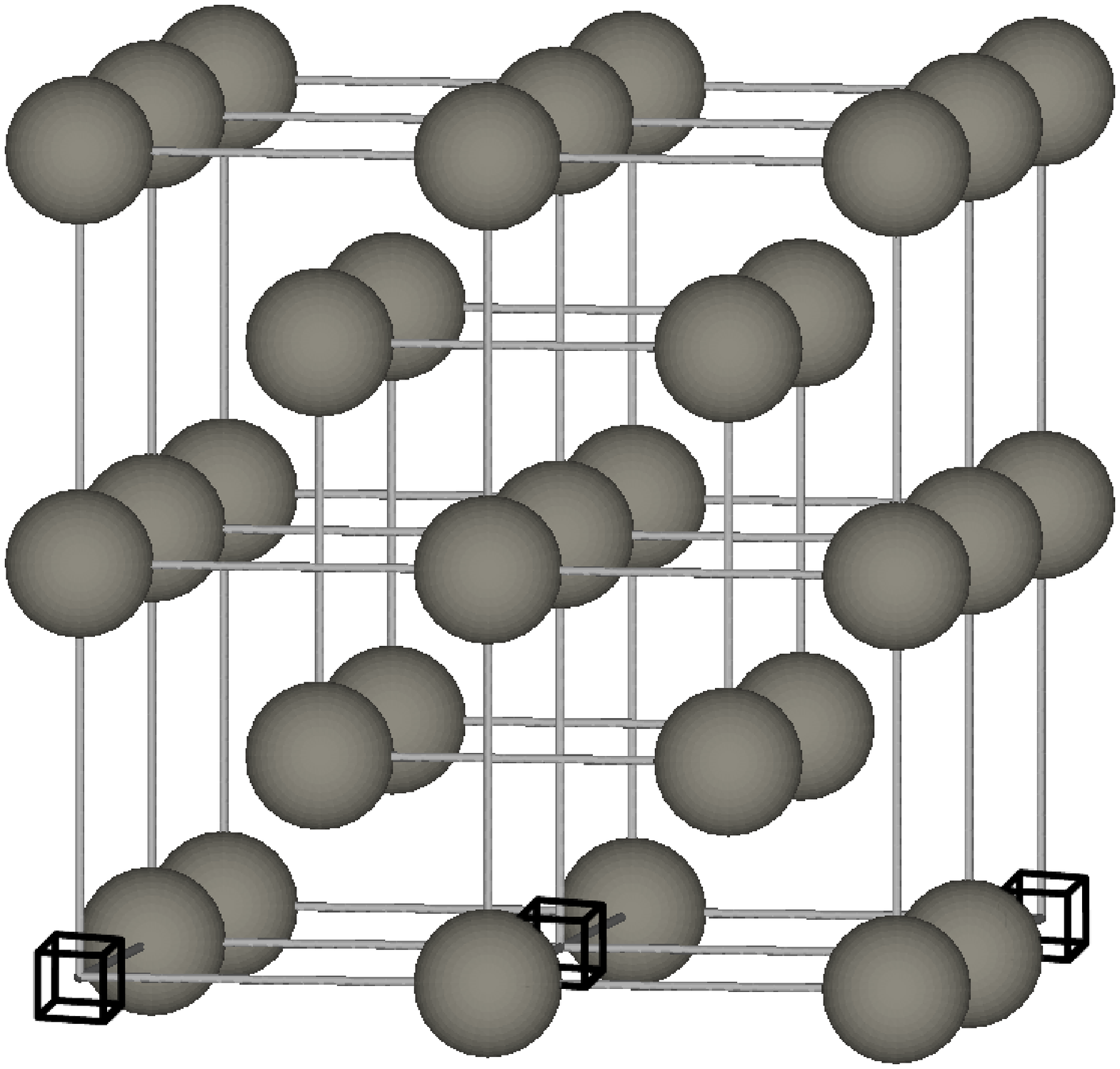}
                        \end{minipage}
                        \begin{minipage}{.22\textwidth}
                                \centering
                                g) \includegraphics[width=.9\linewidth]{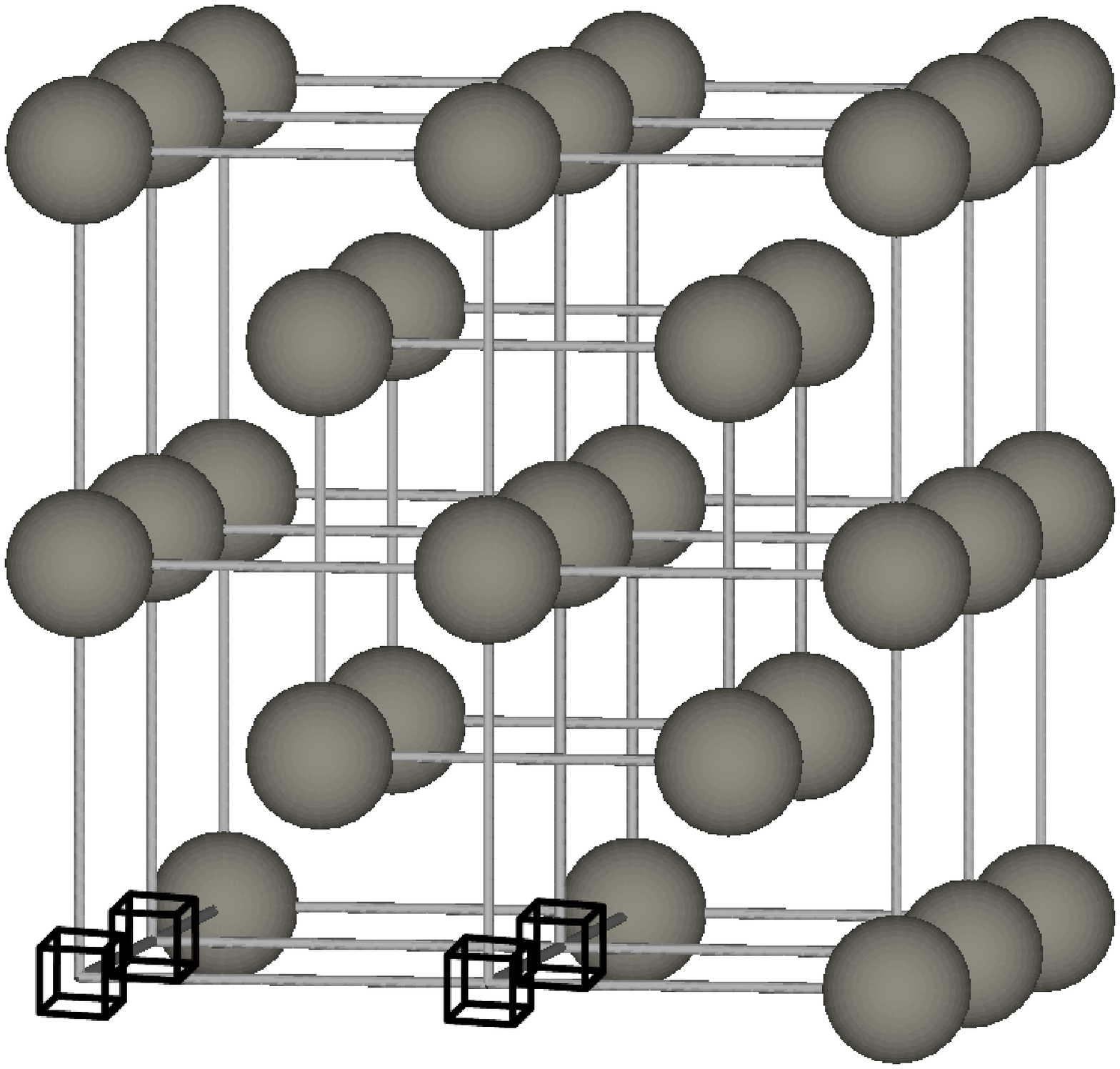}
                        \end{minipage}
                        \begin{minipage}{.23\textwidth}
                                \centering
                                h) \includegraphics[width=.9\linewidth]{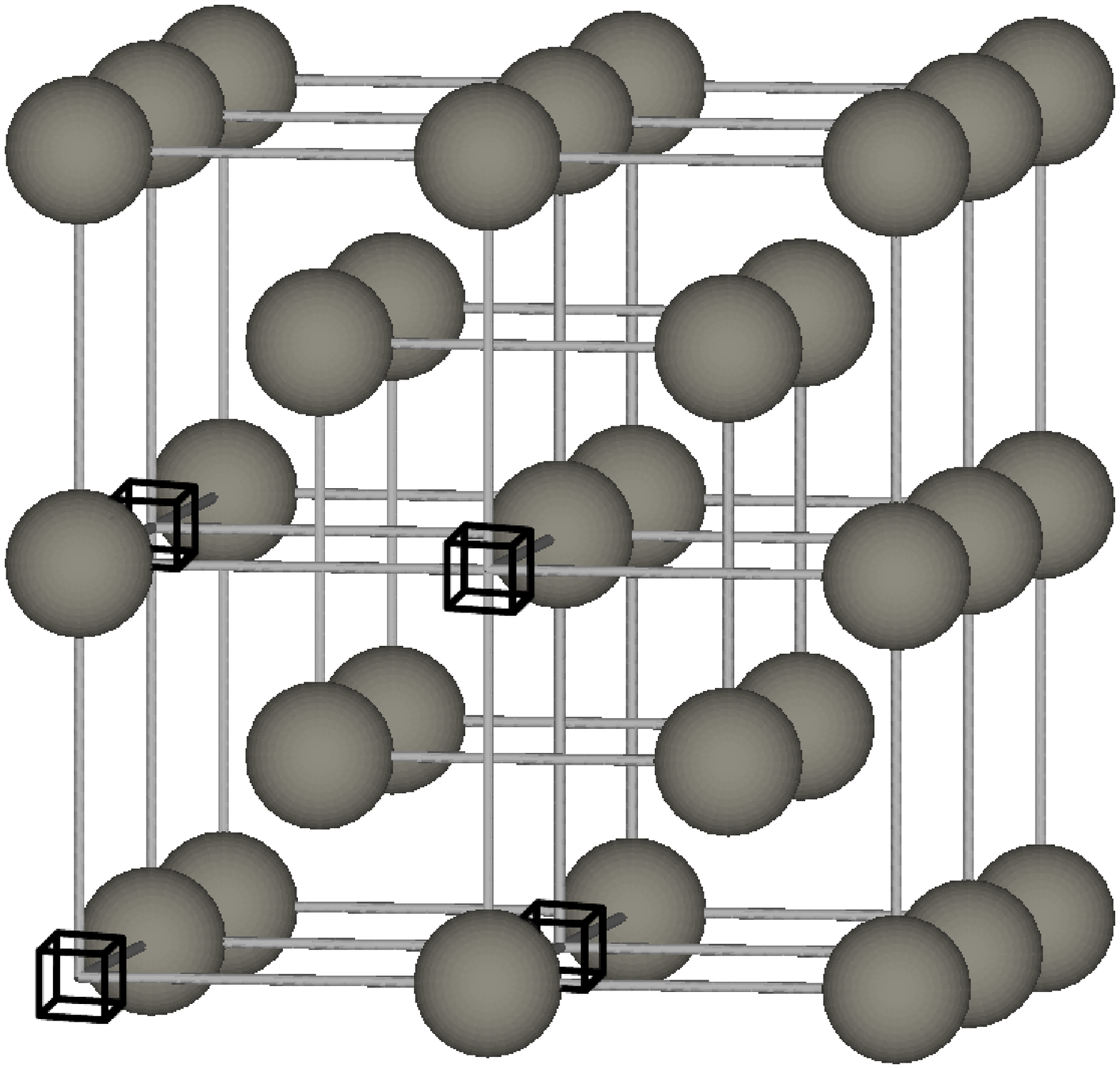}
                        \end{minipage}
                        \newline
                        \begin{minipage}{.22\textwidth}
                                \centering
                                i) \includegraphics[width=.9\linewidth]{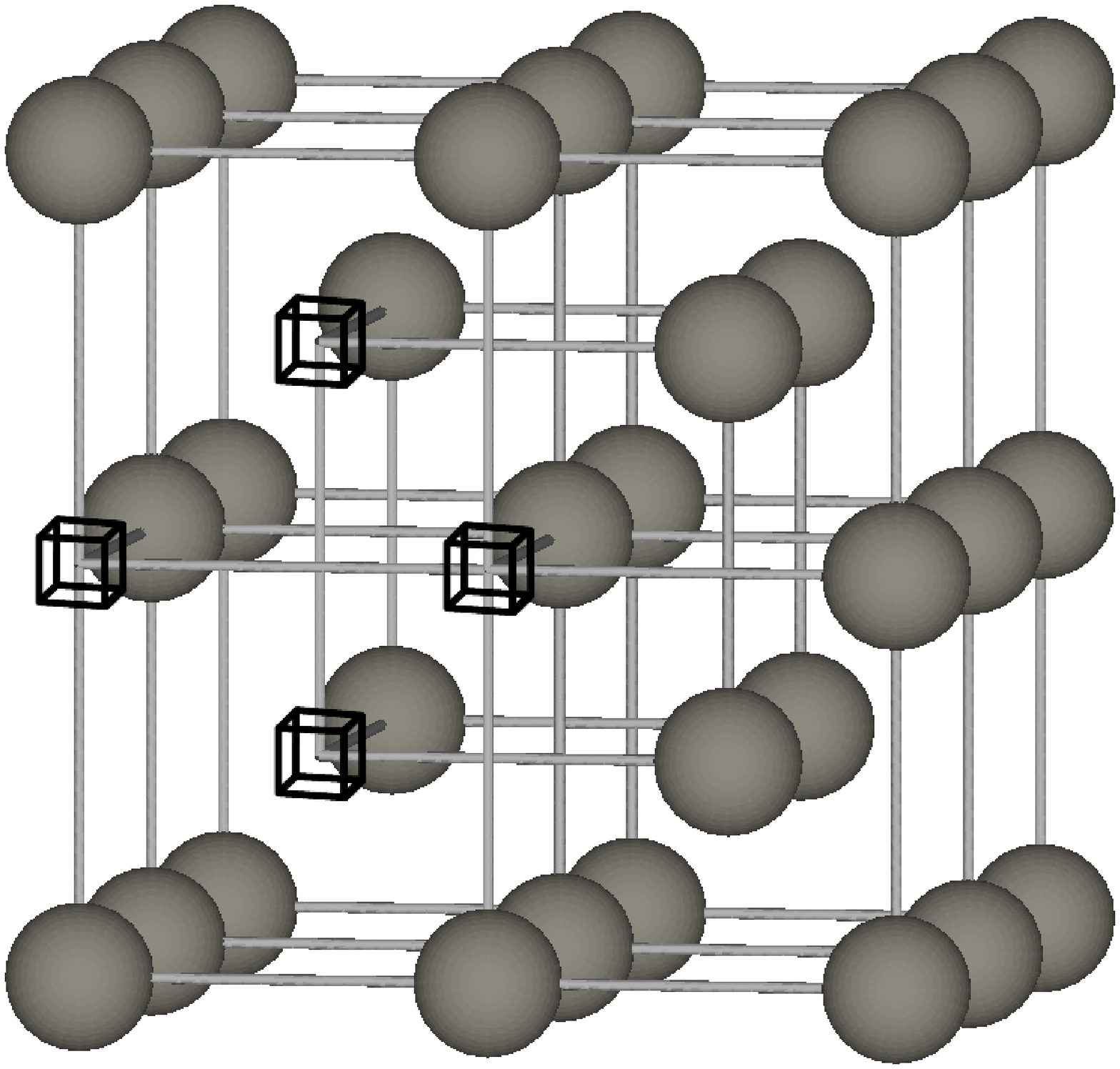}
                        \end{minipage}
                        \begin{minipage}{.22\textwidth}
\centering
                                j) \includegraphics[width=.9\linewidth]{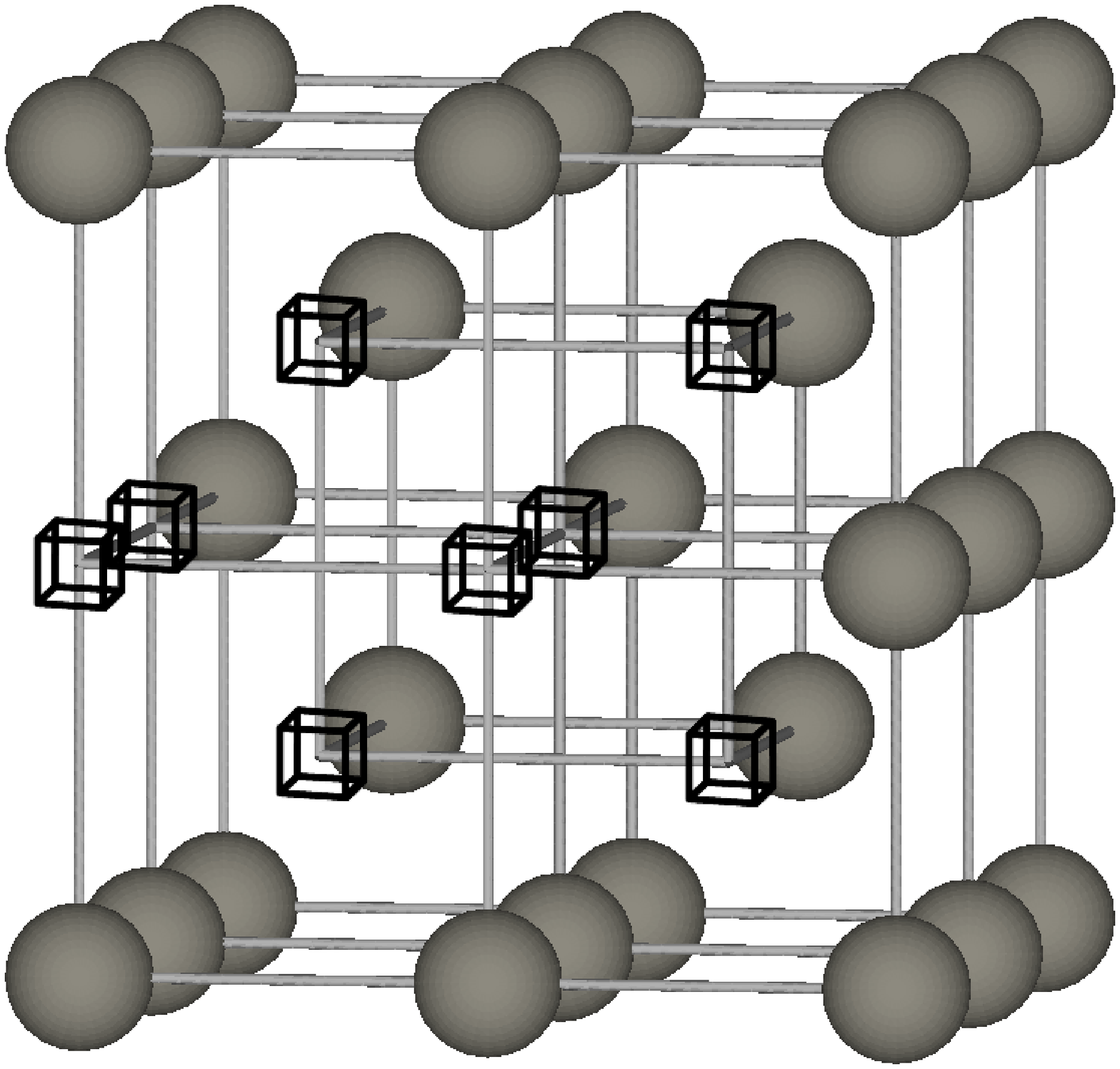}
                        \end{minipage}%
                        \begin{minipage}{.22\textwidth}
                                \centering
                                k) \includegraphics[width=.9\linewidth]{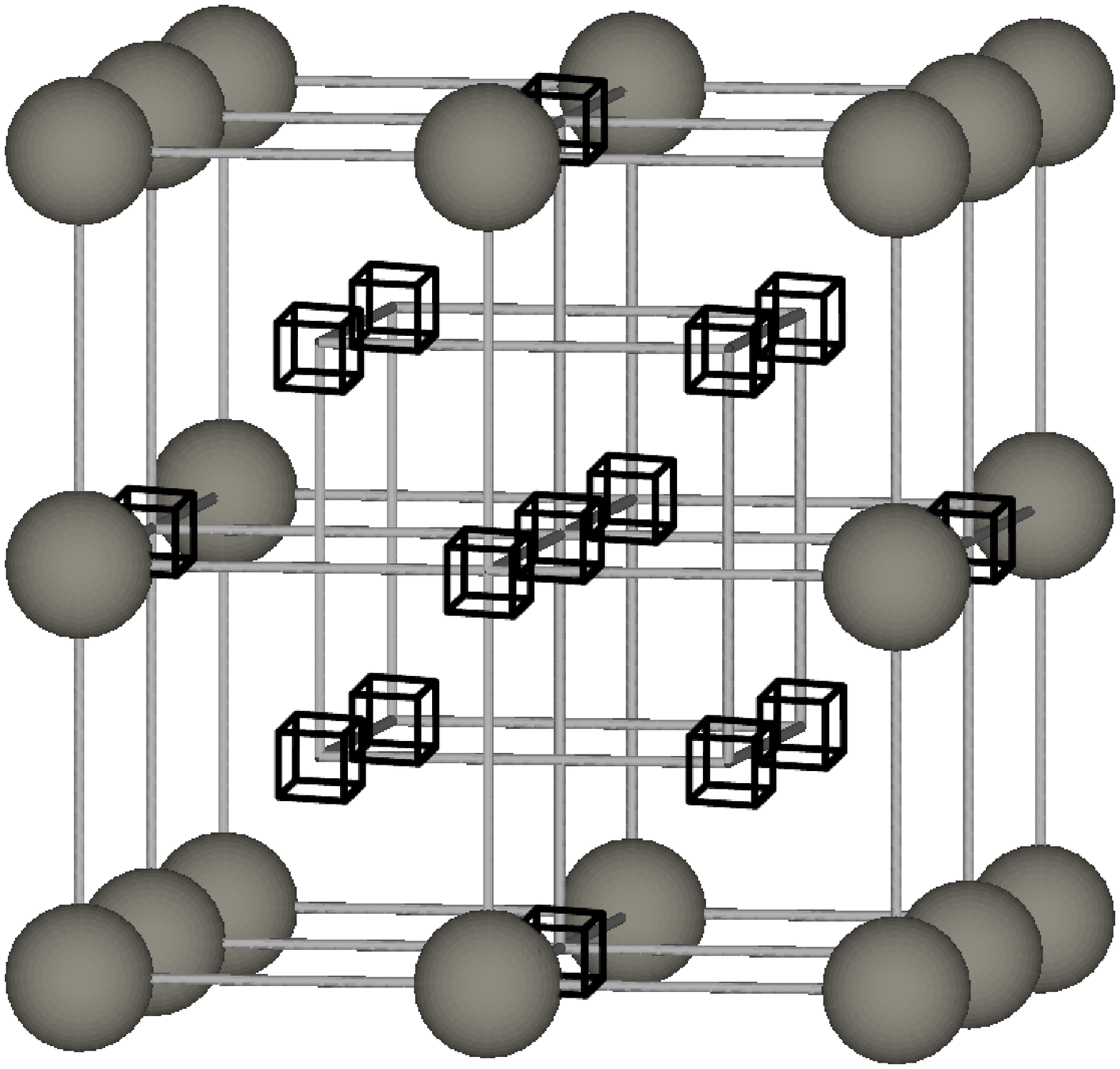}
                        \end{minipage}
                        \newline
                        \caption{ Schematic representation of vacancy clusters; tri-vacancy configurations: a) Vac x3(a), b) Vac x3(b), c)Vac x3(c), d) Vac x3(d), e) Vac x3(e), f) Vac x3(f); quarto-vacancy configurations: g) Vac x4(g), h) Vac x4(h), i) Vac x4(i); void configurations: i) with 4 vacancies - Vac x4:Void(i), j) with 8 vacancies - Vac x8:Void(j), k) with 15 vacancies - Vac x15:Void(k)
                }
                \label{fig:Schematic_vacancy_clusters}
\end{figure*}

\subsection{Interaction of vacancy clusters with Re}

Binding energies of di-vacancies as functions of Re/vacancy ratio for di-vacancies in the first, second, third, fourth and fifth nearest neighbourhood indicated as 1NN, 2NN, 3NN, 4NN and 5NN, respectively, are shown in Fig. \ref{fig:Binding_energy}a and Table \ref{tab:Binding_energy}. In pure W matrix, the binding energies of all the configurations apart from 4NN di-vacancy configuration, are negative, meaning that vacancies repel each other. The most negative is the binding energy of a 2NN di-vacancy configuration that is equal to -0.177 eV per vacancy, which is slightly less negative than -0.190 eV per vacancy reported earlier \cite{Becquart2007,Muzyk2011}. The binding energy of the 1NN di-vacancy configuration equal to -0.012 eV per vacancy is small but it is negative.

In order to investigate the effect of Re atoms on the binding energies of various di-vacancy configurations, W atoms in 1NN and 2NN positions were systematically replaced by Re atoms. Since the binding energy primarily depends on the number of Re atoms in 1NN and 2NN positions with respect to vacancies, the largest contribution to the binding energy of cluster comes from Re atoms that have two vacancies in their first nearest-neighbour coordination shell. Hence, a systematic investigation of the effect of Re atoms on binding energies of the 3NN di-vacancy configuration starts with a replacement of W atoms by Re atoms in the first nearest neighbourhood of two vacancies, see red spheres in Fig. \ref{fig:Schematic shells}b. Afterwards the replacement of W atoms by Re atoms wa performed in the following sequence: atoms that have one vacancy in 1NN, two vacancies in 2NN and one vacancy in 2NN indicated by blue, green and pink spheres in Fig. \ref{fig:Schematic shells}b, respectively. The same scheme was applied to the replacement of W atoms by Re atoms in other di-vacancy and multi-vacancy configurations, see Figs. \ref{fig:Schematic shells}c and \ref{fig:Schematic shells}d for the case of four-vacancy cluster and a 15-vacancy void, respectively.

Binding energies of all the di-vacancy configurations apart from the 2NN di-vacancy configuration (vacancies ordered in the [100] direction) are positive, meaning that vacancies in the neighbourhood of a Re atom attract each other. In other words, a Re atom provides a centre of nucleation for vacancies. The largest binding energy that is equal to 0.157 eV per vacancy characterizes the 3NN di-vacancy configuration (vacancies ordered in the [110] direction) bound by a Re atom, despite the fact that the binding energy of a 3NN di-vacancy configuration without Re atoms is only slightly more negative that that of the 1NN and 4NN di-vacancy configurations. The 3NN di-vacancy configuration  is the most preferable energetically up to the Re/vacancy ratio approaching 7. For the ratios above 7, all the configurations of vacancies apart from the 1NN di-vacancy configuration, have quite similar binding energies. Moreover, they usually have smaller binding energies than the binding energies between Re atoms and a mono-vacancy, see the black line on Fig. \ref{fig:Binding_energy}a. It means that for the Re/vacancy ratios higher than 7, vacancies may prefer to occupy configurations other than third nearest neighbours.

The binding energies of tri-vacancy configurations are shown in Fig. \ref{fig:Binding_energy}b and Table \ref{tab:Binding_energy} as functions of the Re/vacancy ratio. In a pure W matrix the most stable tri-vacancy configuration is the one with vacancies forming the most compact cluster, see Fig. \ref{fig:Schematic_vacancy_clusters}f. This configuration remains to be most stable in the presence of one Re atom and two Re atoms. For higher Re/vacancy ratios the most stable configuration is the one where vacancies are in the 3NN coordination shell, see \ref{fig:Schematic_vacancy_clusters}h.

Binding energies of quarto-vacancy configurations, forming a (100) surface, see Fig. \ref{fig:Schematic_vacancy_clusters}g, and with vacancies in the 3NN coordination shell to each other, see Fig. \ref{fig:Schematic_vacancy_clusters}h, are shown in Fig. \ref{fig:Binding_energy}c whereas those of the most compact vacancy cluster (void) configurations, see Fig. \ref{fig:Schematic_vacancy_clusters}i, are shown in Fig. \ref{fig:Binding_energy}c. The values of binding energies are given in Table \ref{tab:Binding_energy}.

\subsection{Voids and Vacancy clusters}

Void configurations, starting from three-vacancy clusters in 1NN and 2NN configurations, to a 15-vacancy cluster, are stable in a pure W matrix, according to previous {\it ab initio} calculations \cite{Muzyk2011,Nguyen-Manh2014}. When the Re/vacancy ratio increases, configurations with vacancies in the 3NN coordination shell become more stable, similarly to how it was found in the tri-vacancy case with the same relative positions of vacancies. Similarly to how it was in the case of a tri-vacancy configuration, where the Re/vacancy ratio is higher than approximately 7, binding between Re atoms and a mono-vacancy is stronger than for quarto-vacancy configurations.

Fig. \ref{fig:Binding_energy}d and Table \ref{tab:Binding_energy} show that the binding energy of the most compact vacancy clusters without Re atoms increases as a function of the number of vacancies in a void. This agrees with results presented in earlier work \cite{Muzyk2011}. Results obtained with W atoms replaced by Re atoms show that binding energies increase to over 1.5 eV when the voids are surrounded by Re atoms. Overall, the Re-void configurations containing a larger number of vacancies are more stable than those with smaller number of vacancies over the entire range of Re/vacancy ratios.

In order to compare the relative stability of the most compact vacancy configurations (small voids) with other vacancy cluster configurations, a structure containing 8 vacancies and derived from MC simulations of W-Re alloys with 2\% at. Re and 0.05\% at. vacancies
quenched down from high temperatures (see the sub-section IV.A) is also included in Table \ref{tab:Binding_energy}. The binding energy characterizing that structure, where each vacancy is in the third nearest-neighbour position with respect to others, is smaller than those of the 8-vacancy void configuration for the low ratio of Re/vacancy, but becomes larger when this ratio is $\ge$ 4. In particular, we find that the highest value (1.566 eV) of binding energy between Re atoms and 8-vacancy clusters corresponds to the Re/vacancy ratio of 6.63.

Concluding this Section, we note that binding energies at 0K for various vacancy cluster configurations show similar trends in terms of their stability as functions of Re concentrations. In the limit where the Re/vacancy ratio is small, teh more compact configurations are preferred. As this ratio increases, configurations with vacancies being further apart are becoming more stable. Configurations with vacancies being in the 3NN coordination shell with respect to each other have particularly high stability over a broad interval of Re/vacancy ratios.

\begin{table*}
\caption{Binding energy, $E_{bind}$, in eV units, for various Re-vacancy cluster configurations. $N_W$, $N_{Re}$ and $N_Re/N_{Vac}$ are numbers of W atoms, Re atoms and the ratio of Re atoms to vacancies, respectively. Explanation of abbreviations and schematic representations of vacancy clusters is given in Fig. \ref{fig:Schematic_vacancy_clusters}. Vac x8(MC) refers to results obtained using the configuration of 8 vacancies derived from MC simulations with 2\% at. Re and 0.05\% at. vacancies (see Table \ref{tab:MC_results_v1}).
        \label{tab:Binding_energy}}
\begin{ruledtabular}
    \begin{tabular}{ccccc|ccccc}
        Structure & $N_W$ & $N_{Re}$ & $N_{Re}/N_{Vac}$ & $E_{bind}$ &  Structure & $N_W$ & $N_{Re}$ & $N_{Re}/N_{Vac}$ & $E_{bind}$ \\
    \hline
    Vac x1 & 126   & 1     & 1     & 0.183 & Vac x3 (c) & 125   & 0     & 0.00  & -0.159 \\
          & 125   & 2     & 2     & 0.351 &       & 124   & 1     & 0.33  & -0.025 \\
          & 124   & 3     & 3     & 0.521 &       & 123   & 2     & 0.67  & 0.063 \\
          & 123   & 4     & 4     & 0.675 &       & 119   & 6     & 2.00  & 0.373 \\
          & 122   & 5     & 5     & 0.848 &       & 109   & 16    & 5.33  & 0.935 \\
          & 121   & 6     & 6     & 1.005 &       & 108   & 17    & 5.67  & 0.910 \\
          & 120   & 7     & 7     & 1.166 &       & 96    & 29    & 9.67  & 0.355 \\
          & 119   & 8     & 8     & 1.304 & Vac x3 (d) & 125   & 0     & 0.00  & 0.012 \\
          & 118   & 9     & 9     & 1.110 &       & 113   & 12    & 4.00  & 0.553 \\
          & 117   & 10    & 10    & 0.890 &       & 105   & 20    & 6.67  & 0.736 \\
          & 116   & 11    & 11    & 0.665 &       & 99    & 26    & 8.67  & 0.440 \\
          & 115   & 12    & 12    & 0.439 & Vac x3 (e) & 125   & 0     & 0.00  & -0.227 \\
          & 114   & 13    & 13    & 0.218 &       & 117   & 8     & 2.67  & 0.511 \\
          & 113   & 14    & 14    & 0.009 &       & 109   & 16    & 5.33  & 0.955 \\
    Vac x2 (1NN) & 126   & 0     & 0     & -0.012 &       & 96    & 29    & 9.67  & 0.486 \\
          & 125   & 1     & 0.5   & 0.129 & Vac x3 (f) & 125   & 0     & 0.00  & -0.159 \\
          & 124   & 2     & 1     & 0.274 &       & 123   & 2     & 0.67  & 0.214 \\
          & 123   & 3     & 1.5   & 0.351 &       & 121   & 4     & 1.33  & 0.442 \\
          & 122   & 4     & 2     & 0.367 &       & 105   & 20    & 6.67  & 1.343 \\
          & 121   & 5     & 2.5   & 0.385 &       & 101   & 24    & 8.00  & 1.113 \\
          & 120   & 6     & 3     & 0.407 &       & 91    & 34    & 11.33 & 0.531 \\
          & 119   & 7     & 3.5   & 0.484 & Vac x4 (g) & 124   & 0     & 0.00  & -0.136 \\
          & 118   & 8     & 4     & 0.499 &       & 114   & 10    & 2.50  & 0.549 \\
          & 112   & 14    & 7     & 0.682 &       & 106   & 18    & 4.50  & 0.876 \\
          & 106   & 20    & 10    & 0.235 &       & 90    & 34    & 8.50  & 0.515 \\
    Vac x2 (2NN) & 126   & 0     & 0     & -0.177 & Vac x4 (h) & 124   & 0     & 0.00  & -0.074 \\
          & 125   & 1     & 0.5   & -0.039 &       & 117   & 7     & 1.75  & 0.630 \\
          & 124   & 2     & 1     & 0.088 &       & 101   & 23    & 5.75  & 1.321 \\
          & 122   & 4     & 2     & 0.339 &       & 97    & 27    & 6.75  & 1.275 \\
          & 114   & 12    & 6     & 1.018 &       & 89    & 35    & 8.75  & 0.172 \\
          & 104   & 22    & 11    & 0.183 & Vac x4:Void (i) & 246   & 0     & 0     & 0.114 \\
    Vac x2 (3NN) & 126   & 0     & 0     & -0.015 &       & 234   & 12    & 3     & 0.769 \\
          & 125   & 1     & 0.5   & 0.157 &       & 226   & 20    & 5     & 1.066 \\
          & 124   & 2     & 1     & 0.322 &       & 218   & 28    & 7     & 0.923 \\
          & 112   & 14    & 7     & 1.318 & Vac x8:Void (j) & 242   & 0     & 0     & 0.426 \\
          & 110   & 16    & 8     & 1.137 &       & 222   & 20    & 2.50  & 1.040 \\
          & 102   & 24    & 12    & 0.366 &       & 214   & 28    & 3.50  & 1.199 \\
    Vac x2 (4NN) & 126   & 0     & 0     & 0.016 &       & 202   & 40    & 5.00  & 1.135 \\
          & 125   & 16    & 8     & 1.181 &       & 170   & 72    & 9.00  & 1.119 \\
          & 100   & 26    & 13    & 0.302 & Vac x15:Void (k) & 235   & 0     & 0     & 0.939 \\
    Vac x2 (5NN) & 126   & 0     & 0     & -0.033 &       & 223   & 12    & 0.80  & 1.275 \\
          & 125   & 1     & 0.5   & 0.068 &       & 199   & 36    & 2.40  & 1.533 \\
          & 111   & 15    & 7.5   & 1.222 &       & 191   & 44    & 2.93  & 1.536 \\
    Vac x3 (a) & 125   & 0     & 0.00  & 0.035 &       & 185   & 50    & 3.33  & 1.528 \\
          & 124   & 1     & 0.33  & 0.126 &       & 137   & 98    & 6.53  & 1.432 \\
          & 123   & 2     & 0.67  & 0.218 & Vac x8(MC) & 242   & 0     & 0     & -0.237 \\
          & 122   & 3     & 1.00  & 0.293 &       & 228   & 14    & 1.75  & 0.683 \\
          & 116   & 9     & 3.00  & 0.572 &       & 202   & 40    & 5.00  & 1.508 \\
          & 108   & 17    & 5.67  & 0.807 &       & 189   & 53    & 6.63  & 1.566 \\
          & 101   & 24    & 8.00  & 0.453 &       & 177   & 65    & 8.13  & 1.458 \\
          & 100   & 25    & 8.33  & 0.401 &       & 169   & 73    & 9.13  & 1.423 \\
    Vac x3 (b) & 125   & 0     & 0.00  & -0.021 &       & 141   & 101   & 12.63 & 1.364 \\
          & 124   & 1     & 0.33  & 0.119 &       &       &       &       &  \\
          & 122   & 3     & 1.00  & 0.289 &       &       &       &       &  \\
          & 116   & 9     & 3.00  & 0.594 &       &       &       &       &  \\
          & 106   & 19    & 6.33  & 0.756 &       &       &       &       &  \\
    \end{tabular}%
\end{ruledtabular}
\end{table*}

\section{Clustering of Re atoms at finite temperatures}

\subsection{Alloys quenched down from high temperatures}

Within the framework of a constrained model for phase stability of alloys containing high concentration of vacancies, the thermodynamic properties of W-Re-Vac system were analyzed using quasi-canonical MC simulations and ECIs derived from DFT calculations.
The dependence of configuration entropy on temperature is given by equation
\begin{equation}
S_{conf}\left(x_W,x_{Re},x_{Vac},T\right)=\int_0^T\frac{C_{conf}\left(x_W,x_{Re},x_{Vac},T'\right)}{T'}dT',
\label{eq:ConfEntropy}
\end{equation}
where the configurational contribution to the specific heat $C_{conf}$ is related to fluctuations of enthalpy of mixing at a given temperature \cite{Newman1999,Lavrentiev2009} through
\begin{equation}
C_{conf}\left(x_W,x_{Re},x_{Vac},T\right)=\frac{\left\langle \left(H_{mix}\left(x_W,x_{Re},x_{Vac},T\right)\right)^2\right\rangle-\left\langle H_{mix}\left(x_W,x_{Re},x_{Vac},T\right)\right\rangle^2}{T^2},
\label{eq:ConfHeatCapacity}
\end{equation}
where $\left\langle H_{mix}\left(x_W,x_{Re},x_{Vac},T\right)\right\rangle$ and $\left\langle \left(H_{mix}\left(x_W,x_{Re},x_{Vac},T\right)\right)^2\right\rangle$ are the mean and mean square average enthalpies of mixing, respectively, computed by averaging over all the MC steps at the accumulation stage for a given temperature and for a given composition of the ternary W-Re-Vac system.

\begin{figure}
  \includegraphics[width=\linewidth]{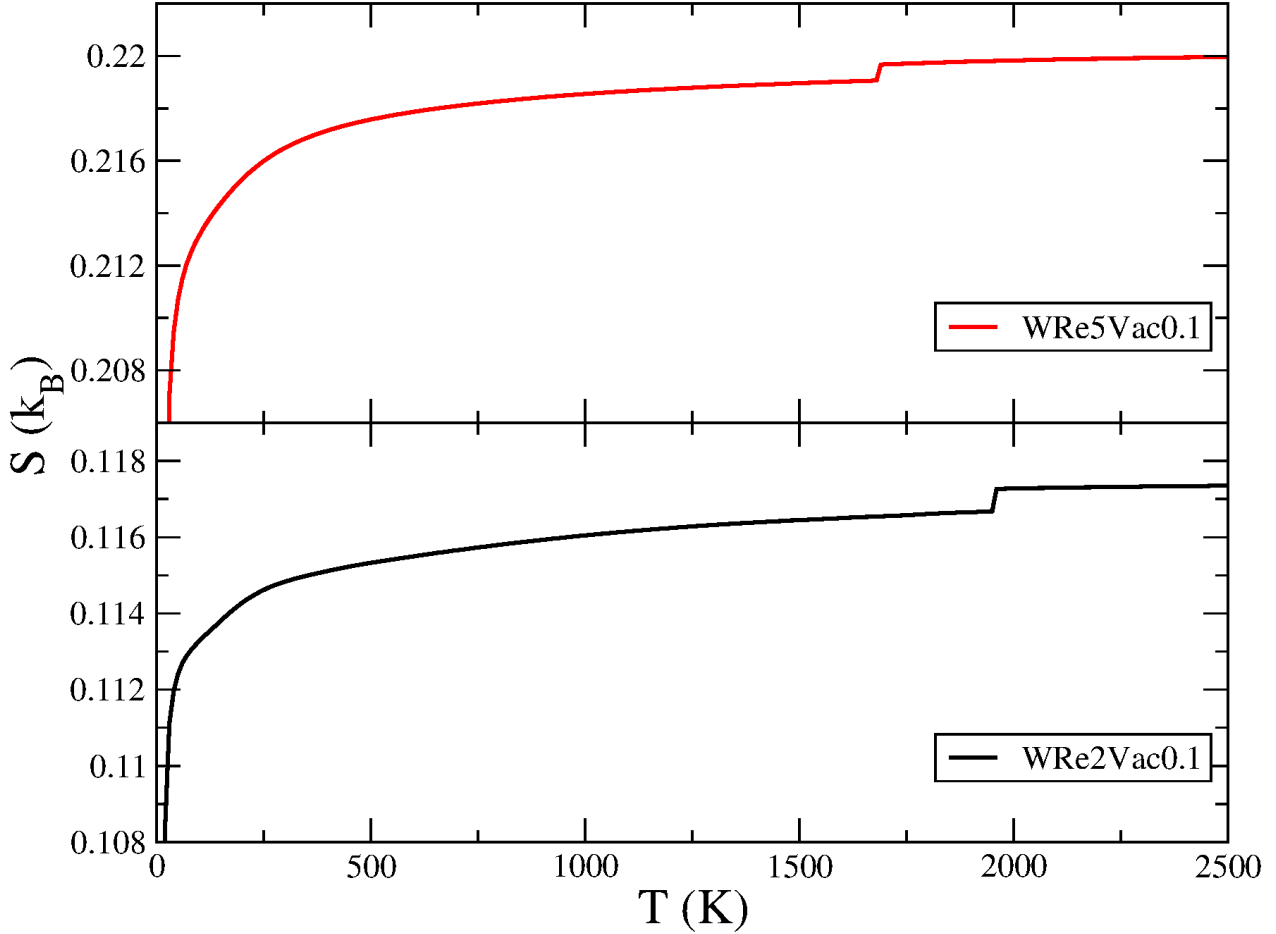}
                        \caption{
                (Color online) Entropy of W-2Re (black line) and W-5Re (red line) alloys with 0.1\% concentration of vacancies as a function of temperature. }
                \label{fig:Entropy_WReVac}
                \end{figure}

The accuracy of evaluation of configuration entropy depends on the temperature integration step in Eq. \ref{eq:ConfEntropy} and on the number of MC steps performed at the accumulation stage. Test simulations show that choosing a sufficiently small temperature integration step is particularly significant. Calculations of configuration entropy for all the alloy compositions considered below were performed with 3000 MC steps per atom at thermalization and accumulation stages, and with temperature step of $\Delta T = 10$ K. Figure \ref{fig:Entropy_WReVac} shows the dependence of configuration entropy as a function of temperature for W-2Re and W-5Re with 0.1$\%$ of vacancy concentration. Transition from ideal random solid solutions into the Re-precipitation regime can be seen clearly from steps in the configuration entropy at T=1700K and T=1950K for the alloys with 5 and 2 at$\%$Re, respectively.




\begin{table*}
\caption{Monte Carlo results as functions of Re and vacancy concentration for alloys quenched down from high temperatures. MC simulations were performed starting from 2500K. Alloys were cooled down with the temperature step of 100 K to the temperature of 100 K with 3000 MC steps per atom performed both at thermalization and accumulation stages.
        \label{tab:MC_results_v1}}
\begin{ruledtabular}
    \begin{tabular}{|c|c|c|c|c|}
              & 1\% at. Re & 2\% at. Re & 5\% at. Re  & 10\% at. Re   \\
     \hline
     0.05\% at. Vac & \includegraphics[width=.2\linewidth]{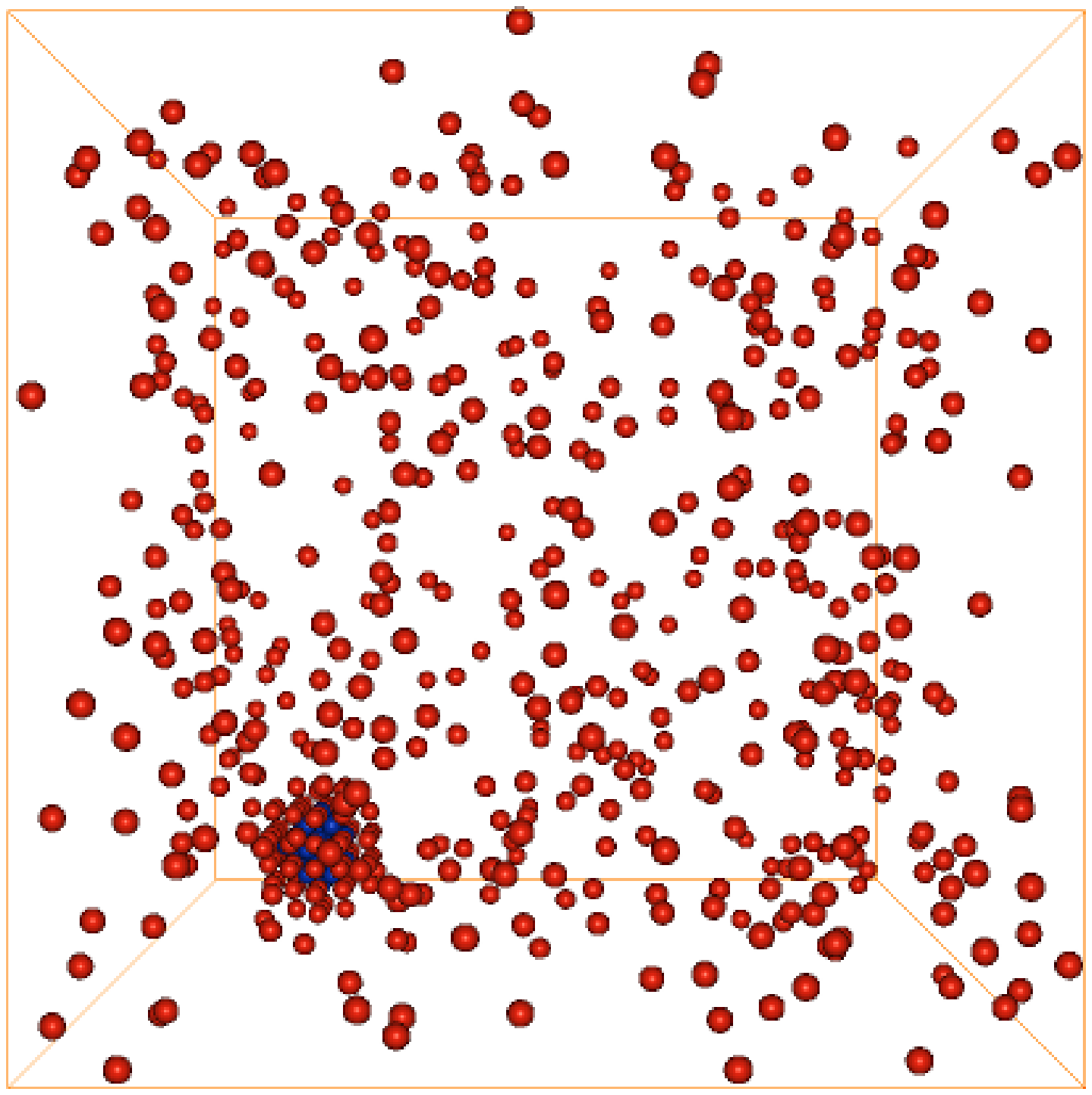} & \includegraphics[width=.2\linewidth]{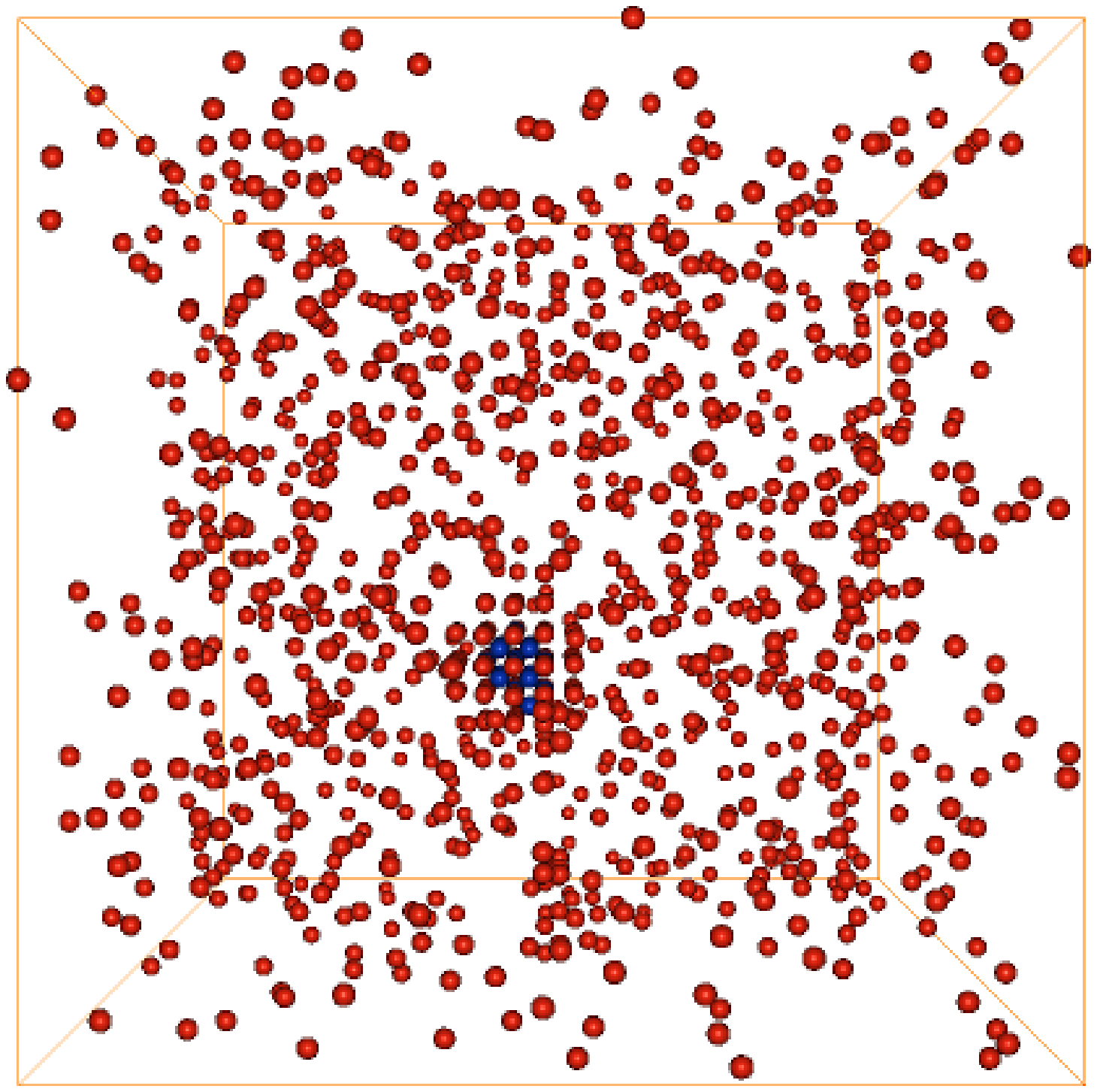} & \includegraphics[width=.2\linewidth]{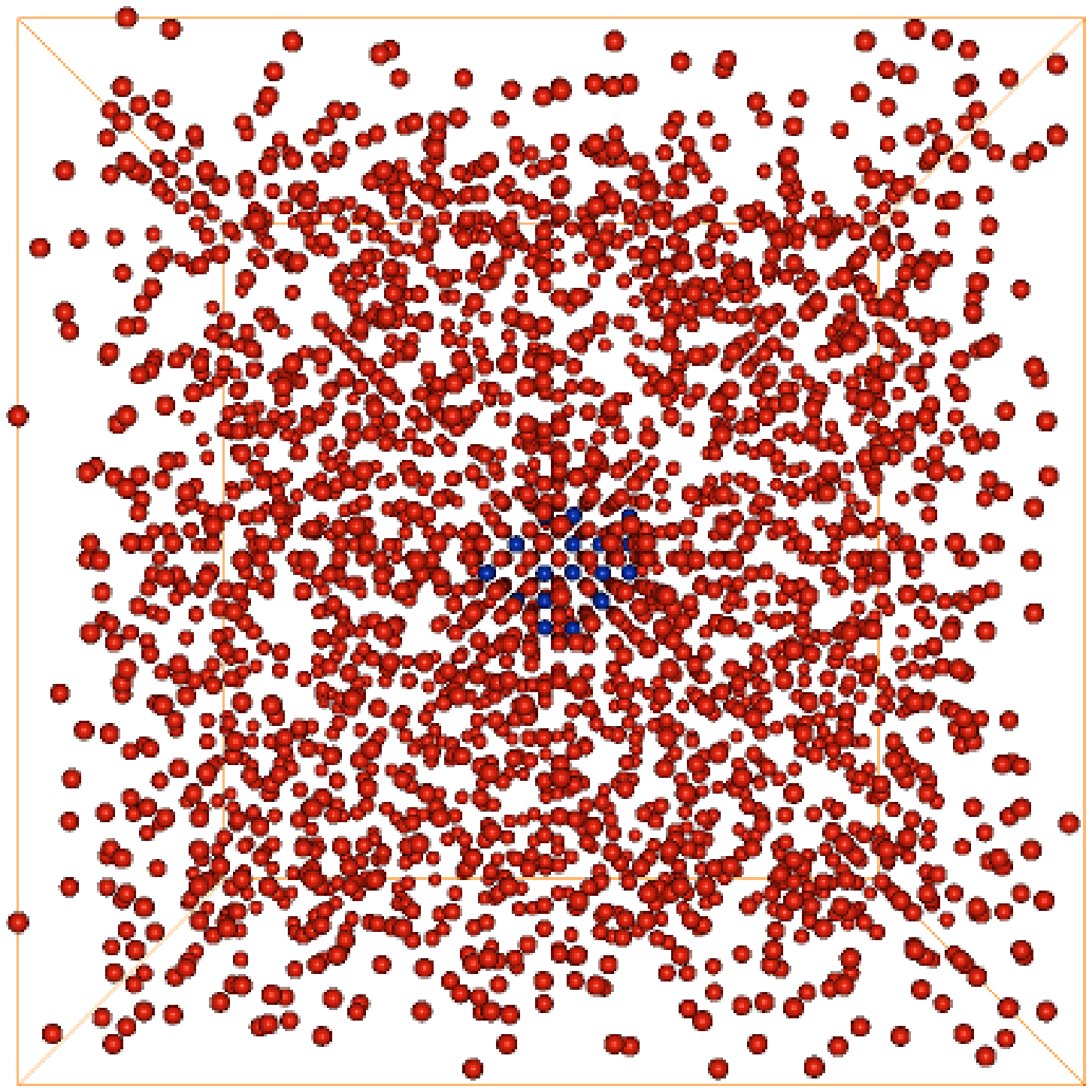} & \includegraphics[width=.2\linewidth]{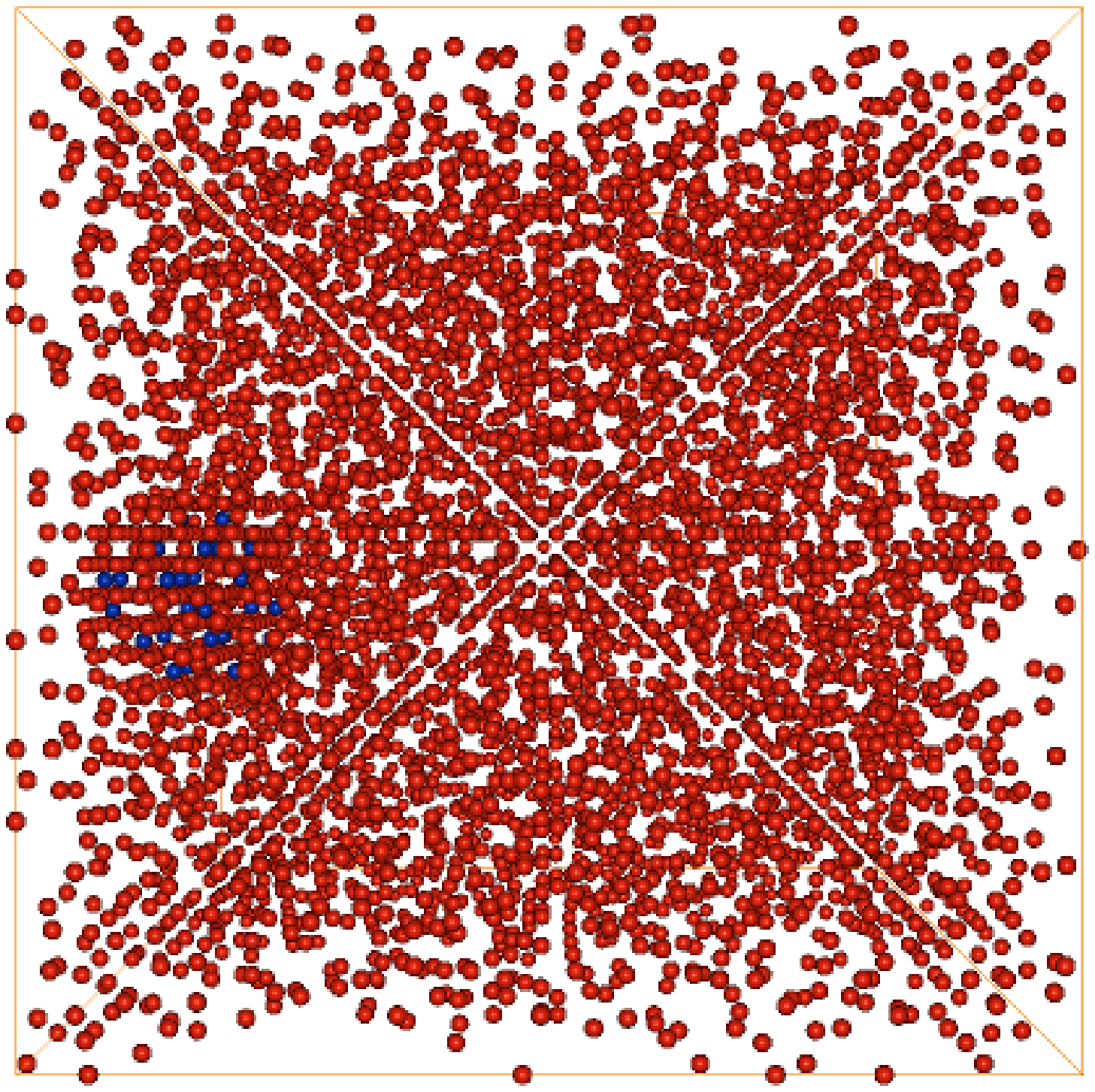} \\
     \hline
     0.1\% at. Vac & \includegraphics[width=.2\linewidth]{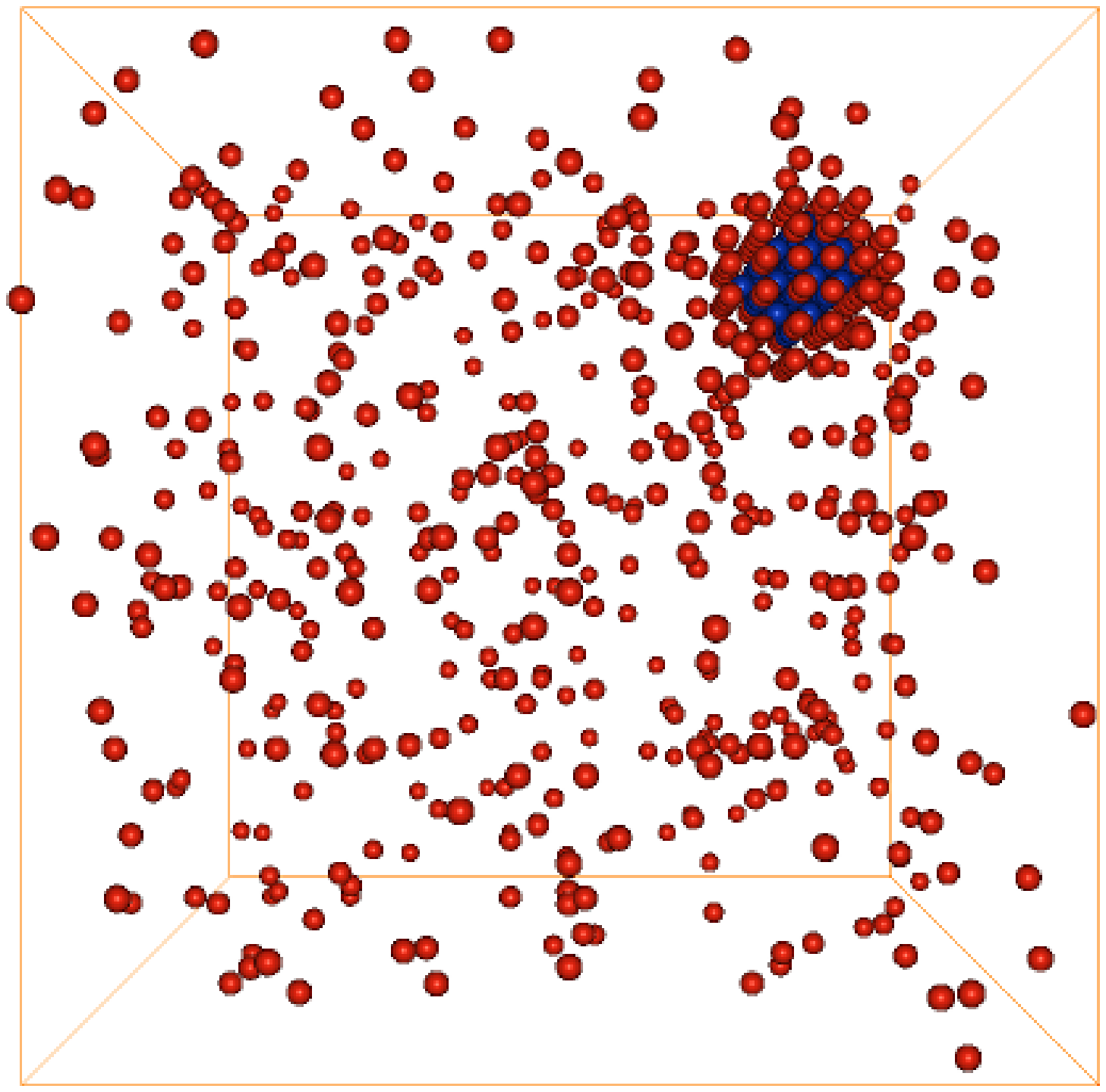} & \includegraphics[width=.2\linewidth]{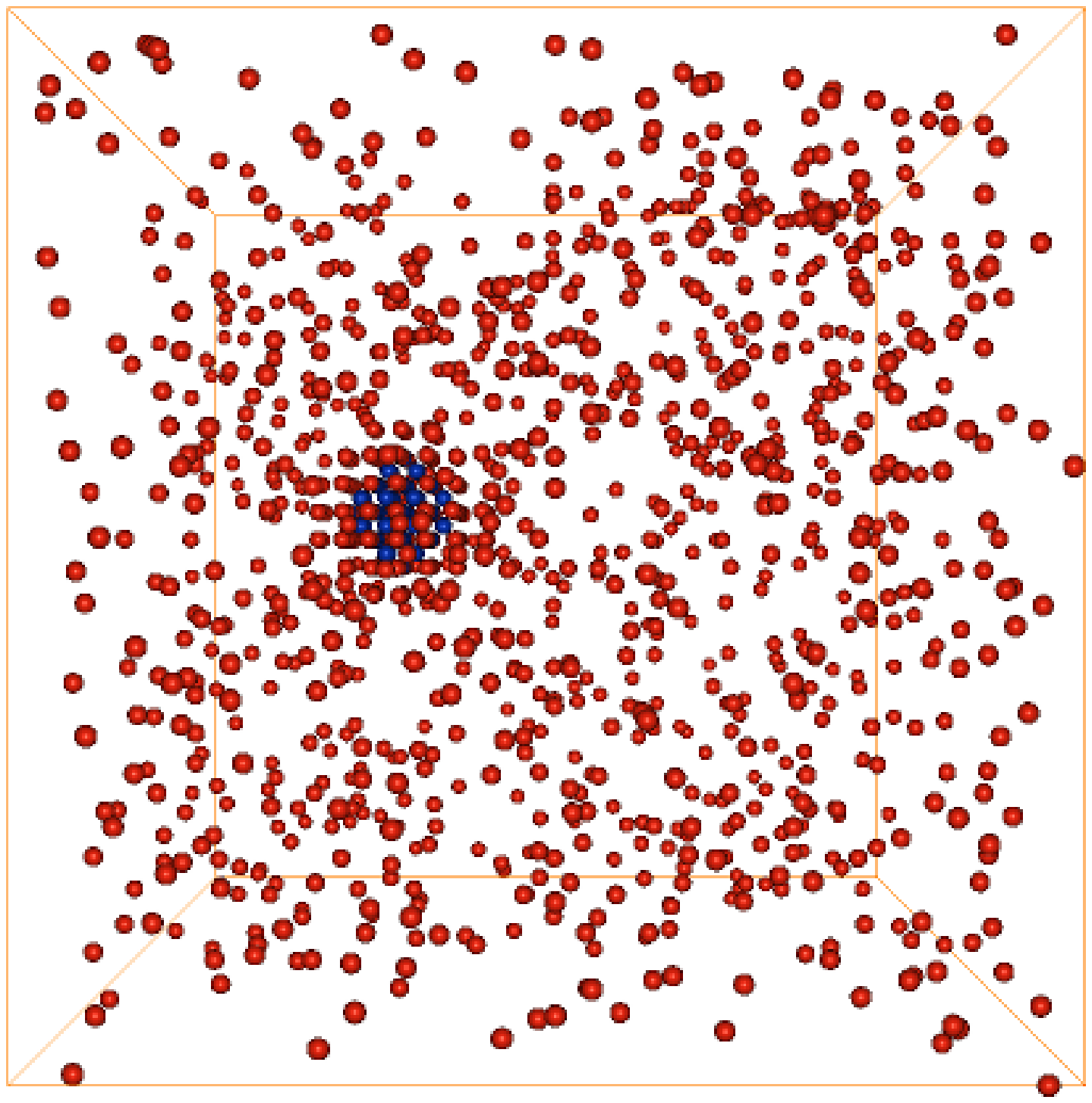} &  \includegraphics[width=.2\linewidth]{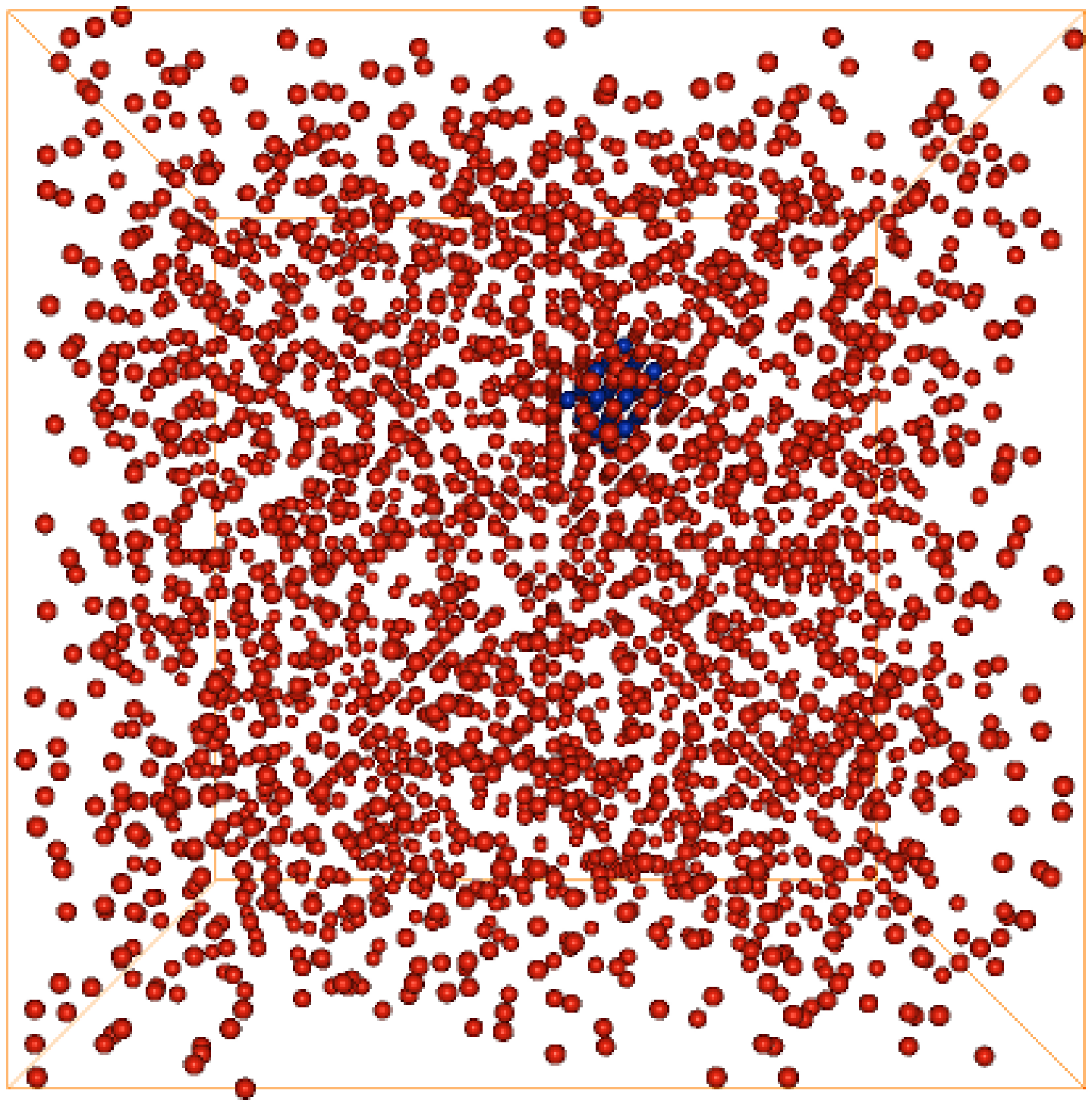}  & \includegraphics[width=.2\linewidth]{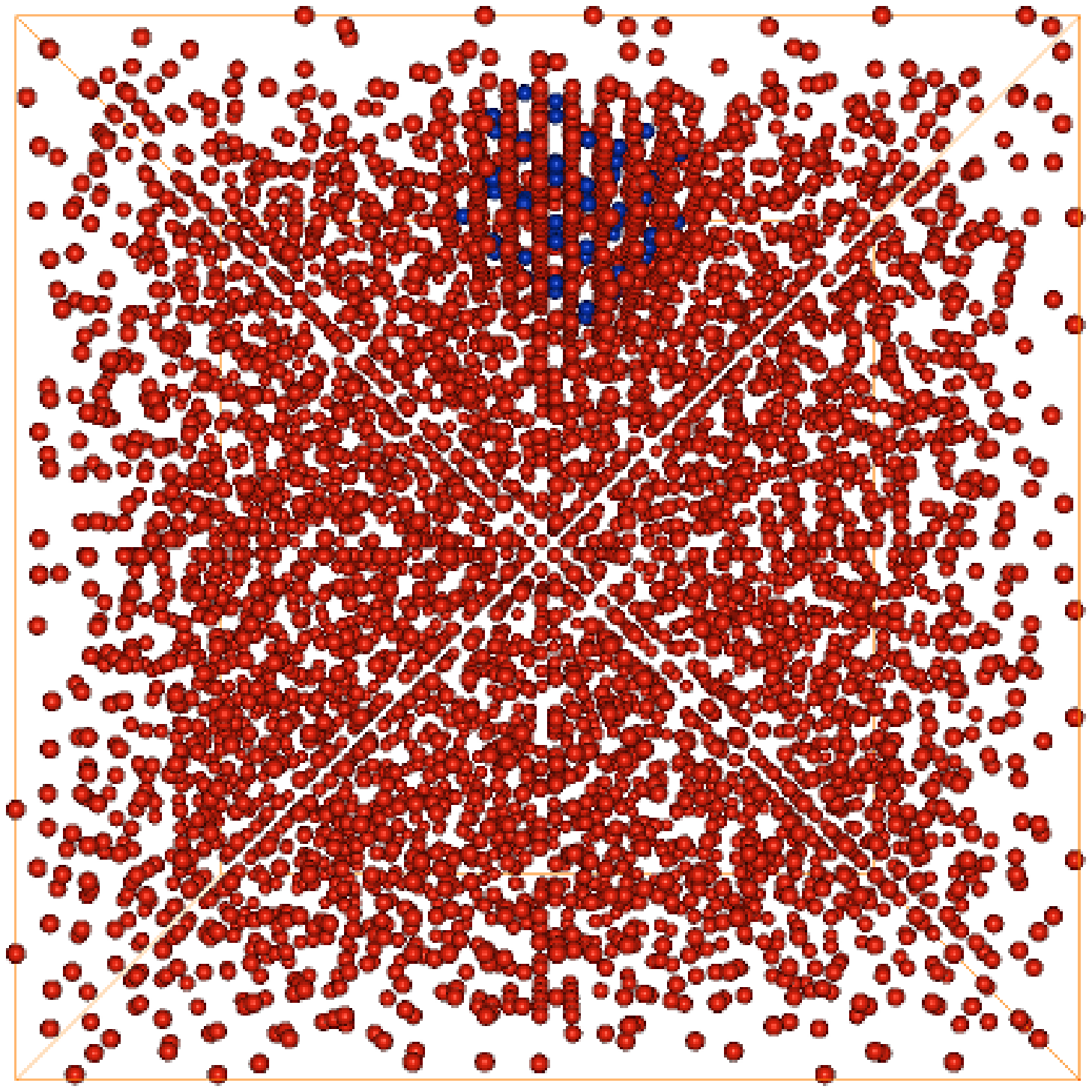} \\
     \hline
     0.1\% at. Vac & \includegraphics[width=.2\linewidth]{WRe1Vac01_100K} & \includegraphics[width=.2\linewidth]{WRe2Vac01_100K} &  \includegraphics[width=.2\linewidth]{WRe5Vac01_100K}  & \includegraphics[width=.2\linewidth]{WRe10Vac01_100K} \\
     \hline
     0.2\% at. Vac & \includegraphics[width=.2\linewidth]{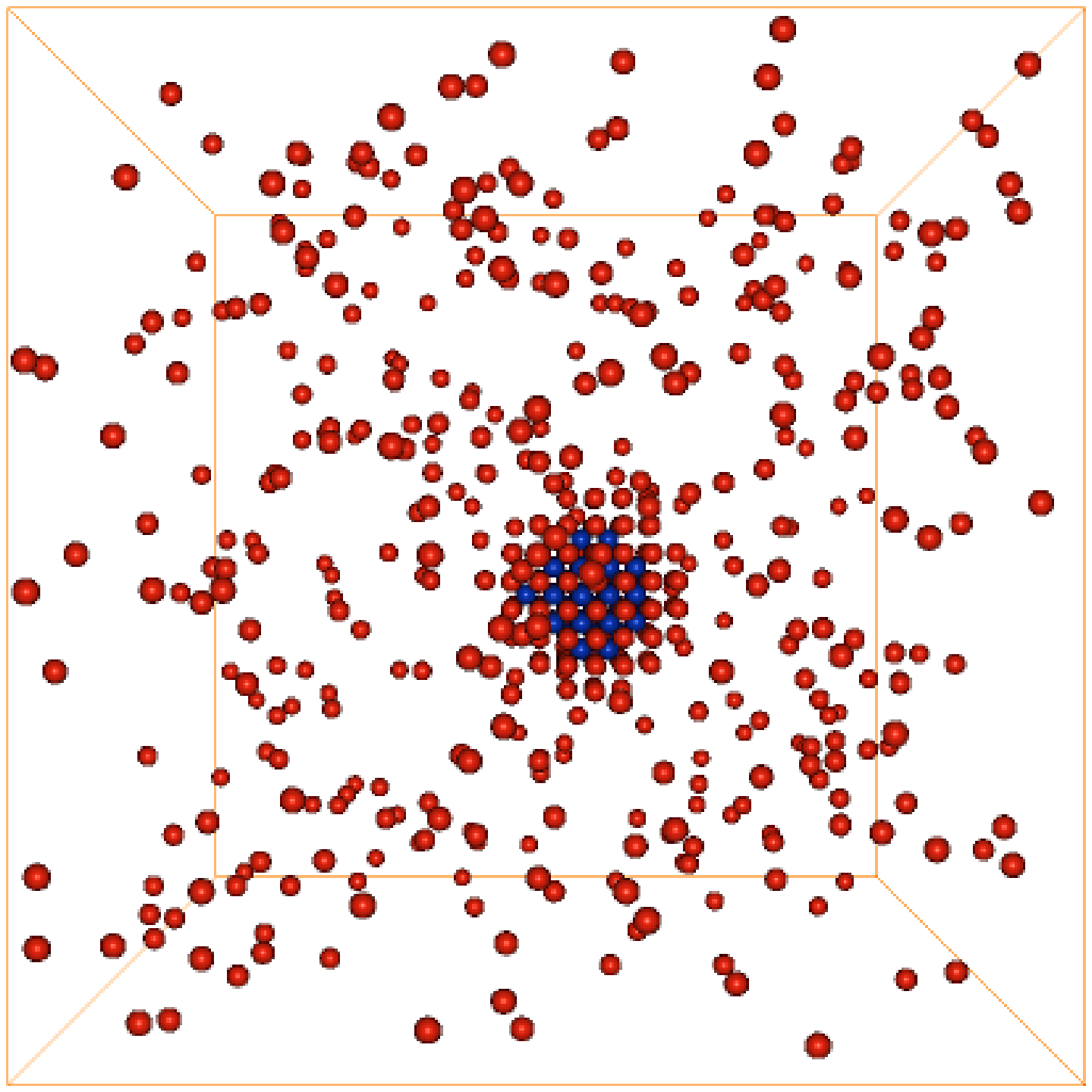} & \includegraphics[width=.2\linewidth]{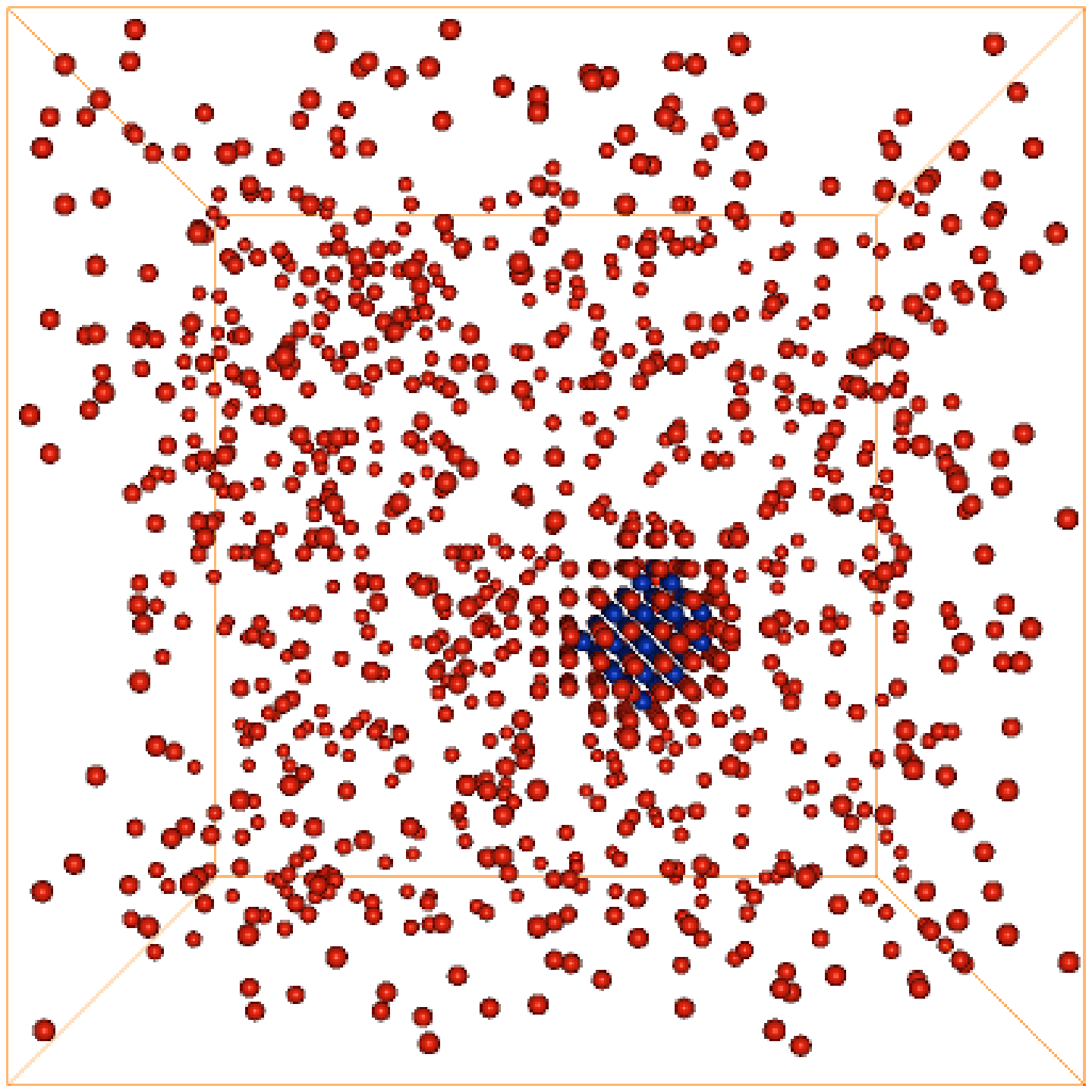} &  \includegraphics[width=.2\linewidth]{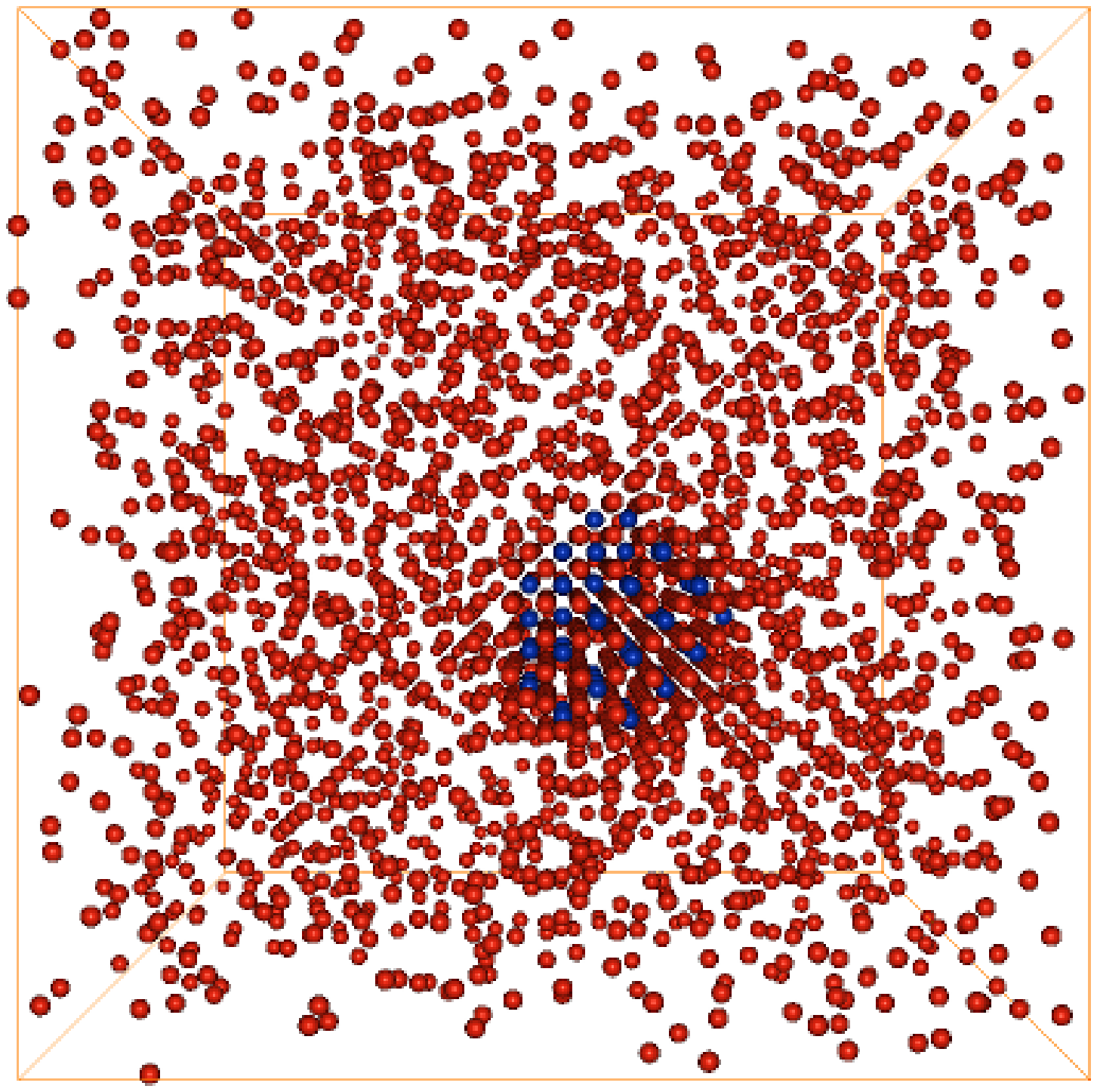} & \includegraphics[width=.2\linewidth]{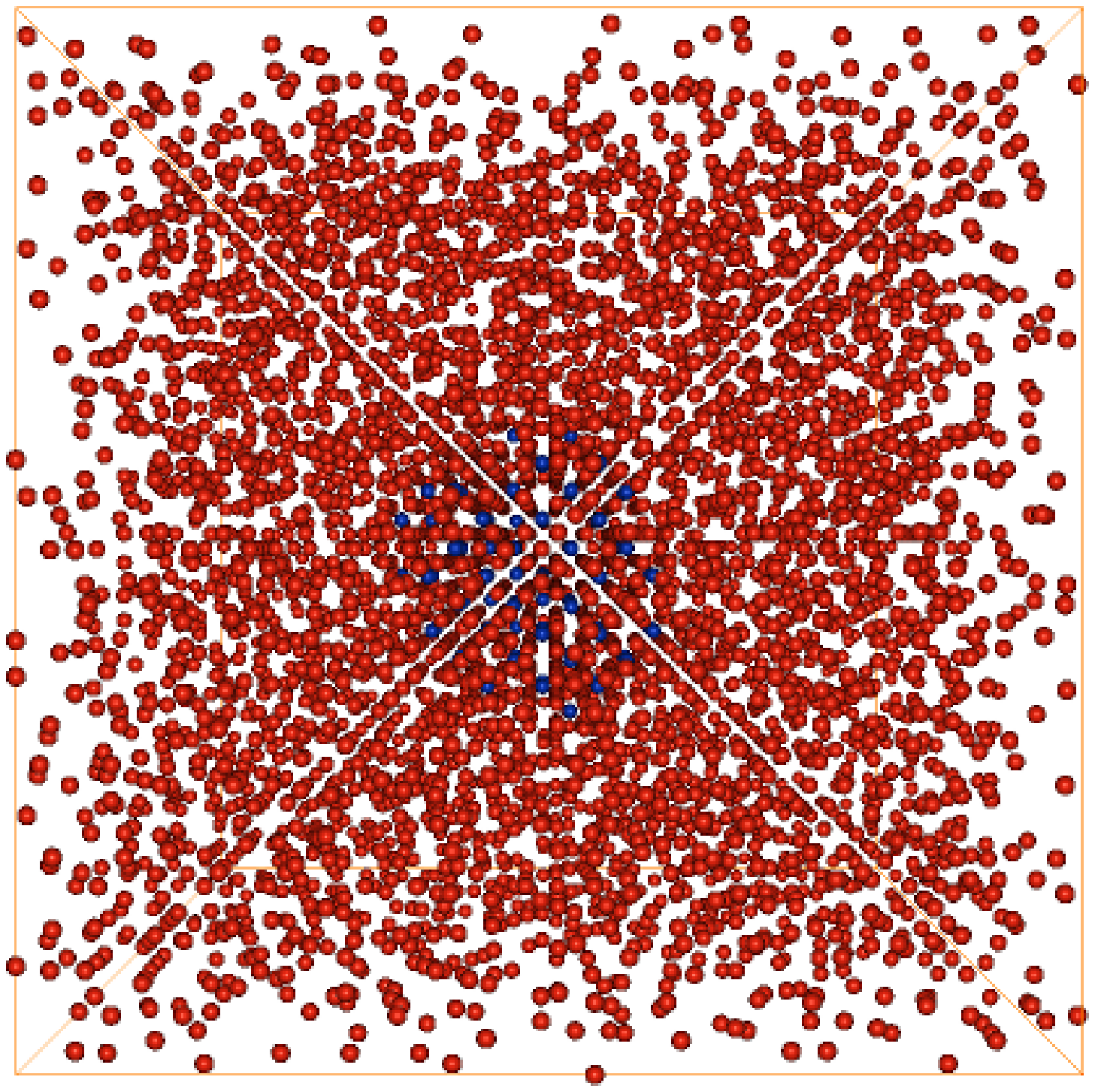} \\
    \end{tabular}%
\end{ruledtabular}
\end{table*}

Table \ref{tab:MC_results_v1} describes structures of W-Re-vacancy alloys at 100K containing 54000 sites derived from MC simulations by quenching down from 2500 K with the temperature step of 100K and shown as functions of concentration of Re atoms and vacancies. The most stable configuration of Re-vacancy clusters at 100K depends on the concentration of Re atoms. For each vacancy concentration, including 0.05 \%, 0.1 \% and 0.2 \%, for low concentrations of Re atoms, 1 \% or 2 \% at. Re, the most stable configurations are voids surrounded by Re atoms. For larger Re concentrations, 5 \% or 10 \% at. Re, vacancies prefer to be further apart from each other, and the most stable configurations are sponge-like Re-vacancy clusters. Those results are in agreement with the analysis of binding energies of Re-vacancy configurations at 0K treated as a function of Re/vacancy ratio described in Section III.

\subsection{Meta-stable finite temperature configurations}

In this sub-section, to understand the anomalous segregation of rhenium occurring in W-Re alloys under irradiation condition where alloys are not quenched from high temperatures but are irradiated at a constant temperature, we performed MC simulations at various temperatures assuming a certain concentration of vacancies and Re atoms.

Table \ref{tab:MC_results_const_Re} shows structures of W-Re alloys with 2 \% at. Re obtained from MC simulations performed assuming $T=$300 K, 800 K, 1600 K and 2500 K and four different concentrations of vacancies: 0.05 \%, 0.1 \%, 0.2 \% and 0.5 \%, which can be related to various values of irradiation dose rates. At the smallest concentration of vacancies of 0.05 \%, which corresponds to the lowest irradiation dose rate, clustering of vacancies in the form of sponge-like Re-vacancy clusters is observed only at 800 K. At 300K Re atoms are bound to vacancies but do not aggregate. Formation of Re-vacancy clusters becomes significantly more pronounced as the concentration of vacancies increases. If it exceeds 0.1\%, at 1600 K vacancies form voids, whereas at 300 K they aggregate into sponge-like Re-vacancy clusters. An interesting result is that in WRe2 alloy at 600 K we observe either the formation of voids surrounded by Re atoms or sponge-like Re-vacancy clusters. The former ones are observed in alloys where vacancy concentration is equal to 0.2\% and 0.5\%, whereas the latter ones form in alloys with lower concentration of vacancies in the range from 0.05 \% to 0.1 \%. This can be explained by the fact that a high concentration of vacancies corresponds to low Re/vacancy ratio, which is the key parameter controlling the stability of Re-vacancy configurations  described in Section III. Another important conclusion is that on the border between the regions of stability of voids and sponge-like Re-vacancy clusters it is possible to find both types of configurations. At 2500K both vacancies and Re atoms are fully dissolved.

\begin{table*}
\caption{Results of Monte Carlo simulations as functions of vacancy concentration and temperature. Concentration of Re atoms is fixed and equal to 2\%. MC simulations were performed at various temperatures and screen shots were taken after 20000 MC steps per atom.
        \label{tab:MC_results_const_Re}}
\begin{ruledtabular}
    \begin{tabular}{|c|c|c|c|c|}
              & 0.05\% at. Vac & 0.1\% at. Vac & 0.2\% at. Vac  & 0.5\% at. Vac   \\
     \hline
     300 K & \includegraphics[width=.2\linewidth]{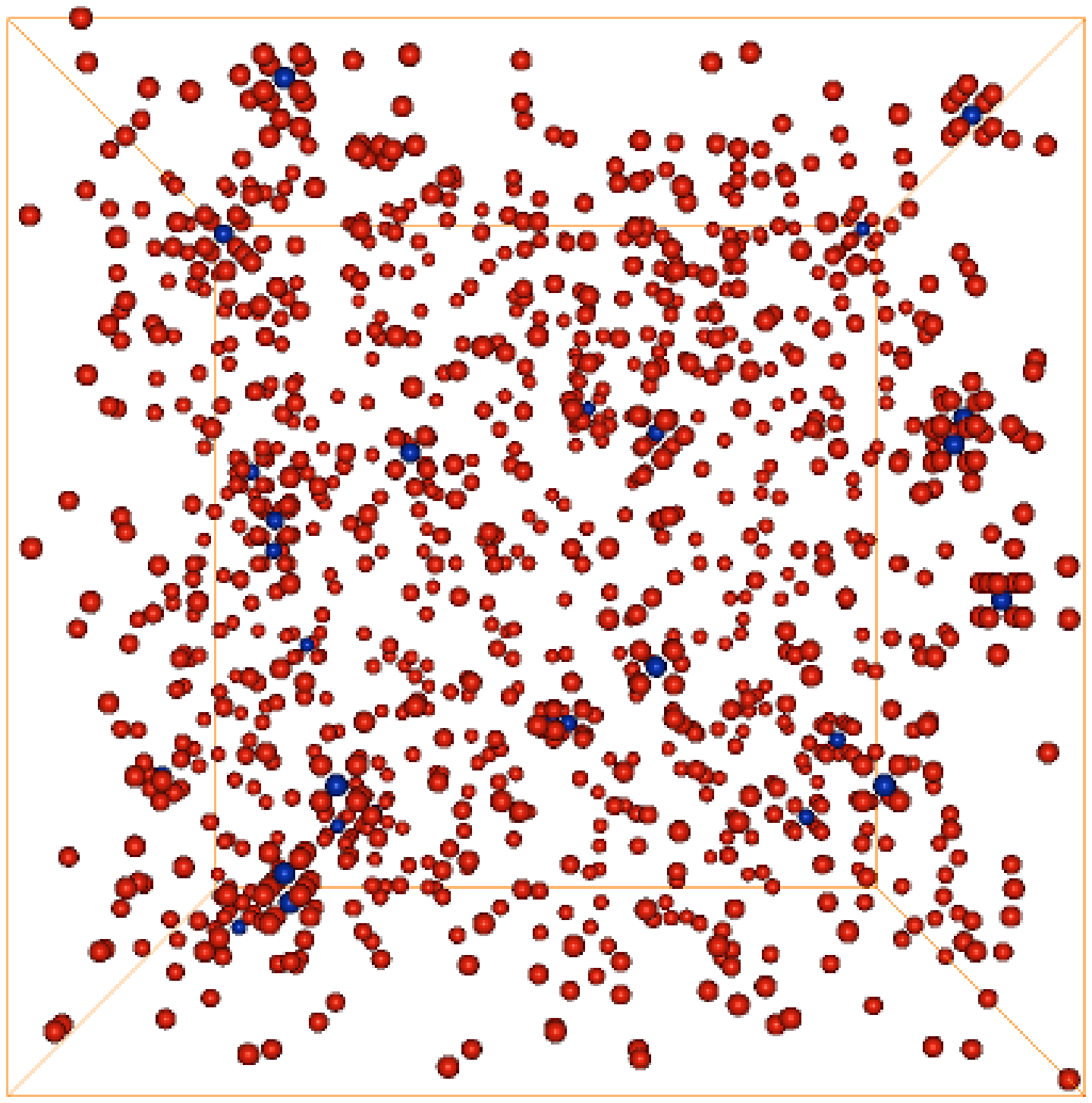} & \includegraphics[width=.2\linewidth]{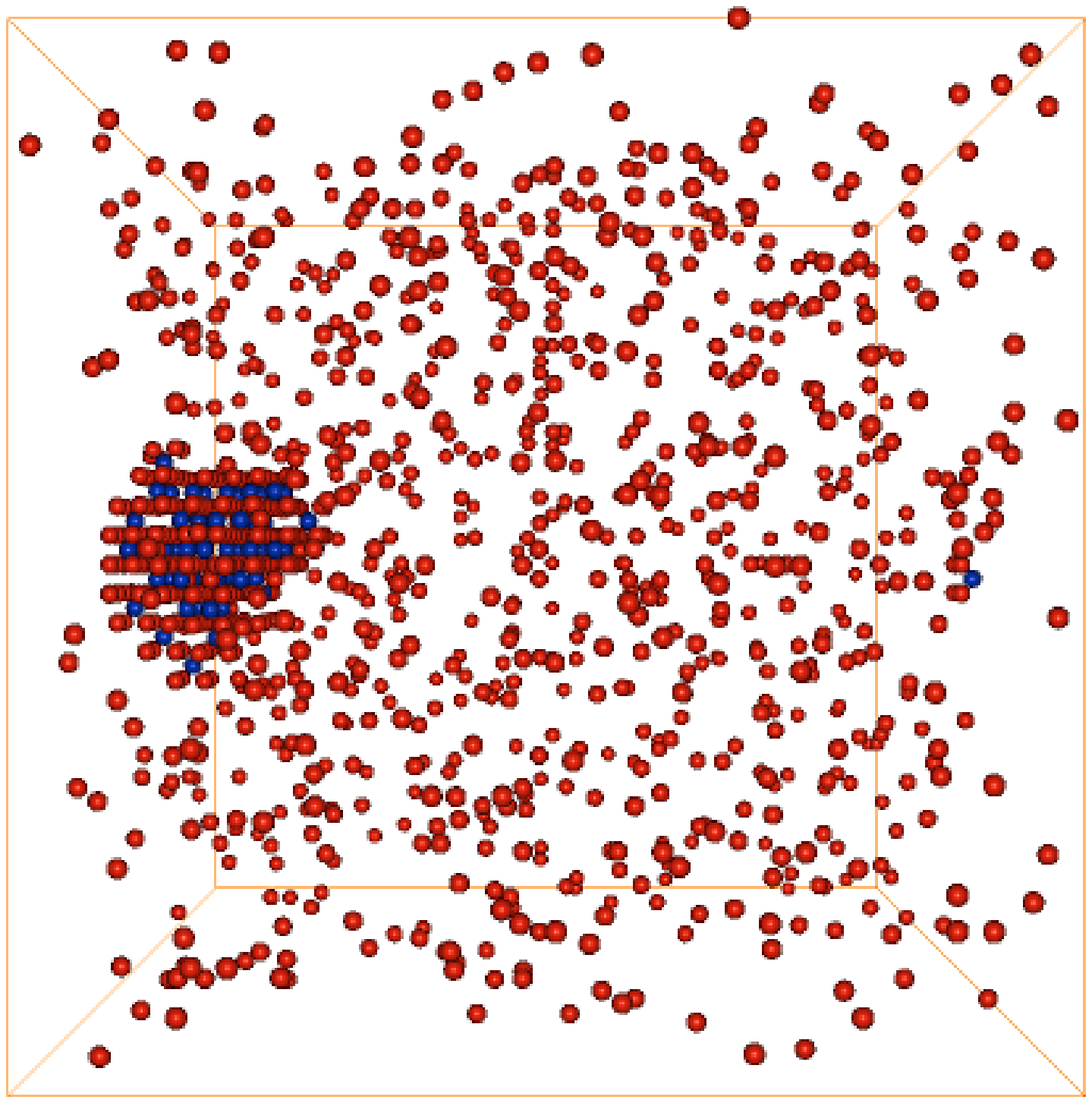} & \includegraphics[width=.2\linewidth]{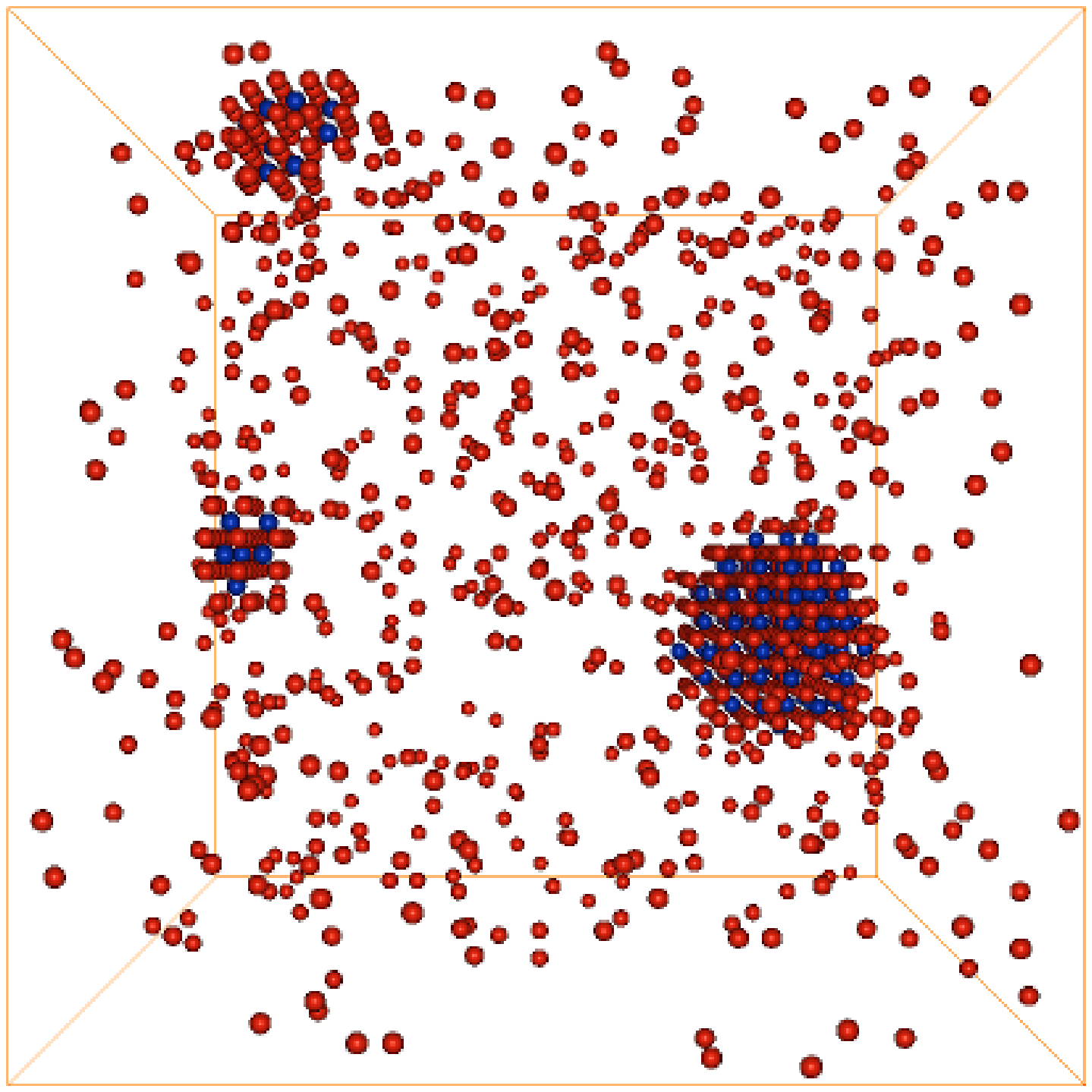} & \includegraphics[width=.2\linewidth]{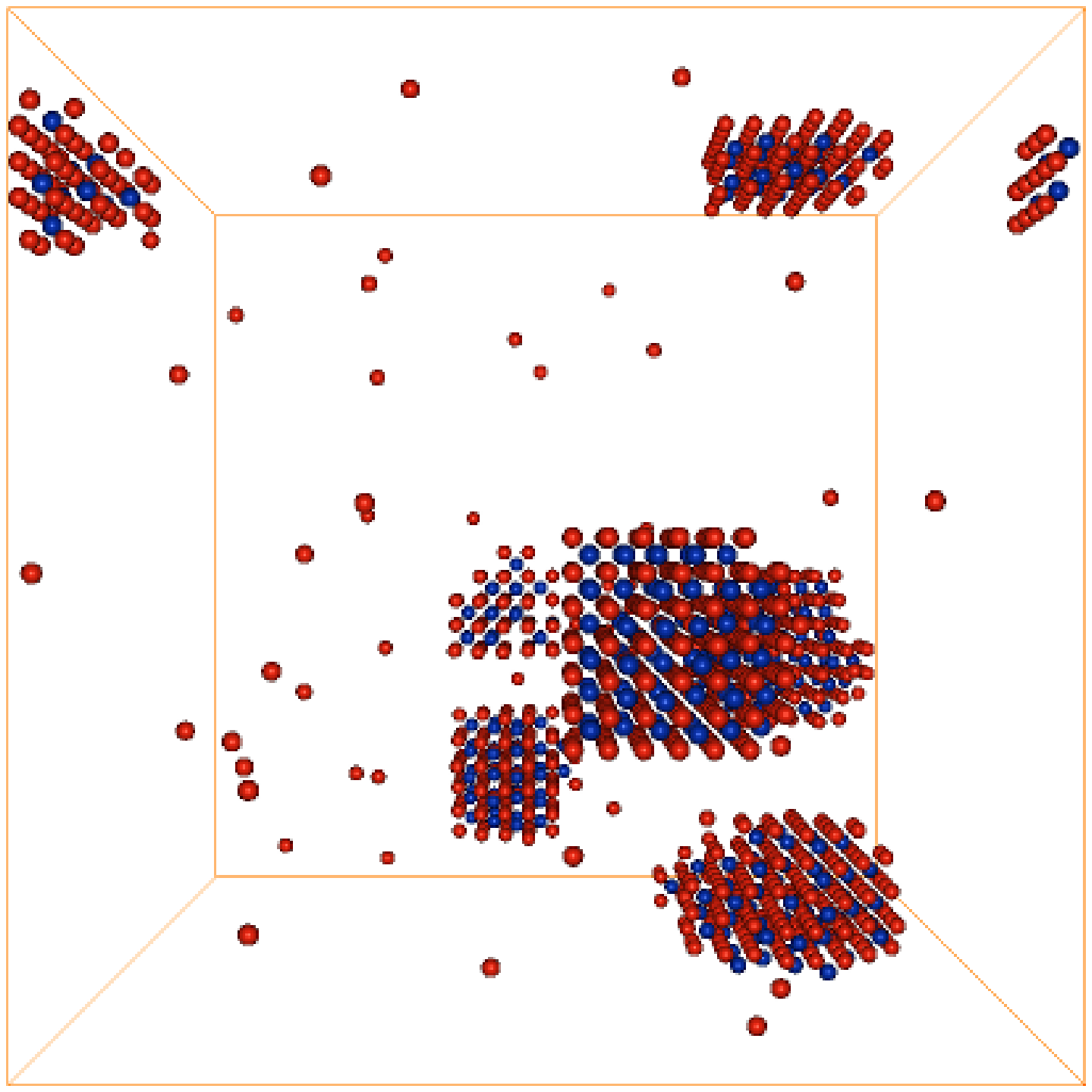} \\
     \hline
     800 K & \includegraphics[width=.2\linewidth]{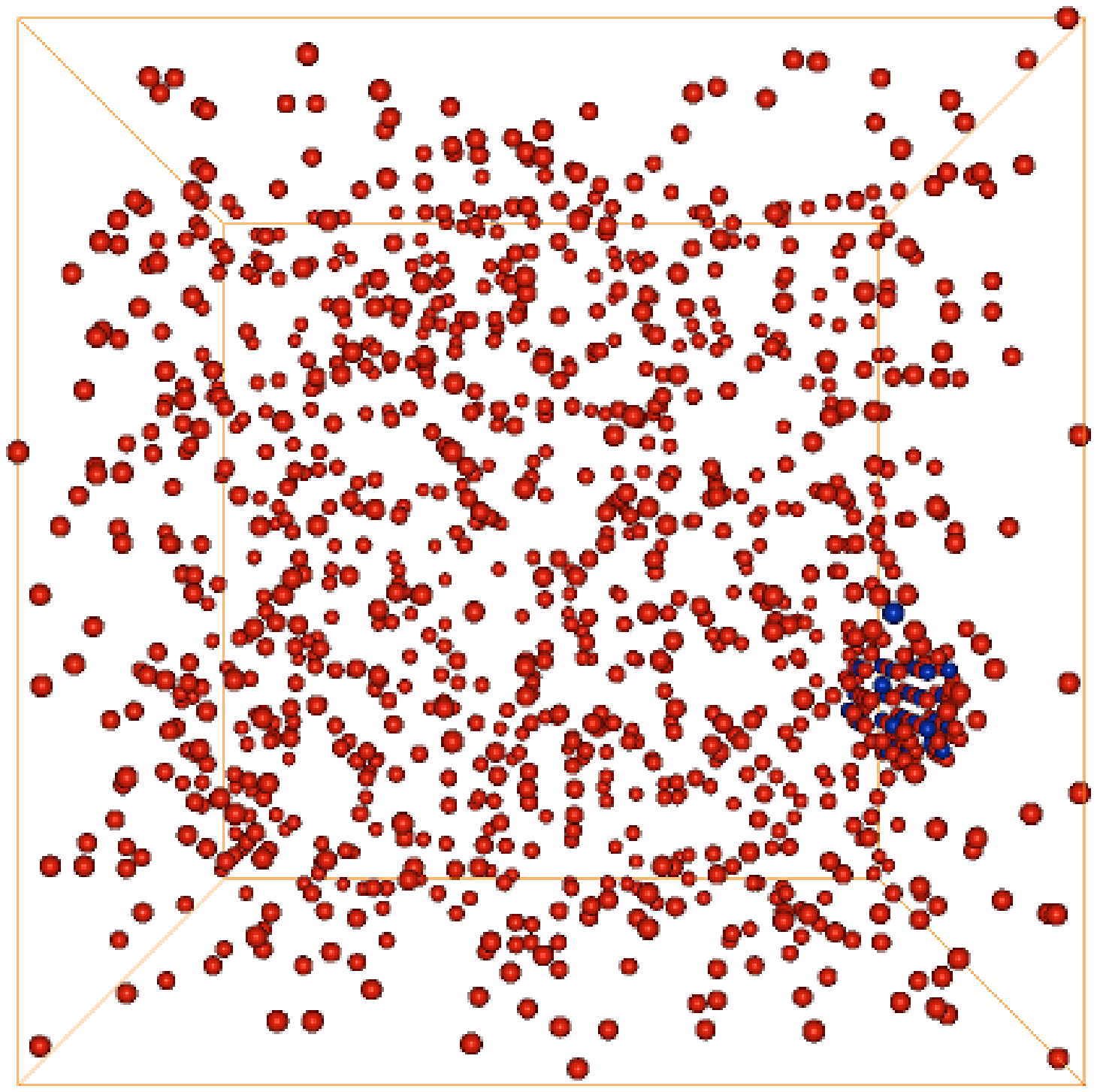} & \includegraphics[width=.2\linewidth]{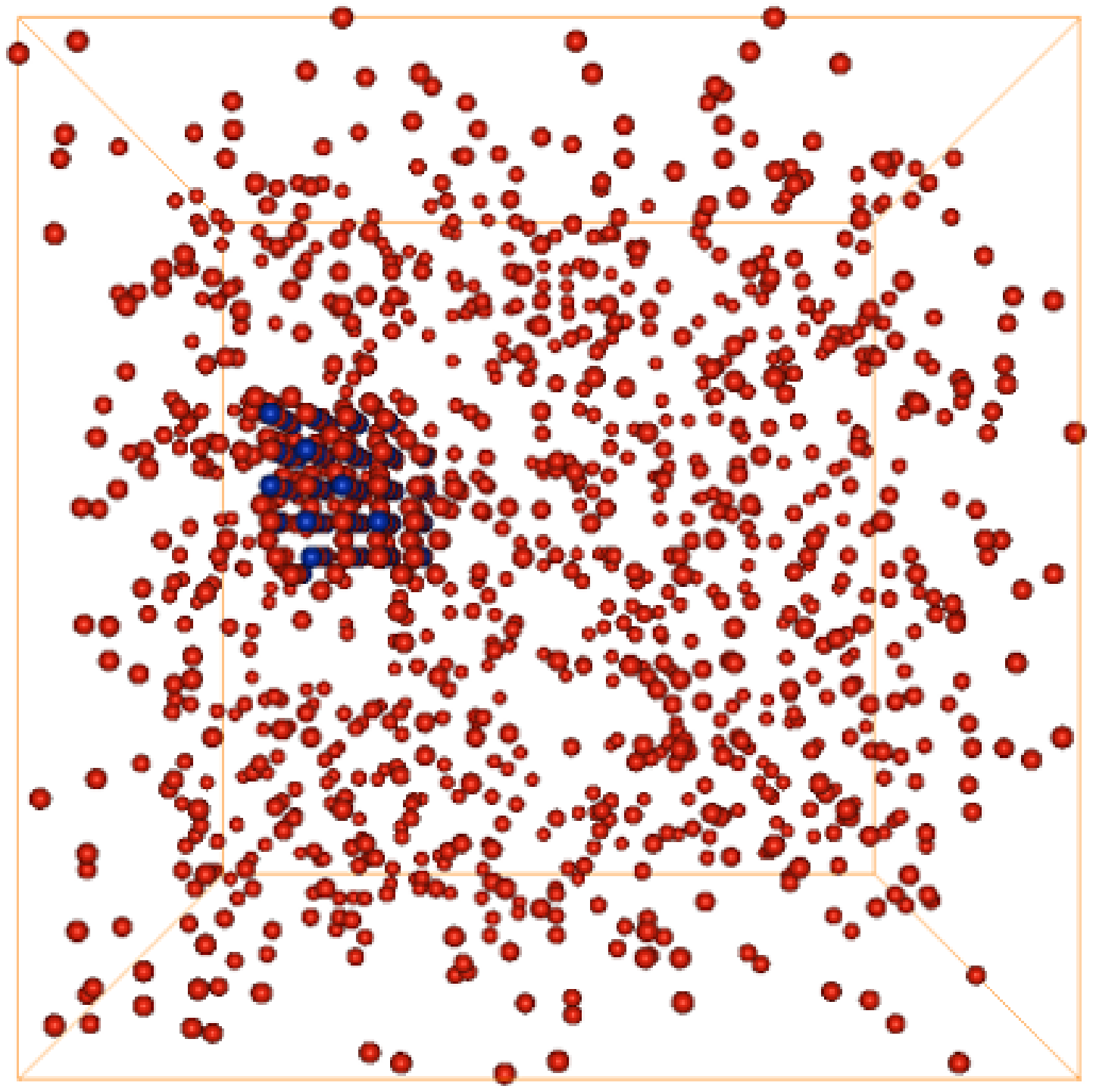} &  \includegraphics[width=.2\linewidth]{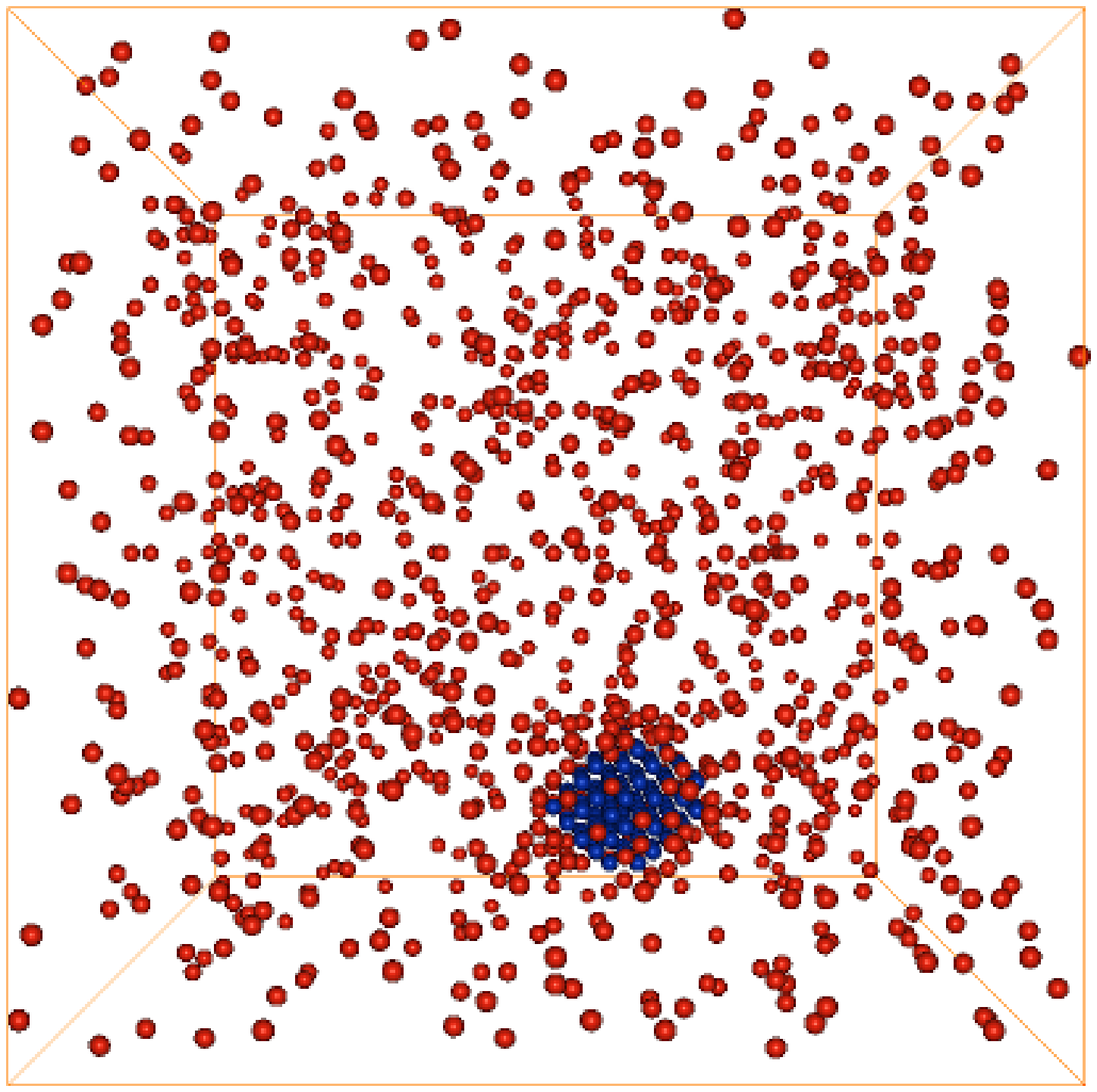}  & \includegraphics[width=.2\linewidth]{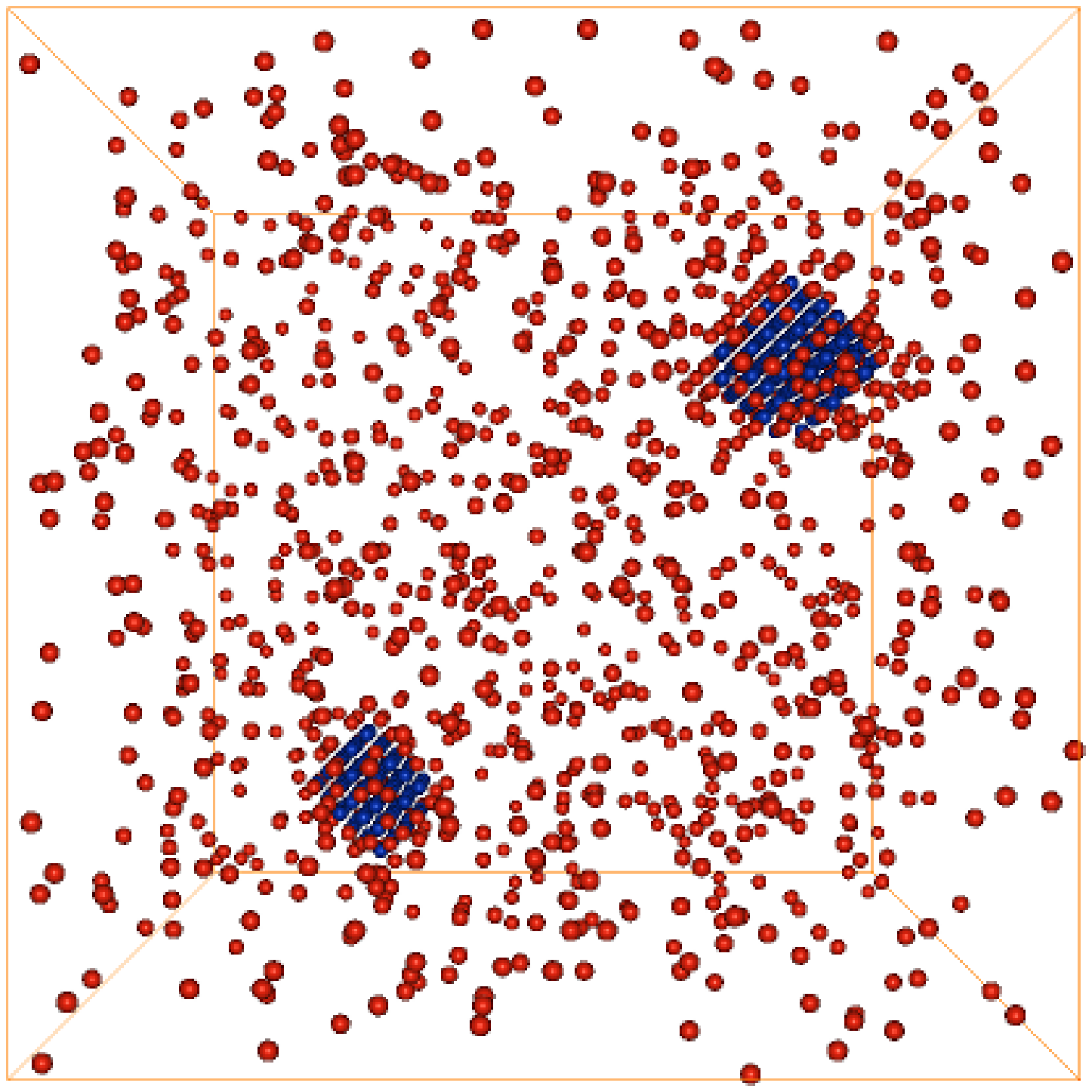} \\
     \hline
     1600 K & \includegraphics[width=.2\linewidth]{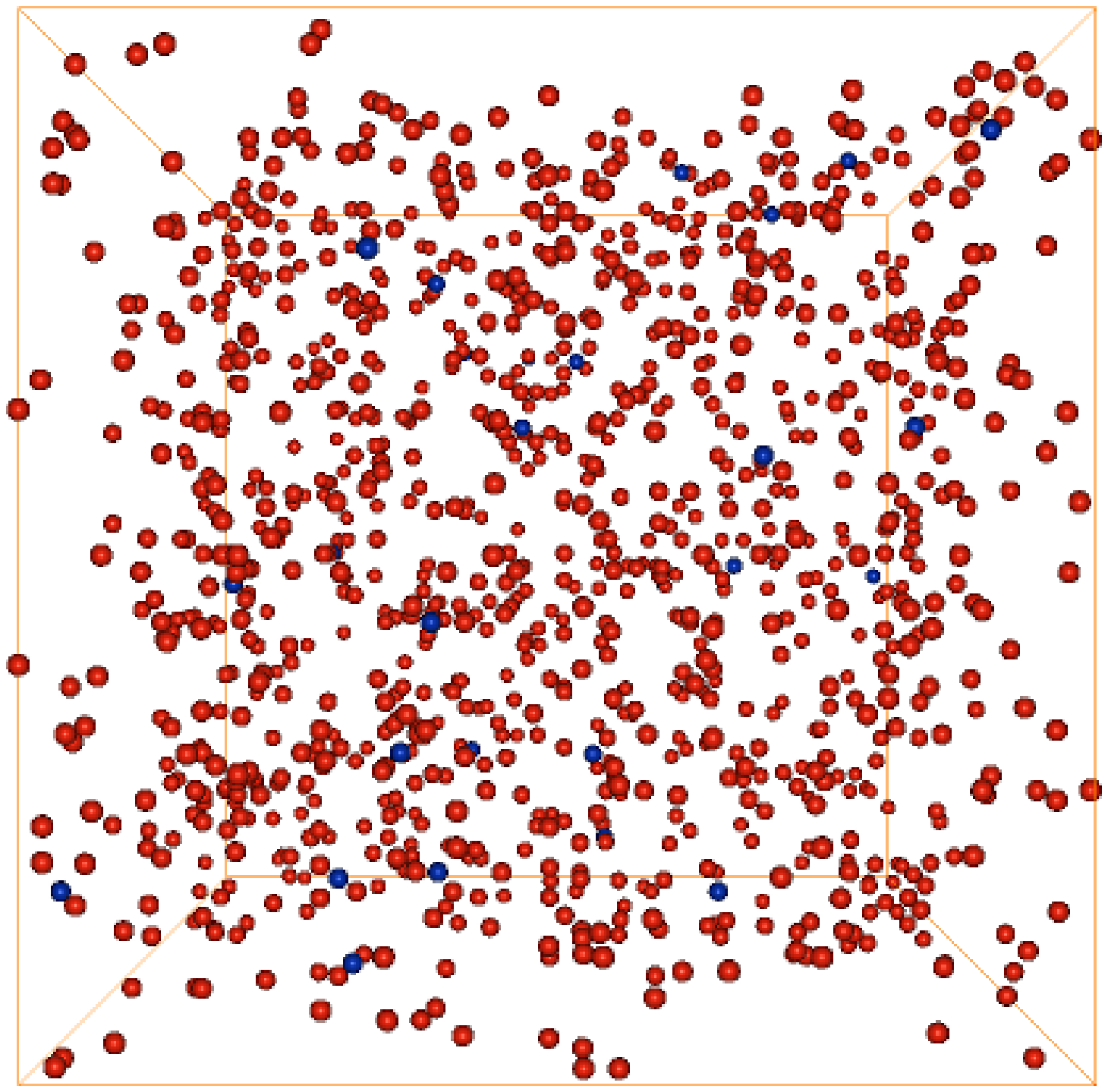} & \includegraphics[width=.2\linewidth]{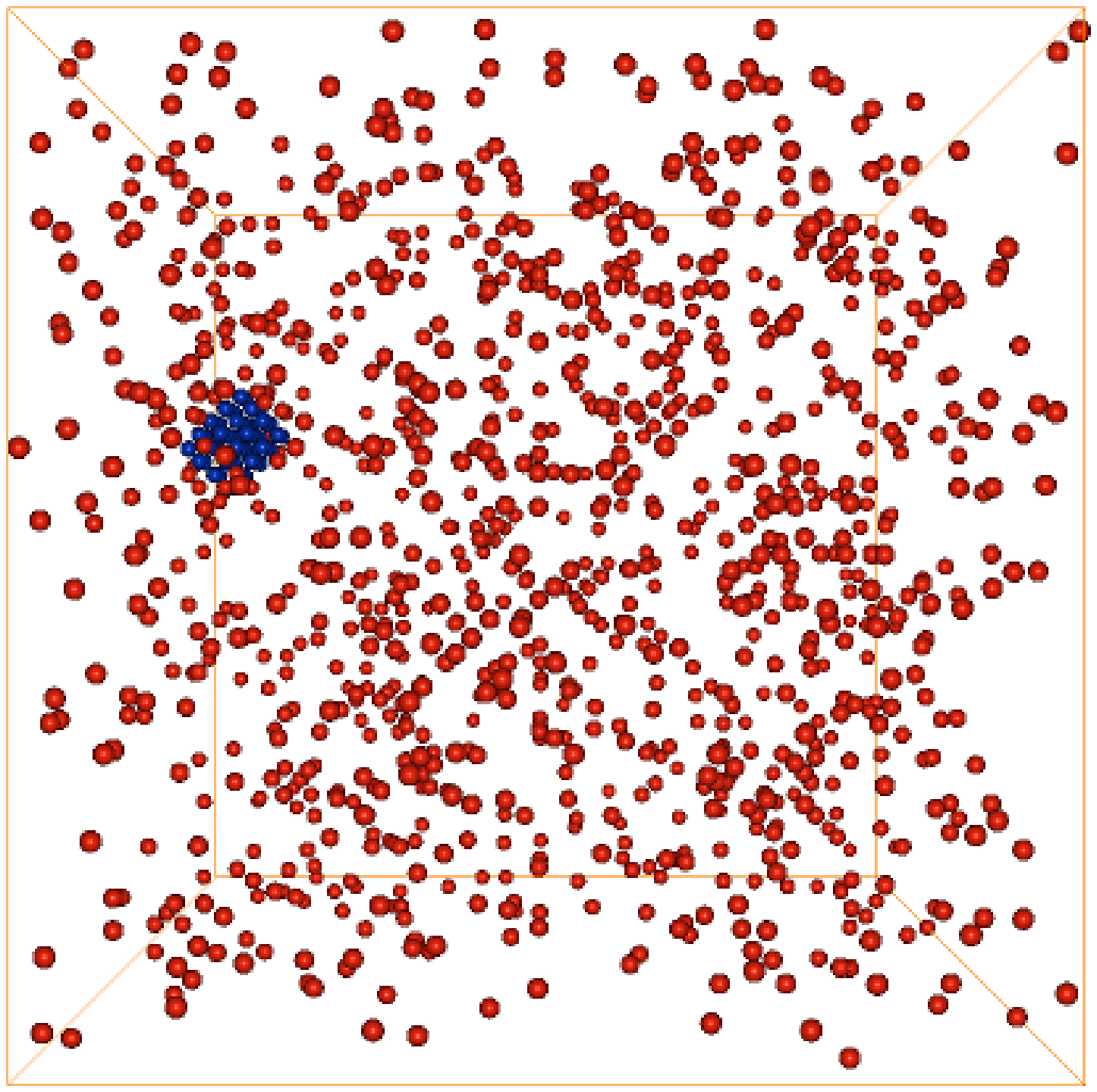} & \includegraphics[width=.2\linewidth]{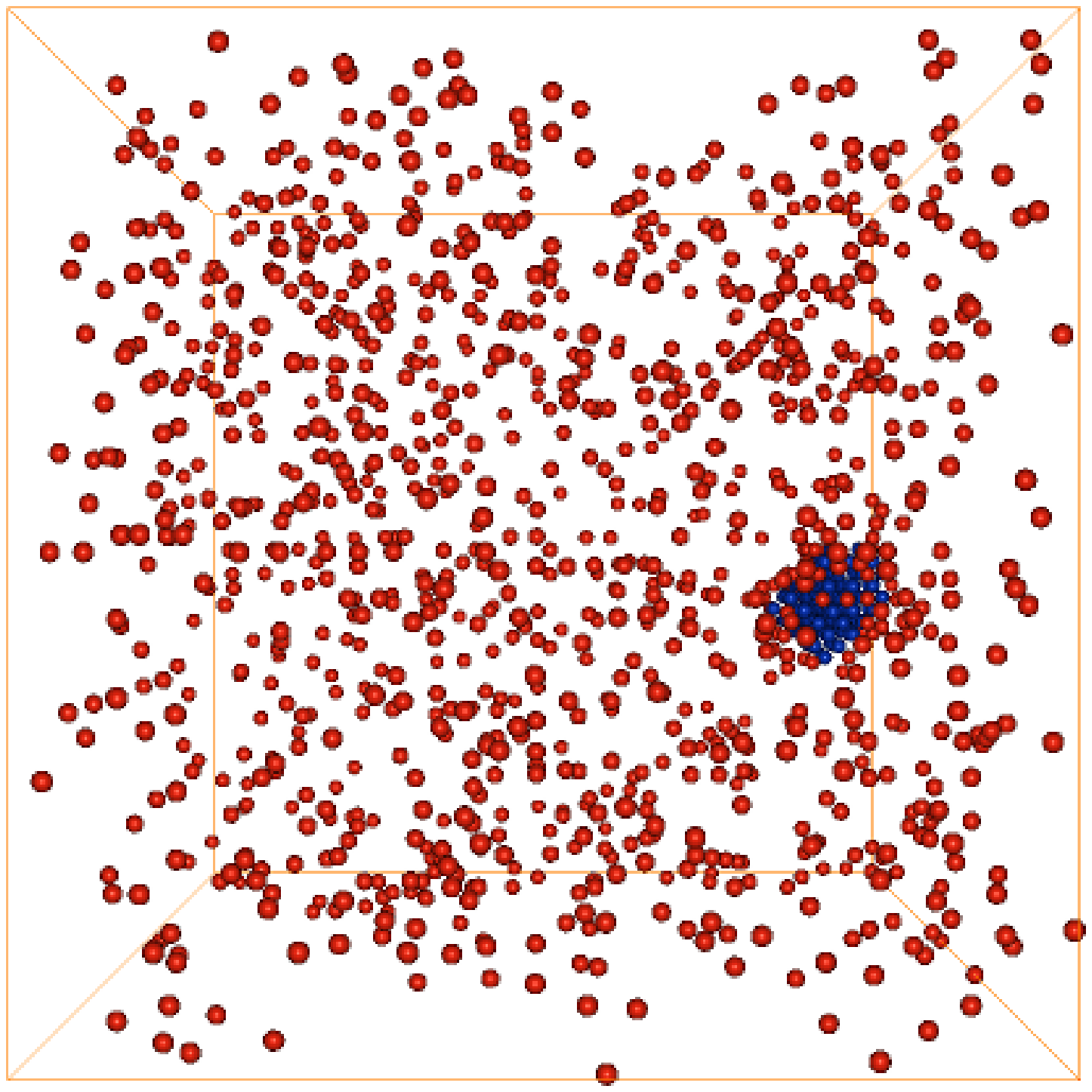} & \includegraphics[width=.2\linewidth]{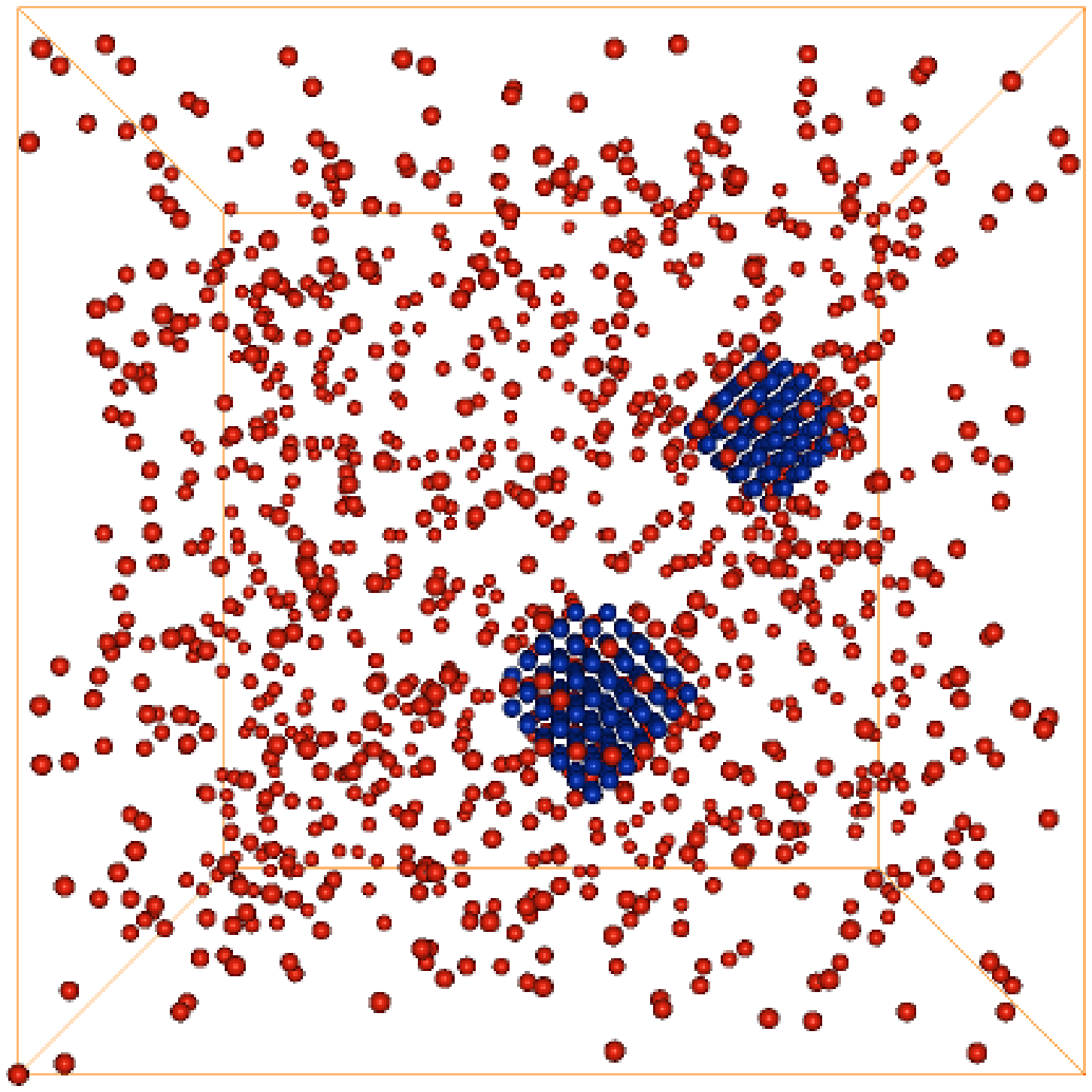} \\
     \hline
     2500 K & \includegraphics[width=.2\linewidth]{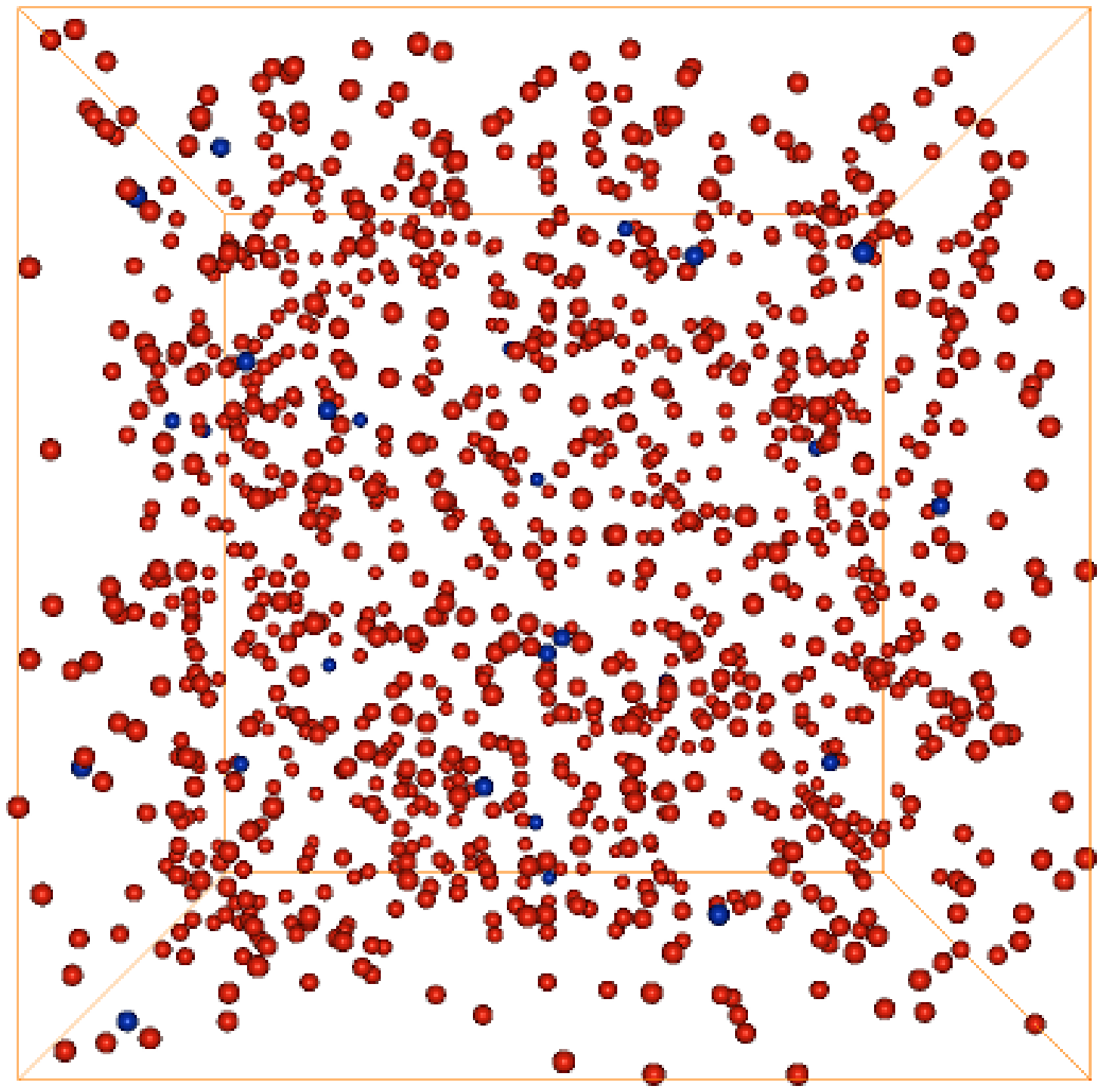} & \includegraphics[width=.2\linewidth]{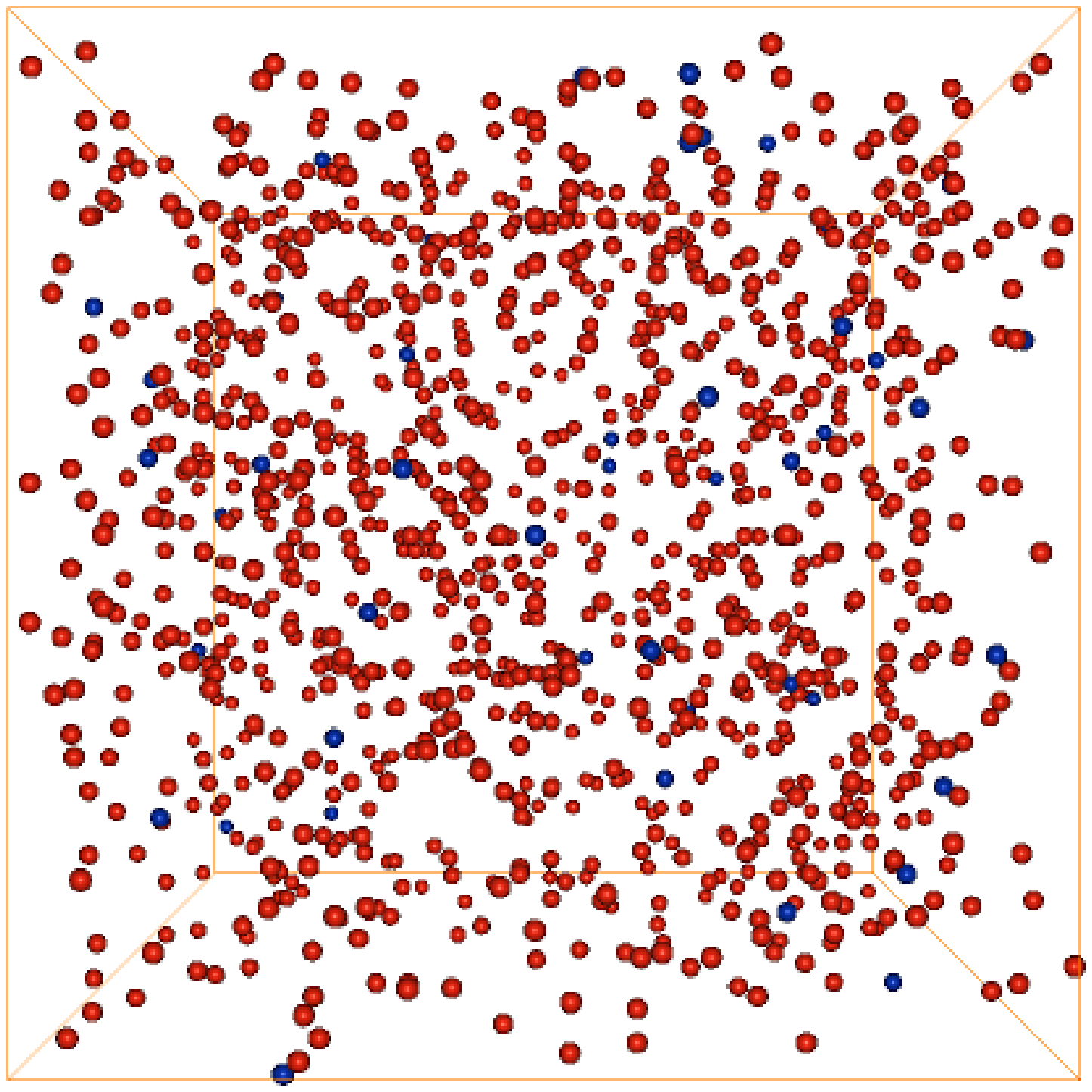} & \includegraphics[width=.2\linewidth]{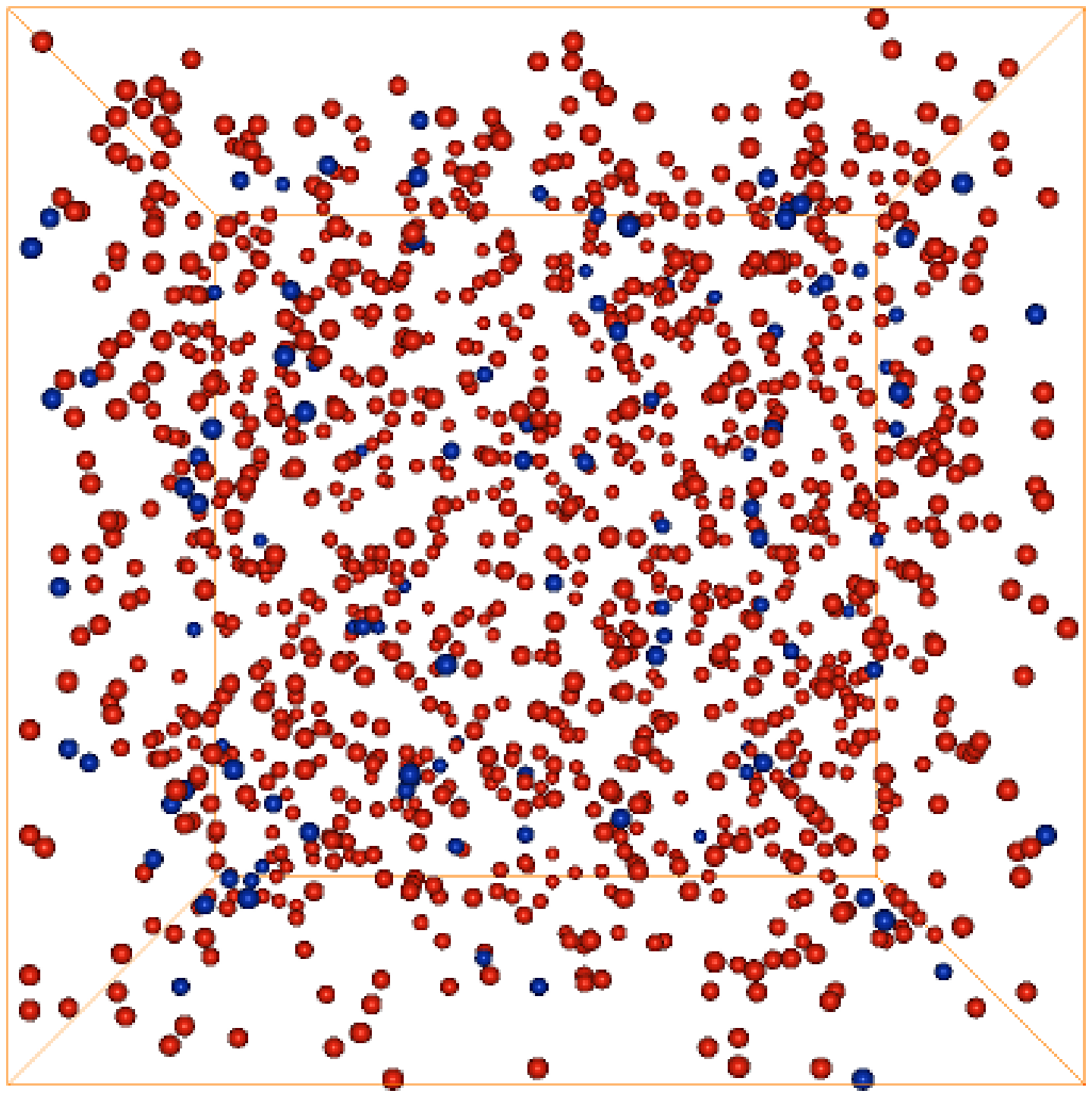}  & \includegraphics[width=.2\linewidth]{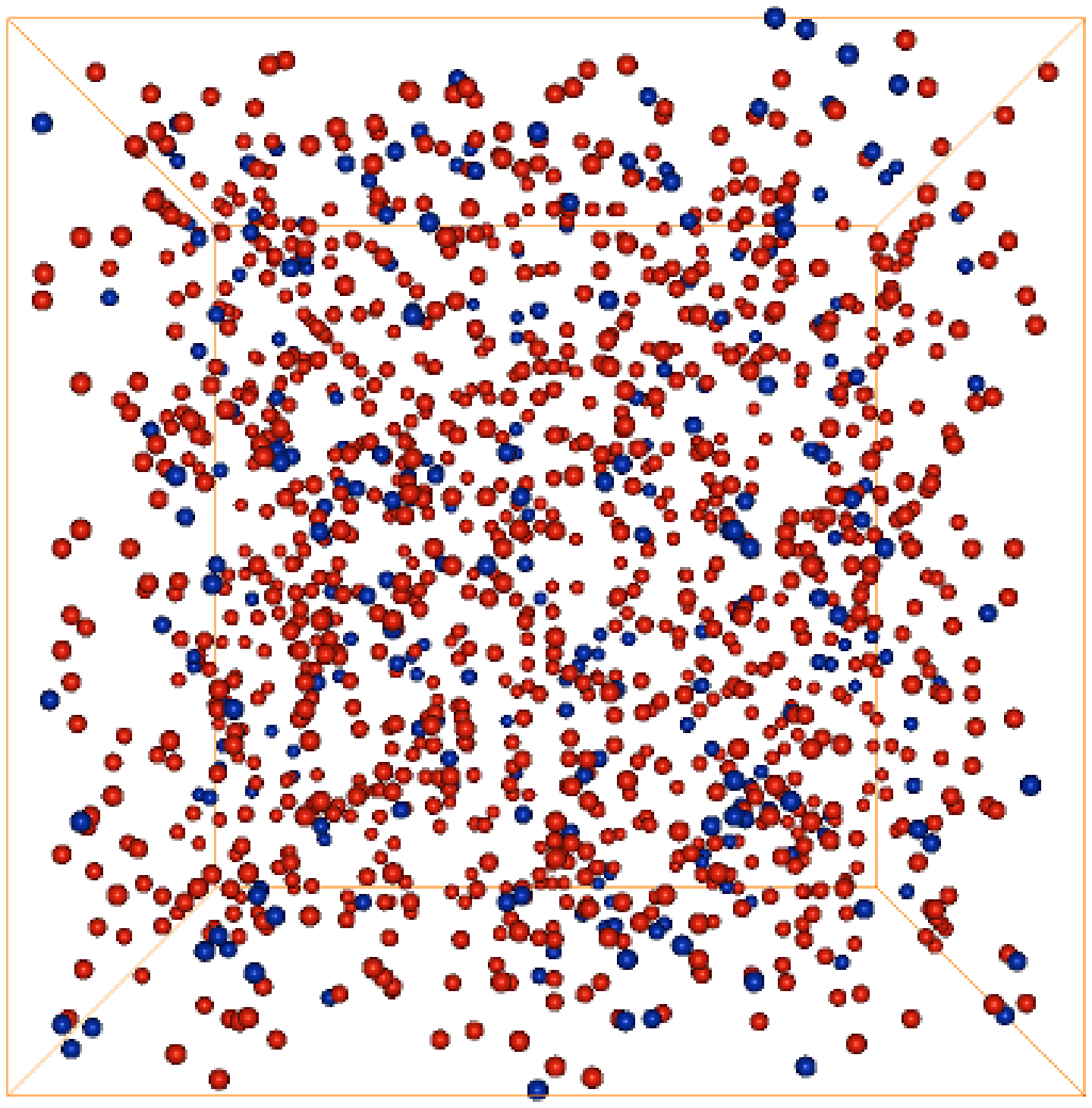} \\
    \end{tabular}%
\end{ruledtabular}
\end{table*}

Table \ref{tab:MC_results_const_Vac} shows structures of W-Re alloys with vacancy concentration of 0.1 \% obtained from MC simulations performed for temperatures of 300 K, 800 K, 1600 K and 2500 K for four different concentrations of Re atoms: 1 \%, 2 \%, 5 \% and 10 \% at. Re. Similarly to the case of W-2\%Re alloys with different concentrations of vacancies, clustering of vacancies for each concentration of Re atoms is only observed at 800 K. At 2500 K all vacancies and Re atoms are fully dissolved. For Re concentrations of 5\% and 10\%, vacancies do not aggregate at 300K and 1600K. In W-Re alloys with atomic 1\% and 2\%  Re, sponge-like Re-vacancy clusters form at 300 K and voids form at 1600 K. Stability of Re-vacancy configurations at 800 K also depends on the Re/vacancy ratio. For low Re concentration close to 1 \%  the Re/vacancy ratio is small enough to favour the formation of a void surrounded by Re atoms. At higher Re concentrations and higher values of the Re/vacancy ratio, sponge-like Re-vacancy clusters are more stable.

\begin{table*}
\caption{Results of Monte Carlo simulations shown as functions of Re concentration and temperature. Concentration of vacancies is fixed and equal to 0.1\%. MC simulations were performed at fixed temperatures and the screen shots were taken after 20000 MC steps per atom.
        \label{tab:MC_results_const_Vac}}
\begin{ruledtabular}
    \begin{tabular}{|c|c|c|c|c|}
              & 1\% at. Re & 2\% at. Re & 5\% at. Re  & 10\% at. Re   \\
     \hline
     300 K & \includegraphics[width=.2\linewidth]{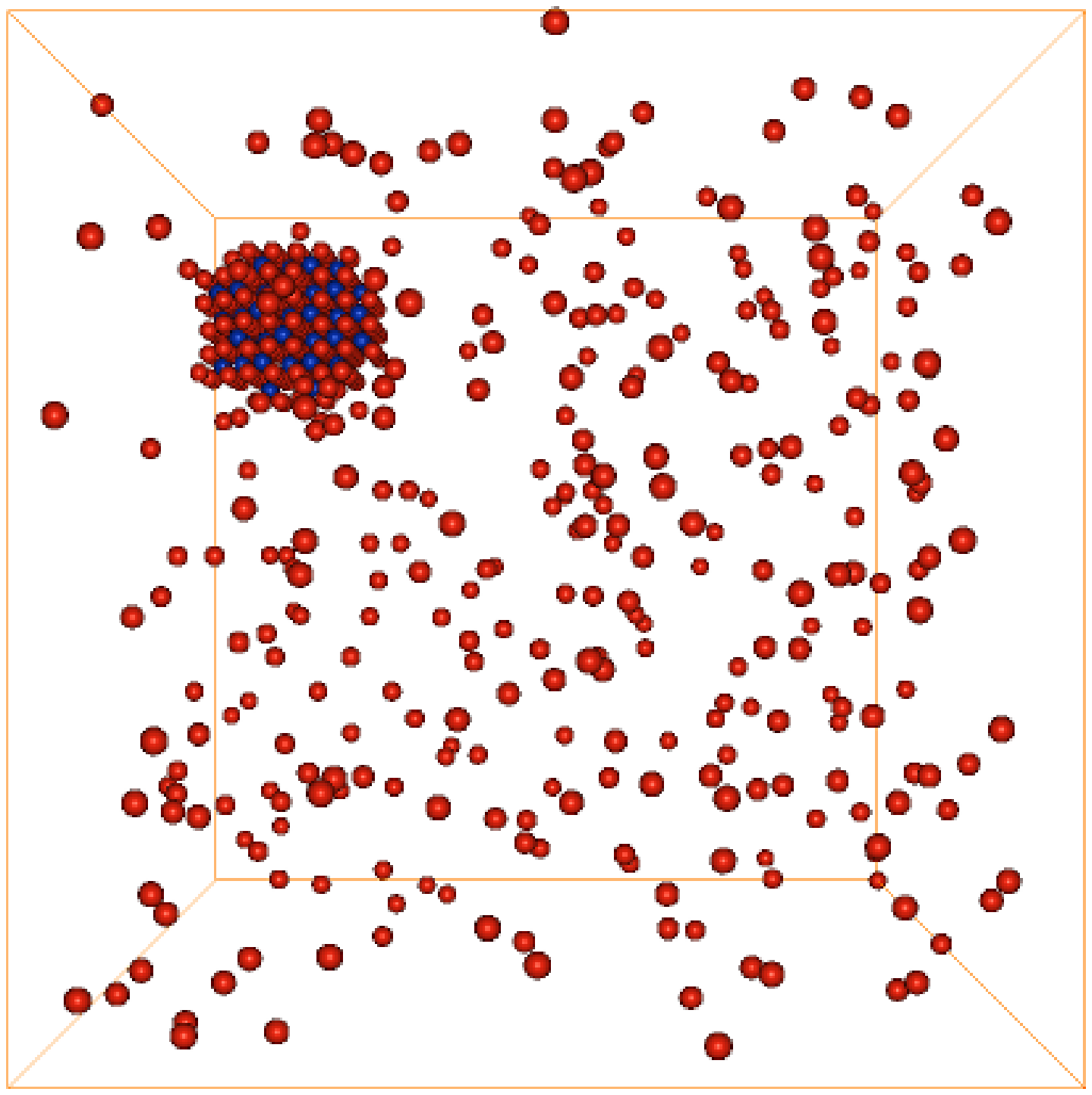} & \includegraphics[width=.2\linewidth]{WRe2Vac01_300K} & \includegraphics[width=.2\linewidth]{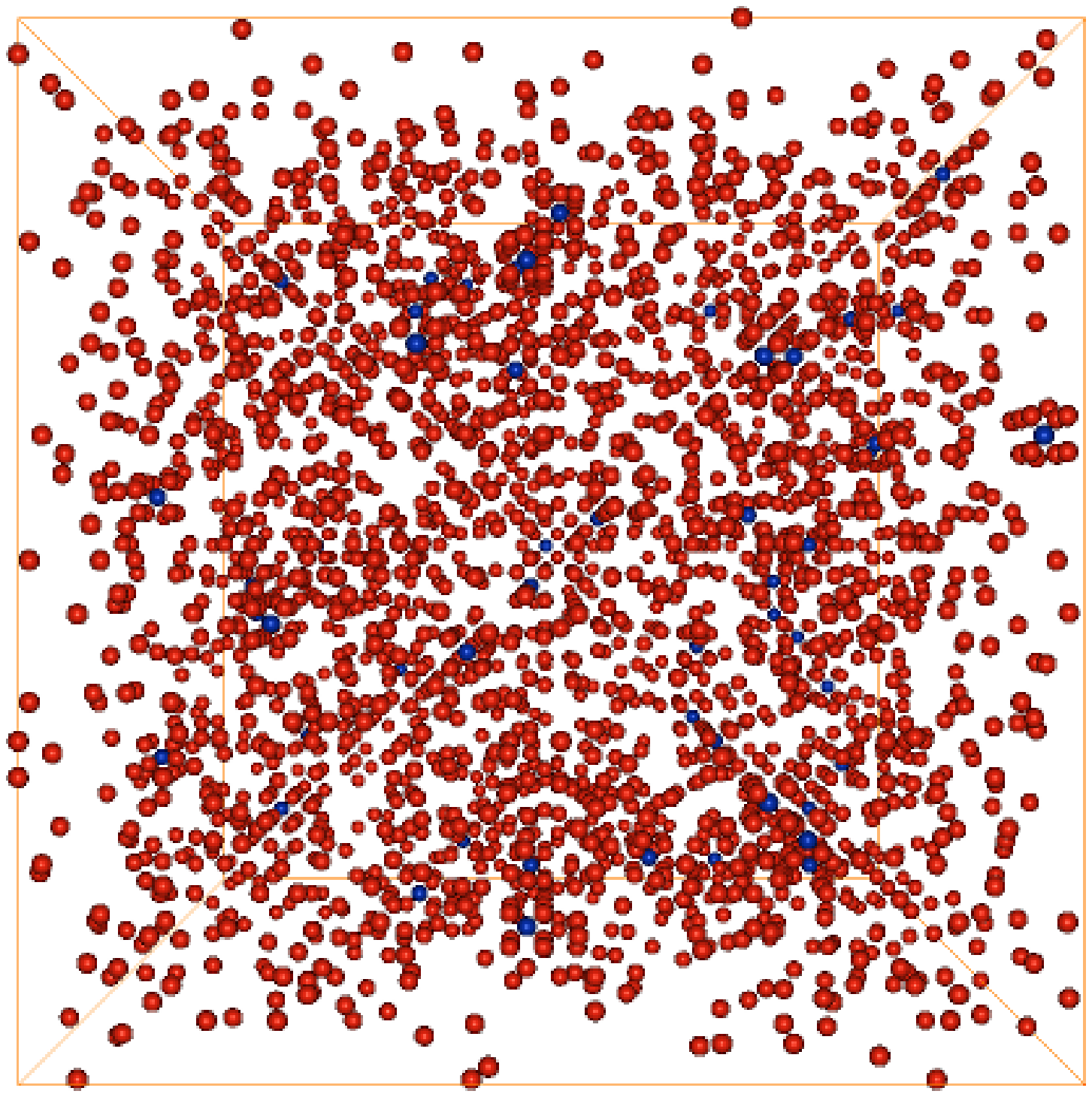} & \includegraphics[width=.2\linewidth]{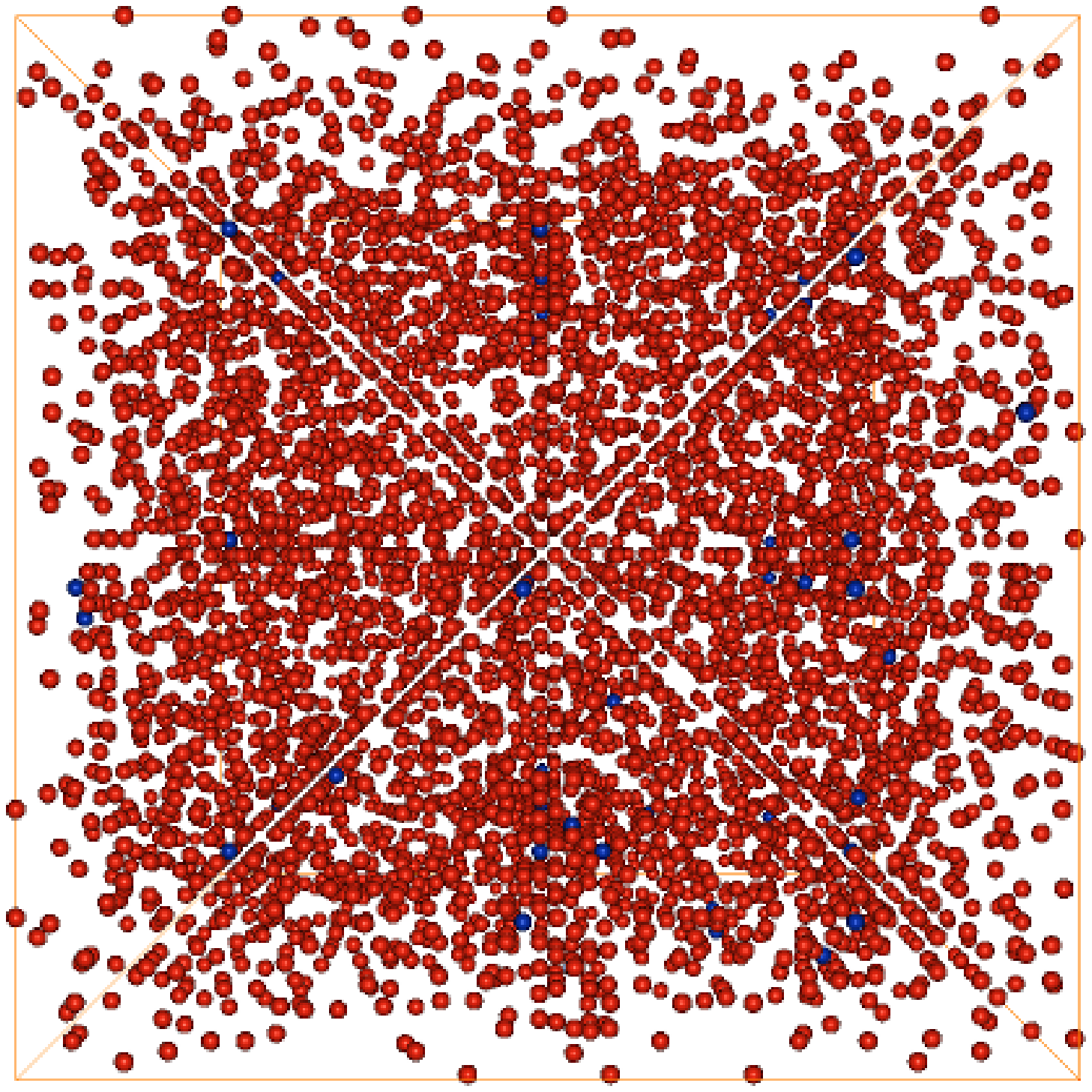} \\
     \hline
     800 K & \includegraphics[width=.2\linewidth]{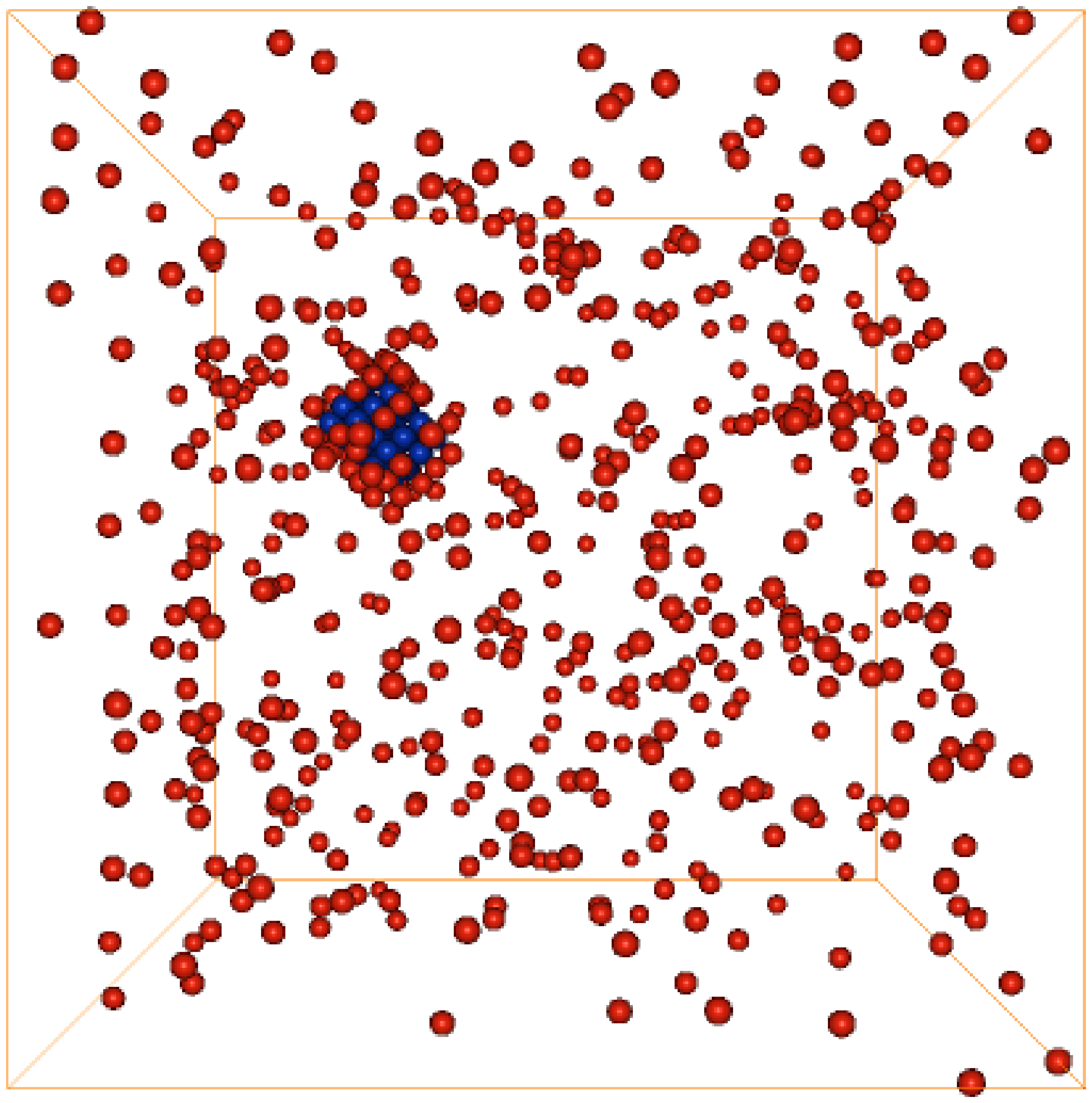} & \includegraphics[width=.2\linewidth]{WRe2Vac01_800K} &  \includegraphics[width=.2\linewidth]{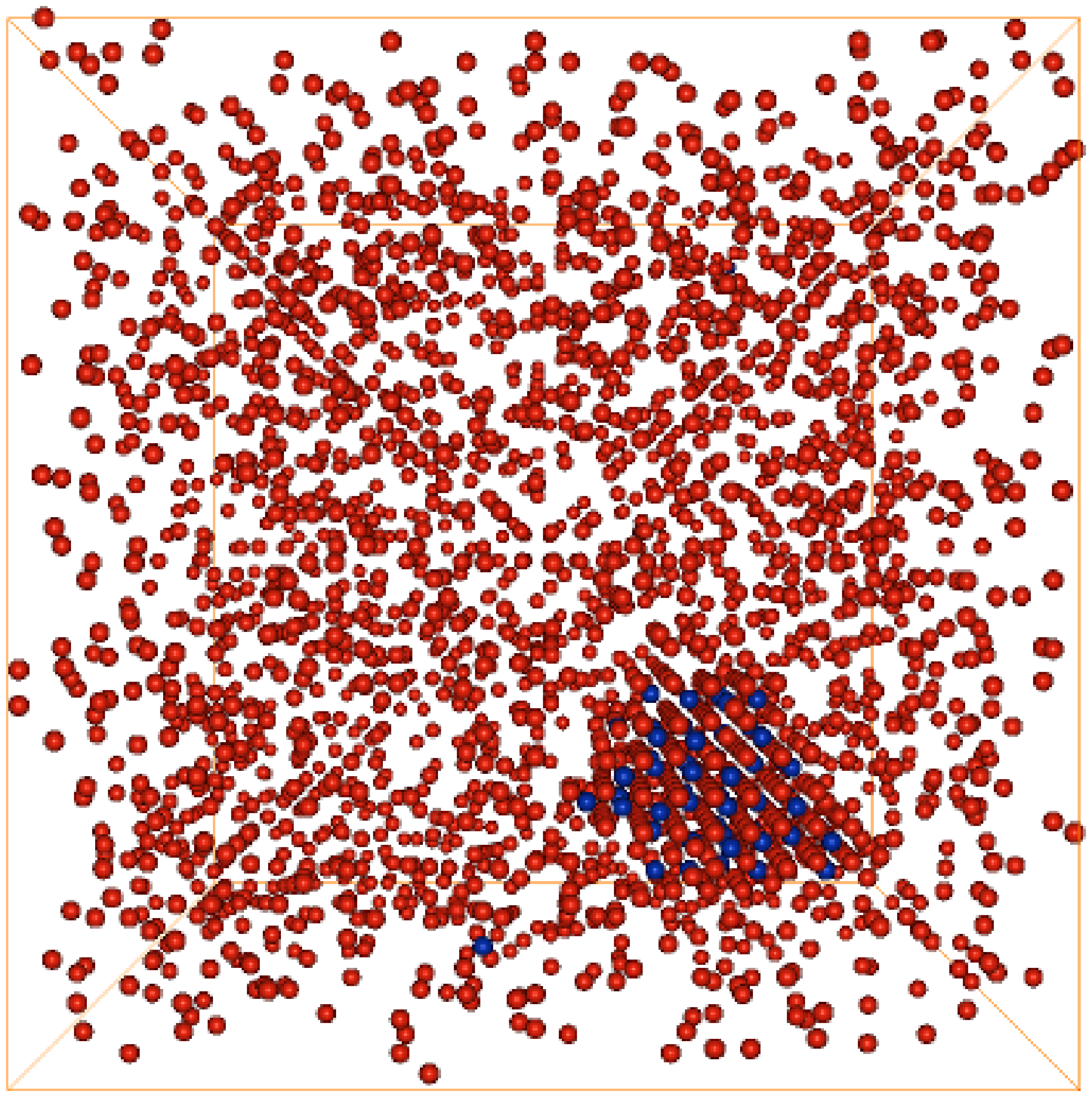}  & \includegraphics[width=.2\linewidth]{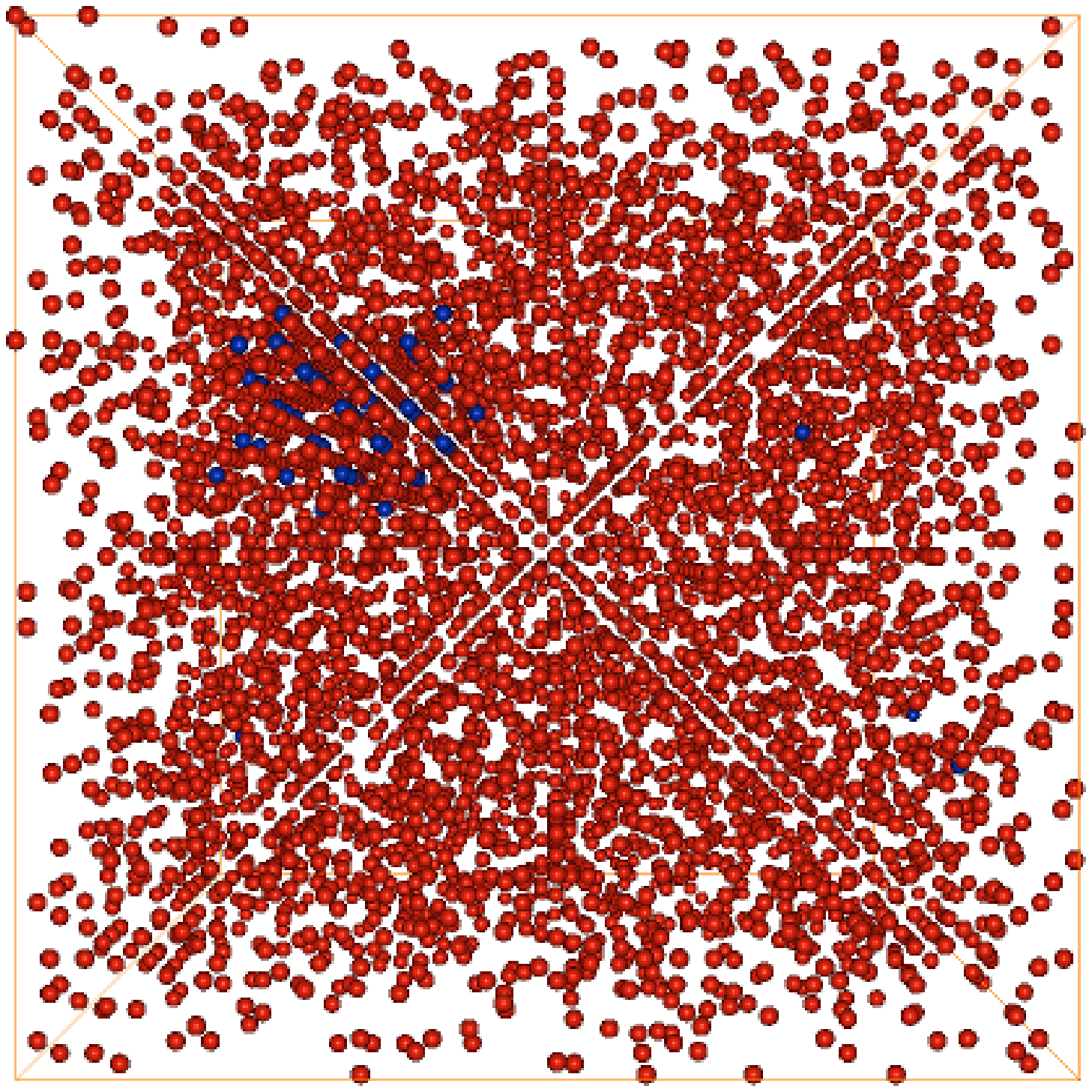} \\
     \hline
     1600 K & \includegraphics[width=.2\linewidth]{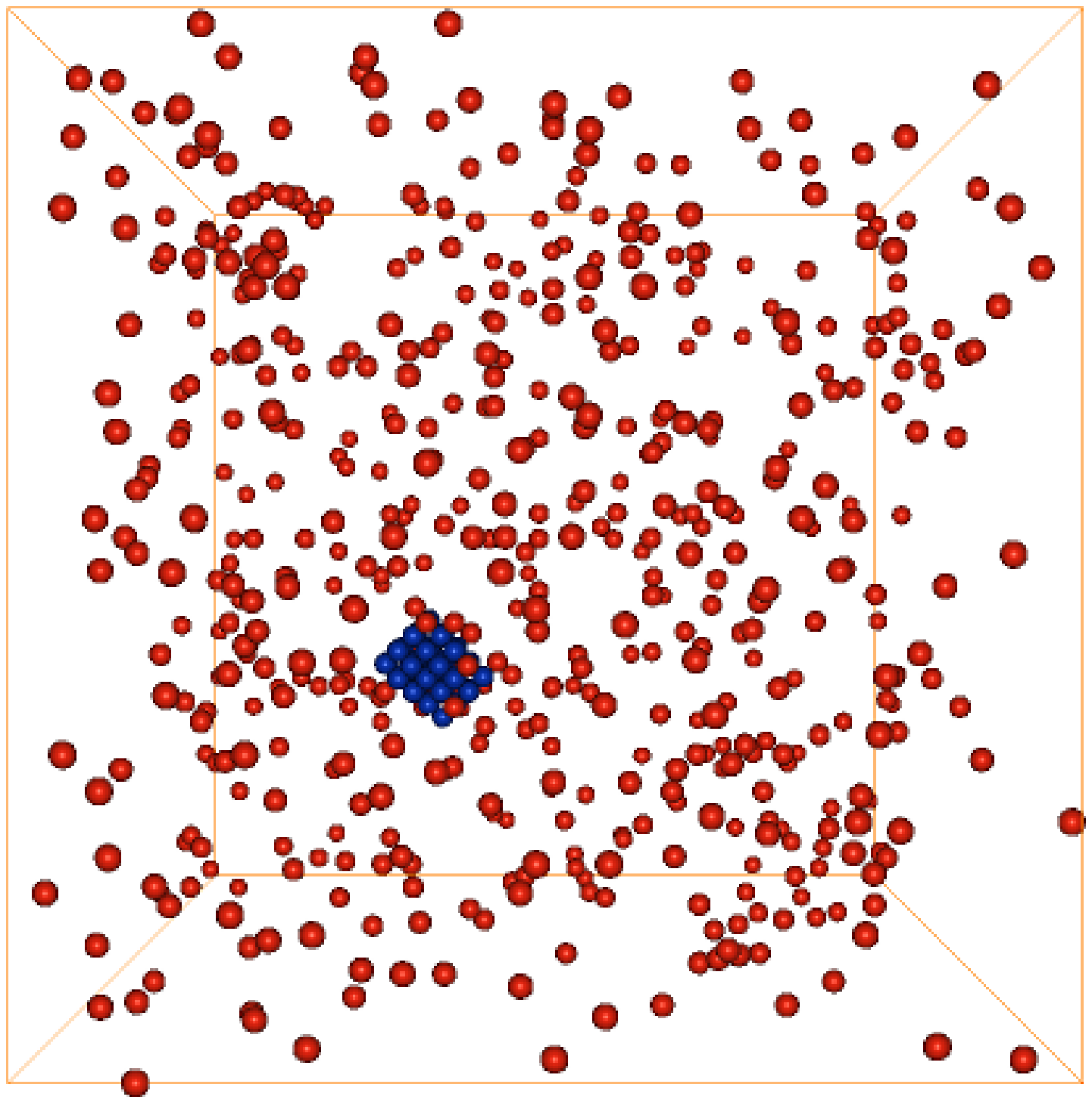} & \includegraphics[width=.2\linewidth]{WRe2Vac01_1600K} & \includegraphics[width=.2\linewidth]{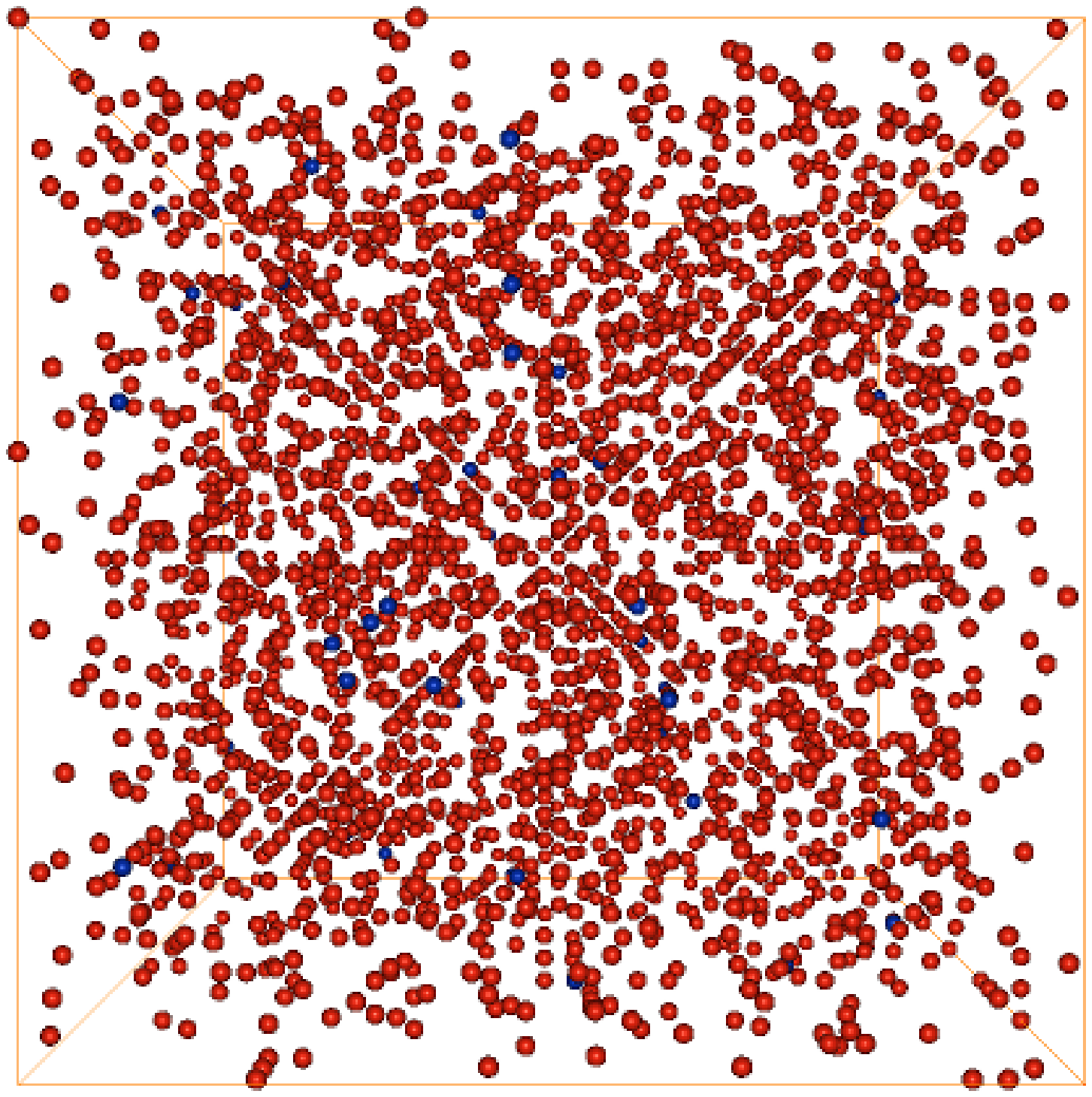} & \includegraphics[width=.2\linewidth]{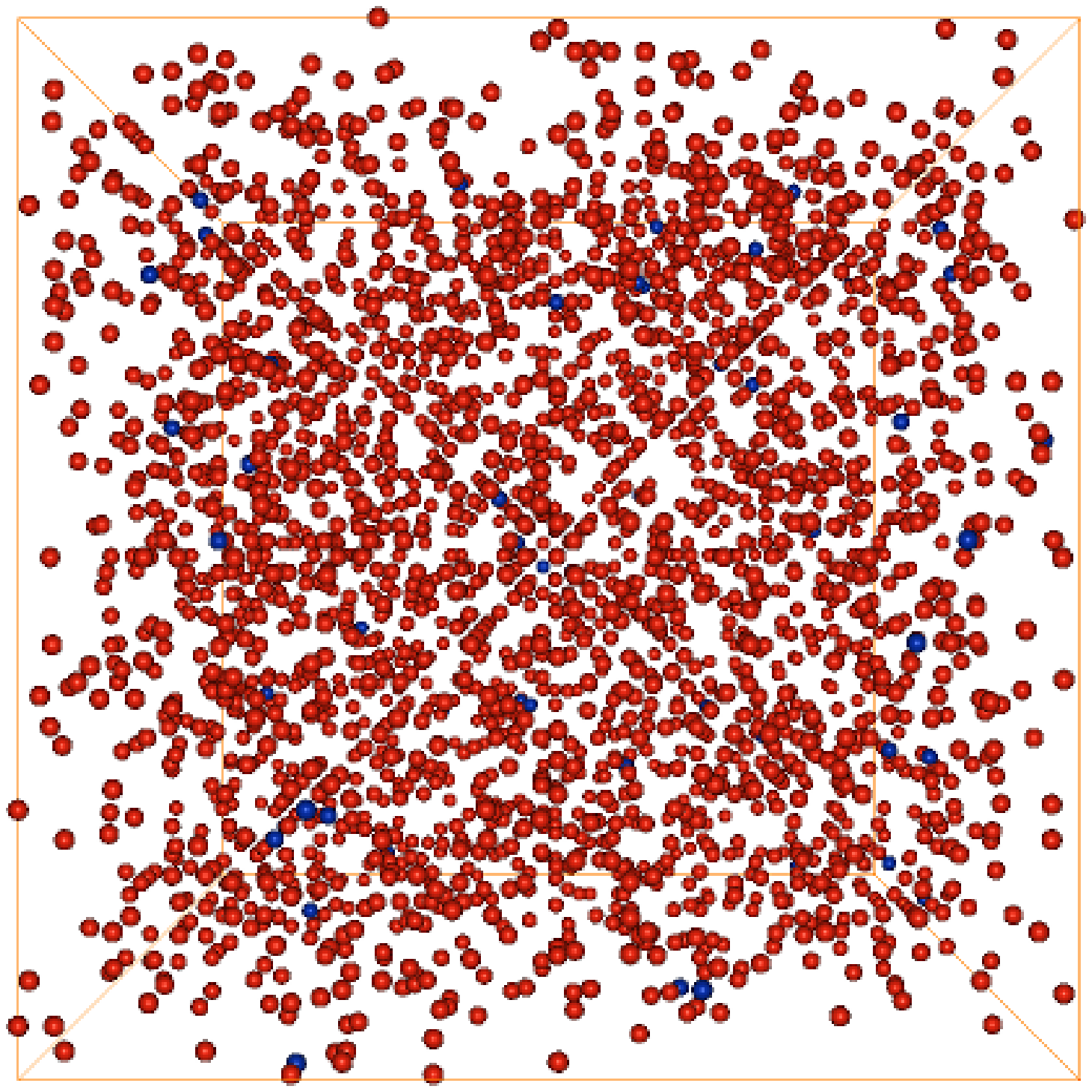} \\
     \hline
     2500 K & \includegraphics[width=.2\linewidth]{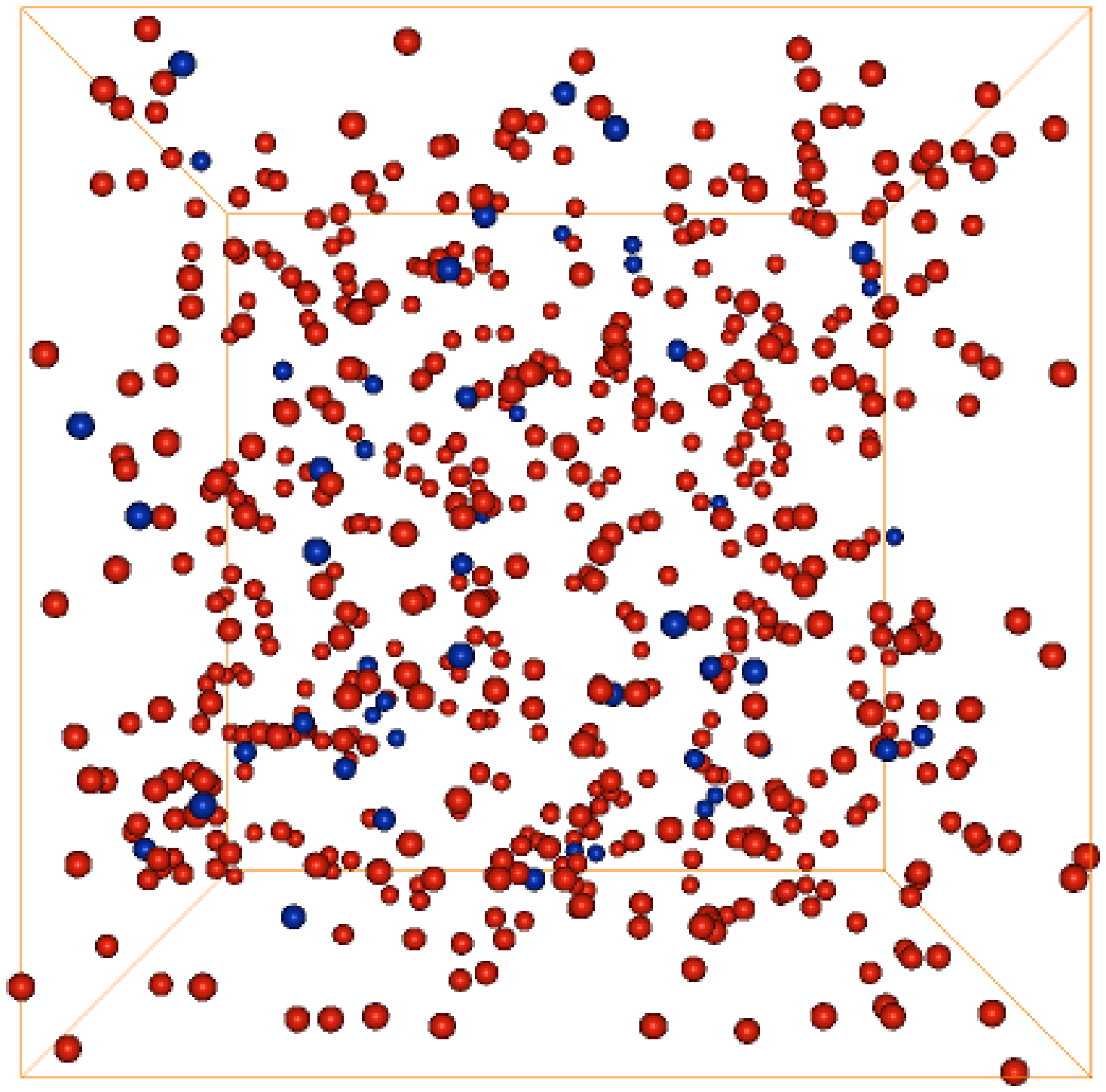} & \includegraphics[width=.2\linewidth]{WRe2Vac01_2500K} &  \includegraphics[width=.2\linewidth]{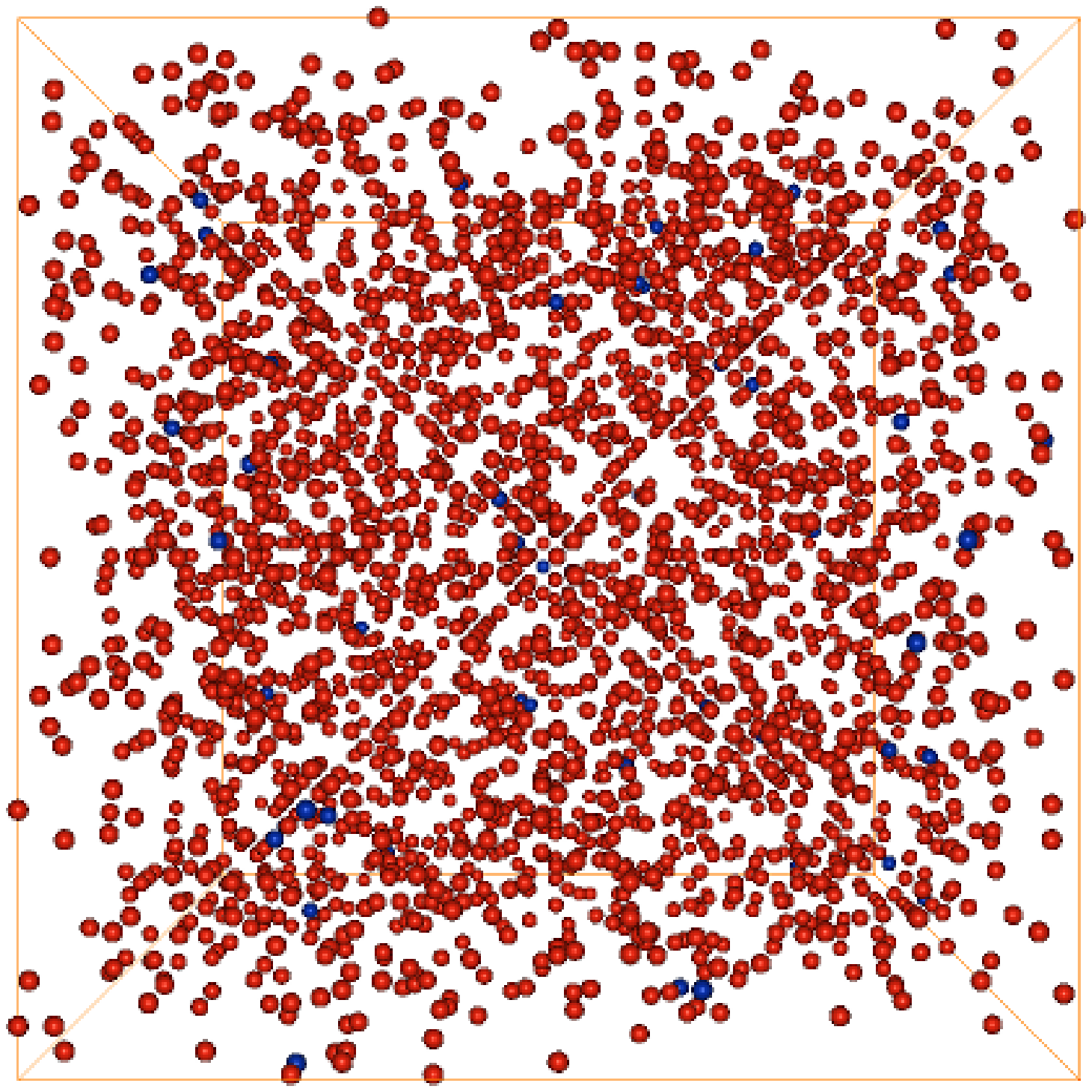} & \includegraphics[width=.2\linewidth]{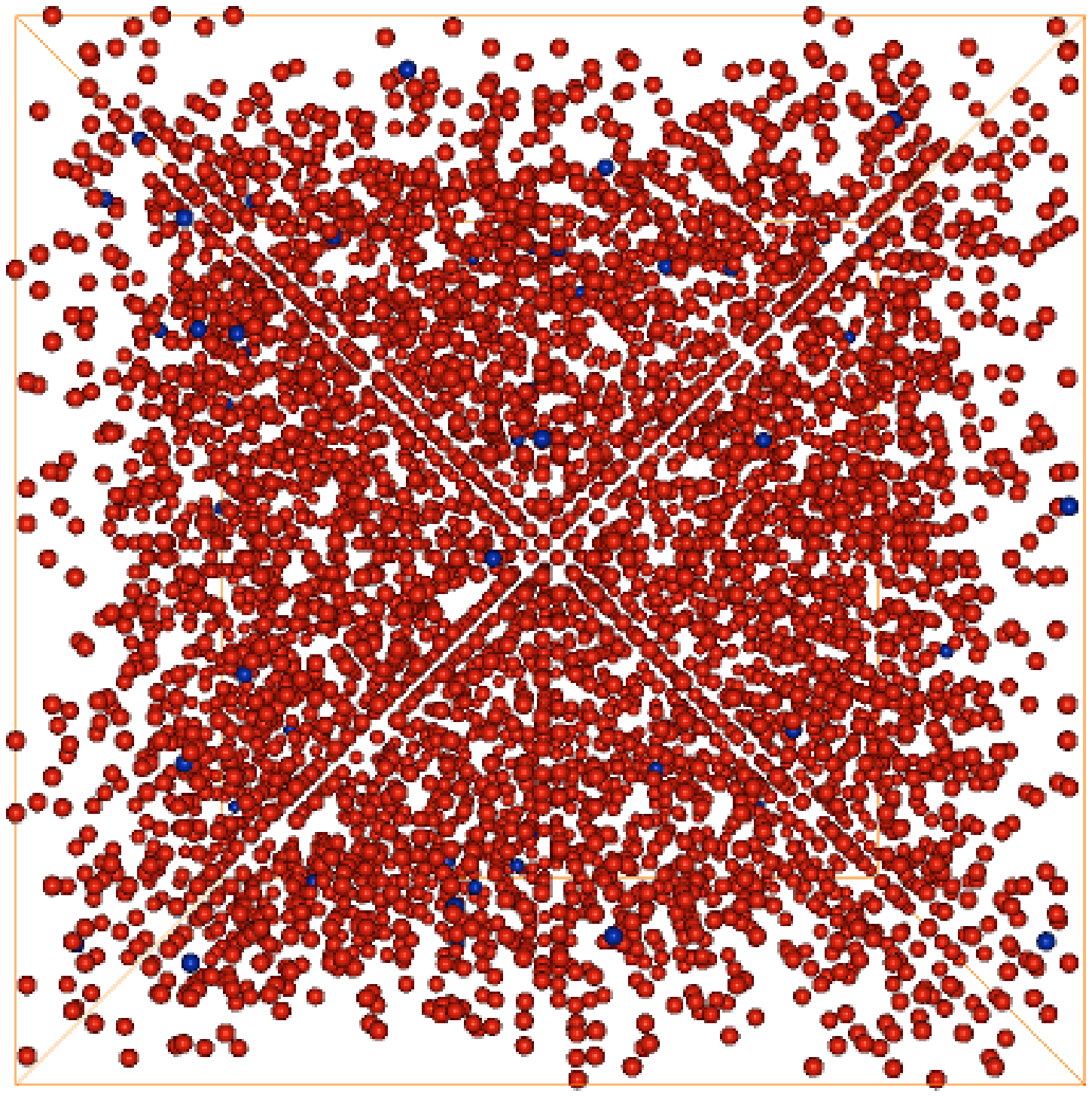} \\
    \end{tabular}%
\end{ruledtabular}
\end{table*}

\section{Discussion and Conclusion}

To rationalize DFT and MC simulations of ternary W-Re-Vac alloys and their application to interpretation of observations of 
radiation-induced precipitation of Re solute atoms in irradiated binary W-Re alloys, we analyzed the results of simulations in terms of
chemical pairwise interactions $V_n^{ij}$ derived from ECIs. Mapping of DFT data to CE was performed by retaining
only the effective interactions between different pairs of atoms within the 5NN range, corresponding to 15 two-body interactions.
The DFT-CE cross-validation error  is now 7.61 eV meV/atom, and although this value is higher than that obtained for Fig. \ref{fig:ECI_WReVac} and Table \ref{tab:ECI_ternary}, it is sufficient for interpreting the Re clustering phenomenon.
With Eq. \ref{eq:CE_expanded} expressed in terms of chemical pairwise interactions,
the configuration enthalpy of mixing of a ternary alloy is\cite{Wrobel2015}
\begin{widetext}
\begin{eqnarray}
\Delta H_{CE}(\vec{\sigma}) &=& J_1^{(0)}+J_1^{(1)}\left(1-3x_A\right) + J_1^{(2)}\frac{\sqrt{3}}{2}\left(x_C-x_B\right) \nonumber \\
&-&4\sum_{n}^{pairs} \left(V_n^{AB}y_n^{AB} + V_n^{AC}y_n^{AC}+V_n^{BC}y_n^{BC}\right) + \sum_{n}^{multibody} \ldots ,
\label{eq:CE_vs_V}
\end{eqnarray}
\end{widetext}
While in the binary alloy case chemical pairwise interactions have a simple meaning: $V_n^{AB}>0$ corresponds to attraction and $V_n^{AB}<0$ to repulsion between atoms $A$ and $B$, in the ternary case, Eq. \ref{eq:CE_vs_V}, they depend not only on $V_n^{AB}$, $V_n^{AC}$, $V_n^{BC}$ but also on the chemical short-range order values associated with the pair probability functions $y_n^{AB}$, $y_n^{AC}$, $y_n^{BC}$.

\begin{table*}
\caption{Effective interactions between different pairs of \textit{atoms}: W-Re ($V_n^{W-Re}$), W-vacancy ($V_n^{W-vac}$) and Re-vacancy ($V_n^{Re-vac}$) in the $n$th nearest-neighbour coordination shell of ternary W-Re-vac alloys, in meV units}
        \label{tab:Vij_ternary}
\begin{ruledtabular}
    \begin{tabular}{cccc}
    $n$     & $V_n^{W-Re}$  & $V_n^{W-vac}$ & $V_n^{Re-vac}$ \\
    \hline
    1     & 10.229 & -22.690 & 53.673 \\
    2     & -0.464 & 72.062 & 129.124 \\
    3     & 0.409 & -19.984 & -4.649 \\
    4     & -0.316 & -91.634 & -77.856 \\
    5     & -1.130 & 82.784 & 89.576 \\
    \end{tabular}
\end{ruledtabular}
\end{table*}

\begin{figure}
  \includegraphics[width=\linewidth]{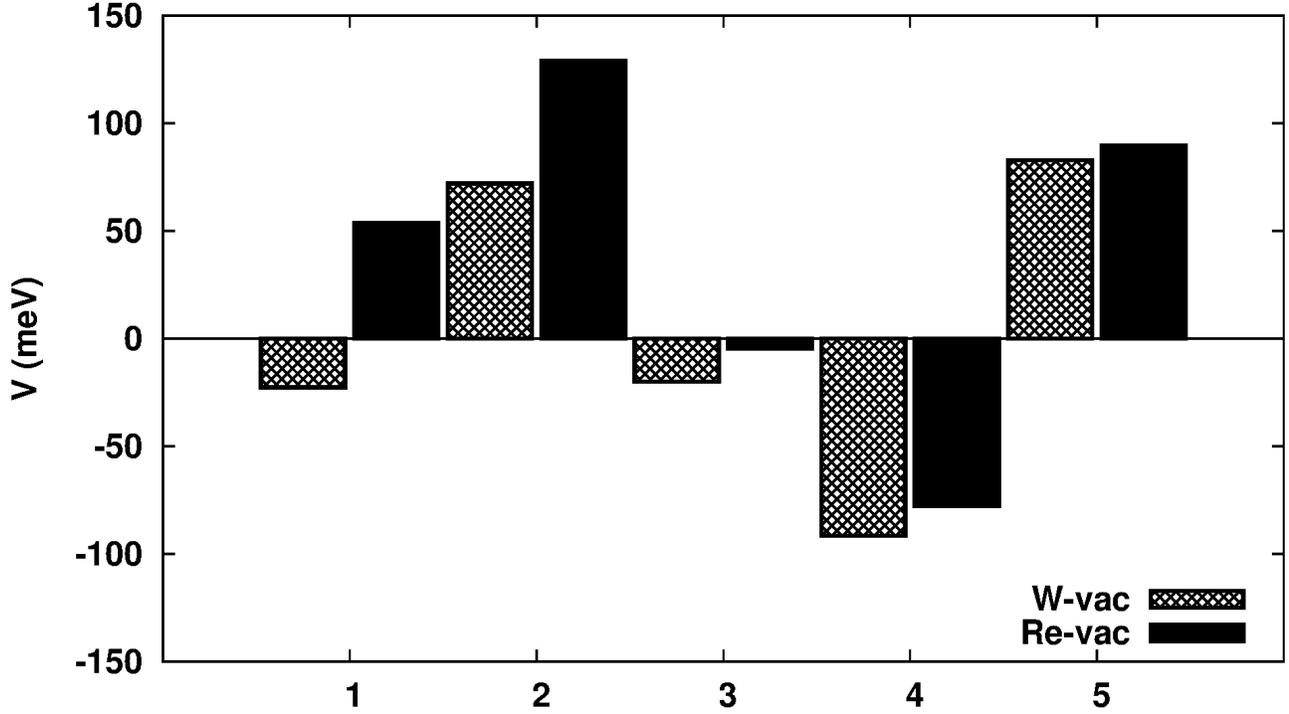}
                        \caption{
                (Color online) Effective interactions of vacancies with W and Re atoms in the
$n$th nearest-neighbour coordination shell in ternary W-Re-vac alloys, in meV units. }
                \label{fig:Vij_WReVac}
\end{figure}

Values of effective pair interactions involving W-Re, W-Vac and Re-Vac at distances up to the radius of the 5NN shell in W-Re-Vac alloy are summarized in Table \ref{tab:Vij_ternary}. Because of small mixing enthalpy values of W-Re pairs (see Fig. \ref{fig:Hmix_WRe}
and Table \ref{tab:Hmix_results}), effective pair interactions $V_n^{W-Re}$ have a negligible effect on the order-disorder phase transformation in ternary W-Re-Vac alloys. Comparison between W-Vac and Re-Vac effective pair interactions is given in Fig. \ref{fig:Vij_WReVac}. In the 1NN shell, the ECIs show that the W-Vac interaction is repulsive whereas Re-Vac pairs attract each other, in agreement with the analysis of binding energies given in Section III. ECIs corresponding to 2NN configurations
both have positive sign, showing that Re and W atoms can occupy 2NN positions around a vacancy. For the 3NN and 4NN shells,
the ECIs both have negative sign, meaning that neither Re or W atoms are energetically preferable in the third and fourth NN
coordination around the vacancy. As a result, this favours vacancies occupying positions in the third and fourth NN position, as illustrated in Fig. 3b for the divacancy configuration in the 3NN.

\begin{figure}
  \includegraphics[width=\linewidth]{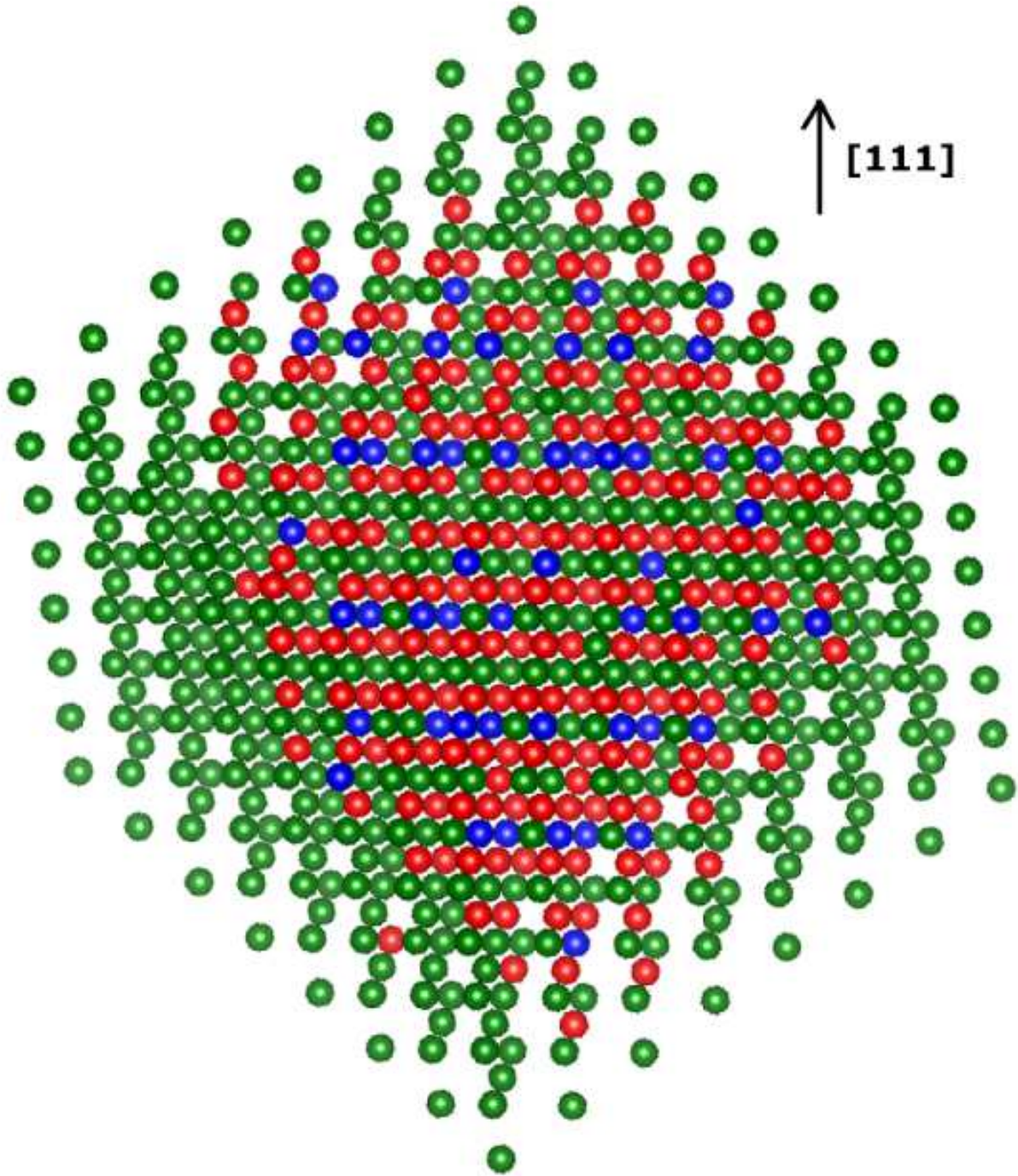}
                        \caption{
                (Color online) Re-Vac cluster obtained from MC simulations for W-1\%Re alloy with 0.1\% concentration of vacancies at 500 K. Green, red and blues spheres indicate W atoms, Re atoms and vacancies, respectively}
                \label{fig:WRe2Vac01_300K_cluster}
                \end{figure}

Fig. \ref{fig:WRe2Vac01_300K_cluster} shows a representative structure of W-Re-Vac cluster along a $[111]$ direction, obtained from MC simulations for W-Re2\%-Vac0.1\% alloy at T=300K (see Table \ref{tab:MC_results_const_Vac}). In all the clusters, vacancies tend to occupy 3NN positions, whereas Re atoms are usually in the 1NN shell with respect to vacancies, with W atoms usually located between the Re atoms. The estimated Re concentration within the cluster shown in Fig. \ref{fig:WRe2Vac01_300K_cluster} is as high as 41\%
with vacancies surrounded by Re atoms. At T=800K, the Re concentration inside the cluster in s W-2\%Re-0.1\%Vac alloy is close to 29$\%$, which is many times higher than the average concentration of Re in the alloy. We note that both the atomic structure and Re
concentration range associated with this cluster agree with the atomistic mechanism of phase transformation into the
the $\sigma$ phase \cite{Kitchingman1968} in the W-Re binary phase diagram \cite{Ekman2000}.

In comparison with experimental ATP observations \cite{Xu2015} of Re clustering in W-2\%Re alloys, induced by self-ion irradiation at 773K, our MC simulations performed for W-2\%Re-0.1\%Vac alloy at 800K show that the size of Re precipitates is close to 2.5 nm in diameter, which is in surprisingly good agreement with rhenium clusters of $\sim$3nm in diameter found in observations. The coherent nature of Re-clusters with matrix found using lattice resolution \cite{Xu2015} strongly supports the model proposed above, consistent with the view that clustering is associated with the presence of large, many orders of magnitude above equilibrium, concentration of vacancies, and relatively small elastic distortion of the lattice. Our prediction that precipitates form in undersaturated Re in W at high temperatures (up to 1600K), in the form of voids decorated by rhenium atoms, is not only consistent with TEM observations of the neutron-irradiated W-1.4Re alloy \cite{Klimenkov2015} but also with the observation of small ($\sim$ 2nm) voids forming above 1200$^{o}$C in ion-irradiated tungsten \cite{Ferroni2015}.

In summary, we developed a first-principles model for segregation and decomposition of under-saturated binary alloys under irradiation by treating it as a ternary alloy (A,B, vacancy) under constrained thermodynamic equilibrium at a fixed vacancy concentration. Using an extensive DFT database in combination with CE and quasi-canonical Monte-Carlo simulations, we applied the model to the investigation of Re precipitation in irradiated bcc Re-W alloys. Depending on Re/vacancy ratio, we found that Re precipitates formed at high temperature and at low Re concentrations have the form of small voids decorated with Re atoms. At higher Re concentrations, precipitates adopt a sponge-like structure, containing very high concentration of rhenium approaching 40at.$\%$Re. Predictions derived from the model, are in very good agreement with experimental observations of both neutron and self-ion irradiated tungsten-rhenium alloys. 

\begin{acknowledgments}
This project has received funding from the European Unions Horizon 2020 research and innovation programme under grant agreement
number 633053, and from the RCUK Energy Programme (Grant Number EP/I501045). JSW was funded by the Accelerated Metallurgy Project,
which is co-funded by the European Commission in the 7th Framework Programme (Contract NMP4-LA-2011-263206), by the European Space Agency and by the individual partner organizations. To obtain further information on the data and models underlying this paper please contact PublicationsManager@ukaea.uk. The views and opinions expressed herein do not necessarily reflect those of the European Commission.
DNM would like to acknowledge the International Fusion Energy Research Centre (IFERC) for providing access to a supercomputer (Helios) at Computational Simulation Centre (CSC) at Rokkasho (Japan).
\end{acknowledgments}

\bibliography{WRe_Vac_references_v1}

\end{document}